\documentclass[12pt,letterpaper]{article}

\usepackage[utf8x]{inputenc}
\usepackage[T1]{fontenc}
\usepackage{mathptmx} 
\usepackage{authblk}
\usepackage[pdftex]{graphicx} 
\usepackage[pdftex,linkcolor=black,pdfborder={0 0 0}]{hyperref} 
\usepackage{calc} 
\usepackage{enumitem} 
\usepackage{xcolor}
\usepackage{enumitem}
\usepackage{xurl}
\usepackage{booktabs}
\usepackage{float} 
\usepackage[most]{tcolorbox}
\usepackage[style=numeric-comp, backend=biber, sorting=none, bibstyle=nature, defernumbers=true]{biblatex}
\usepackage[all]{nowidow} 
\usepackage[protrusion=true,expansion=true,patch=none]{microtype} 
\usepackage[letterpaper, lmargin=0.1\paperwidth, rmargin=0.1\paperwidth, tmargin=0.1\paperheight, bmargin=0.1\paperheight]{geometry} 
\usepackage[font=small, labelfont=bf, aboveskip=5pt]{caption} 
\usepackage{makecell}
\usepackage{amssymb,amsmath}
\usepackage{bm}
\usepackage[bb=boondox,bbscaled=.95,cal=boondoxo]{mathalfa}
\usepackage{xspace}
\usepackage[detect-all]{siunitx}
\usepackage{setspace}
\usepackage[left]{lineno}

\usepackage{subcaption}
\usepackage[%
	sort&compress]
    {cleveref}
\Crefname{appendix}{Supplement}{Supplements}

\newcommand\ie{i.e.\xspace}

\newcommand{\X}{$\mathbb{X}$\xspace}

\DeclareUnicodeCharacter{2212}{-}

\newcommand{\Title}{Community-based fact-checking does not detectably reduce subsequent misleading posting by corrected users\xspace}

\begin{document}

\title{\singlespacing 
\textbf{\Title}}

\author[1,2]{Yuwei Chuai}
\author[3]{Thomas Renault}
\author[4]{Nicolas Pröllochs}
\author[1]{\\ Gabriele Lenzini}
\author[2,5,*]{Mohsen Mosleh}

\renewcommand\Affilfont{\small} 
\renewcommand\Authfont{\large}
\renewcommand\Authand{, }
\renewcommand\Authands{, }

\affil[1]{SnT, University of Luxembourg, Luxembourg, Luxembourg}
\affil[2]{Oxford Internet Institute, University of Oxford, Woodstock Road, Oxford, OX2 6GG}
\affil[3]{Université Paris-Saclay, RITM, 54 Boulevard Desgranges, 92330 Sceaux, France}
\affil[4]{Department of Business and Economics, JLU Giessen, Giessen, Germany}
\affil[5]{Sloan School of Management, Massachusetts Institute of Technology, 100 Main Street Cambridge, MA 02142}

\affil[*]{Correspondence to: mohsen.mosleh@oii.ox.ac.uk}

\date{\vspace{−5ex}}

\maketitle

\vspace{10ex}
\begin{center}
    \textit{This manuscript is a preprint and has not yet been peer reviewed.}\\
    \textit{First posted: Aug 27, 2026}\\
    \textit{This version: Nov 18, 2026}
\end{center}

\thispagestyle{empty}

\clearpage

\begin{abstract}
Community-based fact-checking can reduce the spread of annotated misleading posts, but whether it produces lasting behavioral change among corrected authors remains unclear. Here, we conduct a large-scale quasi-experimental study of the Community Notes system on \X (formerly Twitter), tracking four weeks of activity before and after note display for \num{19854} accounts and \num{57935} correction events (noted posts and matched controls), covering \num{11909591} original posts. Difference-in-differences estimates show no statistically significant overall changes in original-post activity or in the production of misleading posts following correction. However, accounts in the highest quartile of GPT-estimated pre-correction misleadingness increase their original-post activity by 10.2\%, with suggestive evidence of an 8.3\% increase in misleading-post production. Community notes can thus constrain individual misleading posts without improving the subsequent behavior of their authors, highlighting that correcting content and changing its producers are distinct objectives for platform governance.

\noindent \textbf{Keywords:} Community-based fact-checking, content moderation, misinformation producers, behavioral responses
\end{abstract}

\thispagestyle{empty}

\clearpage
\begin{center}
    \textbf{Significance statement}
\end{center}

\begin{quote}
    Social media platforms are rapidly adopting community-based fact-checking, in which users collectively annotate misleading posts. Such systems reduce the spread of individual posts, but whether they change the behavior of the accounts that produce them is unknown. Analyzing \num{11909591} posts from \num{19854} accounts on \X, we detect no overall reduction in posting or misleading-content production, while accounts posting more misleading content before correction subsequently produce more of it. These findings suggest that limiting the reach of misleading posts may need to be complemented by interventions addressing their continued production. Correcting posts and changing producers are distinct goals; platforms should evaluate community-based fact-checking on both.
\end{quote}

\clearpage
\section*{Introduction}

Social media platforms increasingly rely on community-based fact-checking to combat the spread of misleading content~\cite{allen2021scaling,pennycook2019fighting,chuai2026community,slaughter2025community,martel2024crowds,ecker2024misinformation,yasseri2023can}. Unlike traditional fact-checking, whose coverage is constrained by the capacity and speed of professional fact-checkers~\cite{chuai2025political}, community-based systems allow users to collaboratively identify potentially misleading posts, provide contextual information, and evaluate the helpfulness of one another's contributions. This participatory model may therefore enable fact-checking at a scale that is difficult to achieve through expert-based approaches alone~\cite{allen2021scaling,martel2024crowds}. The most prominent implementation is \X's Community Notes system, which displays a contextual annotation when contributors who have previously held different viewpoints broadly agree that it is helpful (for an overview of the system and note status, see \cite{wojcik2022birdwatch,cn2026introduction} and \Cref{supp:data_overview}). The model has since expanded beyond \X: Meta has introduced a similar Community Notes system on Facebook, Instagram, and Threads, while YouTube and TikTok are testing analogous note-based systems~\cite{meta2025testing,youtube2024testing,tiktok2025testing}. As community-based fact-checking becomes more widespread, understanding its effectiveness as a platform governance intervention is becoming increasingly important.

Existing research provides substantial evidence for the effectiveness of community-based fact-checking on annotated posts. Quasi-experimental studies show that displaying community notes substantially reduces subsequent reposting and other forms of engagement with annotated posts, alters their diffusion through social networks, and increases the likelihood that their authors delete them~\cite{chuai2024roll,chuai2026community,slaughter2025community}. Experimental evidence further indicates that contextual notes can increase trust in fact-checks, improve the identification of misleading content, and discourage users from resharing annotated posts~\cite{drolsbach2024community,wojcik2022birdwatch}. Together, these findings demonstrate that community-based fact-checking can mitigate the impact of individual misleading posts once notes become visible. Despite this significant promise, the system's reliance on cross-perspective agreement may limit the visibility of potentially helpful but polarizing notes, with only a small proportion of proposed contributions ultimately displayed to the public~\cite{bouchaud2026community}. Moreover, accumulating sufficient cross-perspective ratings to determine note helpfulness can delay note display and constrain the system's overall effectiveness~\cite{chuai2026community,zhang2026needs}.
Beyond these limitations, prior research has primarily focused on the corrected posts themselves. Yet, even deleting an annotated post does not necessarily imply that its author subsequently posts less or produces more reliable content. Whether community notes induce behavioral changes that extend beyond the corrected post therefore remains largely unknown.

Understanding the downstream effect of fact-checks from an author-level perspective is important because suppressing individual posts and changing the behavior of their producers are distinct objectives. Related research has examined downstream effects on audiences' belief accuracy, media trust, and generalized skepticism~\cite{bachmann2023studying,hoes2024prominent}, but not the subsequent content-production behavior of corrected authors. An intervention may successfully interrupt one diffusion cascade while having little influence on the account that generated it, especially if that account compensates by publishing additional content. Moreover, misinformation sharing is highly concentrated: a relatively small minority of accounts contributes disproportionately to the circulation of false or low-quality information online~\cite{guess2019less,grinberg2019fake,baribi2024supersharers,van2024broader}. If corrections make these accounts more cautious, even modest behavioral improvements could reduce a disproportionate amount of misleading content. Conversely, if they continue or increase their activity without improving content quality, strong post-level effects may overstate the broader effectiveness of the intervention. Evaluating community-based fact-checking therefore requires examining both how much corrected accounts subsequently post and what kinds of content they produce.

Research on other content-moderation interventions suggests that platform feedback can alter users' subsequent behavior, but not uniformly. Providing explanations for content removal and deleting rule-breaking content can reduce subsequent violations~\cite{jhaver2019does,horta2023automated}, whereas temporary user bans can increase content production while reducing the quality of posted content~\cite{zhang2026social}. Thus, there are competing expectations regarding the author-level downstream effects of community notes. From a \textit{corrective-learning and deterrence} perspective, community notes provide direct feedback about why a post may be misleading and link to supporting evidence, potentially increasing authors' awareness of factual or contextual problems. Because notes are displayed publicly, they may also impose reputational costs: annotated posts become visibly contested and may attract critical or morally outraged responses from other users~\cite{chuai2025community}. Such informational and reputational signals could prompt authors to reassess their posting practices, reduce their posting activity, or avoid similarly misleading claims in the future. Consistent with this possibility, prior research on professional fact-checking finds that making corrective scrutiny salient can discourage inaccurate statements by political elites and reduce users' subsequent circulation of misinformation~\cite{nyhan2015effect}. Publicly displayed community notes also increase and accelerate authors' voluntary retraction of corrected posts~\cite{chuai2026community}. Under this account, corrected authors should reduce their subsequent posting activity or improve the quality of their content. Alternatively, public corrections may produce \textit{behavioral resistance or counter-responses}. Authors may discount a note, dispute its legitimacy, or experience the correction as a reputational or identity threat. Rather than directing attention toward accuracy, being publicly corrected may evoke defensiveness and shift attention toward social or partisan considerations. A field experiment on \X, for example, finds that directly correcting users who had shared false political news subsequently reduces the quality and increases the partisan slant and toxicity of their reposts, although it does not affect their original posts~\cite{mosleh2021perverse}.

Here, we conduct a large-scale quasi-experimental study of whether displaying community notes (\ie, corrections) changes authors' subsequent original-post activity and content on \X. We track \num{19854} accounts for four weeks before and after each (pseudo-)correction they received, covering \num{11909591} original posts. Using difference-in-differences (DiD) designs, we examine changes in both posting activity and content and assess how these responses differ across author characteristics, correction timing, and pre-correction posting behaviors. 
We find no statistically significant overall changes in the number of original posts or in the production of GPT-classified misleading posts. However, accounts with higher pre-correction misleadingness increase their posting activity and produce more misleading posts following correction.
Thus, community notes do not detectably improve authors' subsequent behavior and may even be associated with counterproductive responses among certain accounts.
Our findings demonstrate that evaluating community-based fact-checking requires considering not only its effects on individual corrected misleading posts but also its downstream effects on the accounts behind them. Correcting misleading content and improving producer behavior represent distinct objectives for platform governance.

\clearpage
\section*{Results}
\subsection*{No evidence of overall changes in posting activity following community corrections}

We start by estimating the overall changes in authors' posting activity following the display of community notes. Our dataset comprises \num{29049} corrected posts (with displayed notes) and \num{28886} matched controls (with non-displayed notes) published by \num{19854} accounts over nearly two years (see \nameref{sec:methods} and \Cref{tab:data_overview}, \Cref{supp:data_overview}). For each corrected post, we track the author's original-post activity (\ie, the number of original posts, excluding replies and reposts) for four weeks before and after note display, which defines the correction event. We track the author of the matched control post over an equivalent observation window around a pseudo-correction time, calculated by adding the interval between publication and first note display for the paired treated post to the control post's publication time. In total, our dataset contains \num{11909591} original posts, corresponding to \num{3244360} daily and \num{463480} weekly count observations.

Because accounts can receive multiple displayed community notes on different posts, the same account may contribute multiple correction events to the analysis. In our dataset, 64.1\% of corrected accounts had only one post annotated by community notes, whereas the remaining 35.9\% accounted for 76.2\% of all correction events (see \Cref{supp:note_frequency}). To reduce potential interference from prior corrections, we initially restrict the analysis to focal events without previously observed corrections, which constitute 34.2\% and 45.7\% of treatment and control events, respectively (see \Cref{supp:prior_corrections,supp:overall_changes} for analysis of correction history). Because 13.8\% of eligible accounts also received additional corrections within the subsequent four weeks (see \Cref{supp:correction_recurrence}), we consider two specifications: (i) excluding first-correction events followed by additional corrections during the post-treatment period and (ii) retaining these events to assess the sensitivity of our estimates to subsequent correction exposure.
We estimate both leads-and-lags and aggregated pre-post DiD models using Poisson pseudo-maximum likelihood regression with correction-event and time fixed effects to accommodate the count nature of the outcome~\cite{correia2020fast}. We transform the DiD coefficients as $e^{\beta}-1$ to express proportional changes in original-post counts relative to matched control events (see \nameref{sec:methods}).

\begin{figure}[ht]
    \centering
    \captionsetup[subfloat]{font={bf, small}, skip=0pt, singlelinecheck=false, labelformat=simple, position=top}
    \subfloat[]{\includegraphics[width = \textwidth]{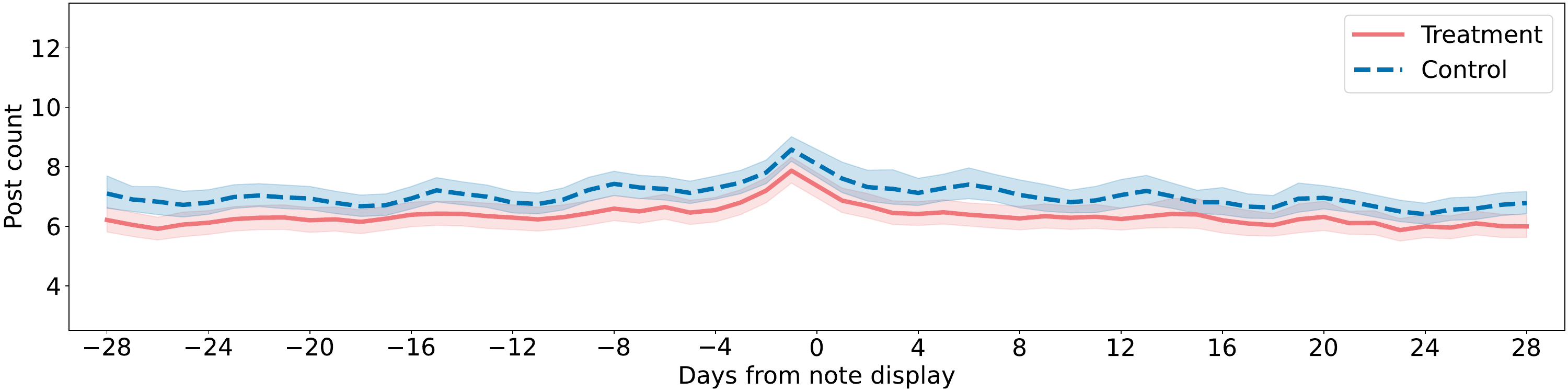}}
    \\
    \subfloat[]{\includegraphics[width = .32\textwidth]{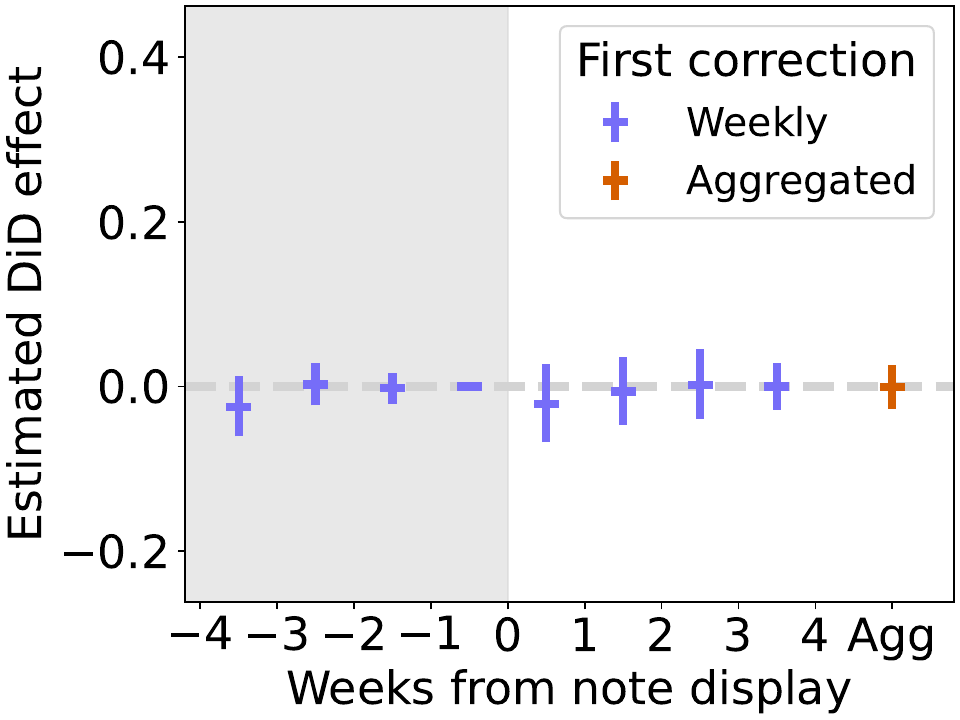}}
    \hfill
    \subfloat[]{\includegraphics[width = .32\textwidth]{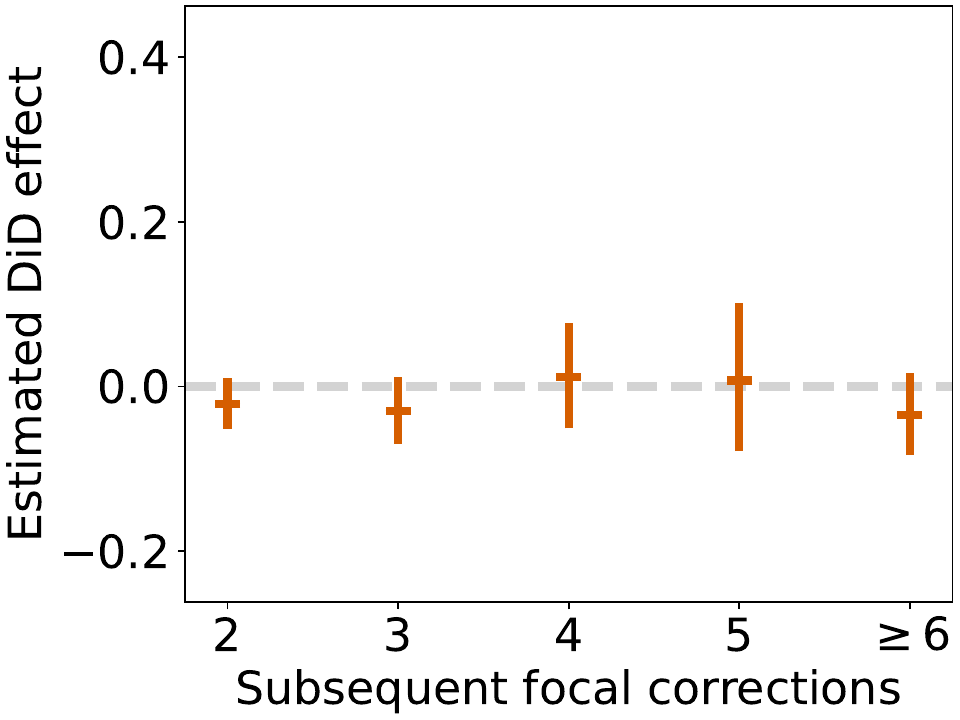}}
    \hfill
    \subfloat[]{\includegraphics[width = .32\textwidth]{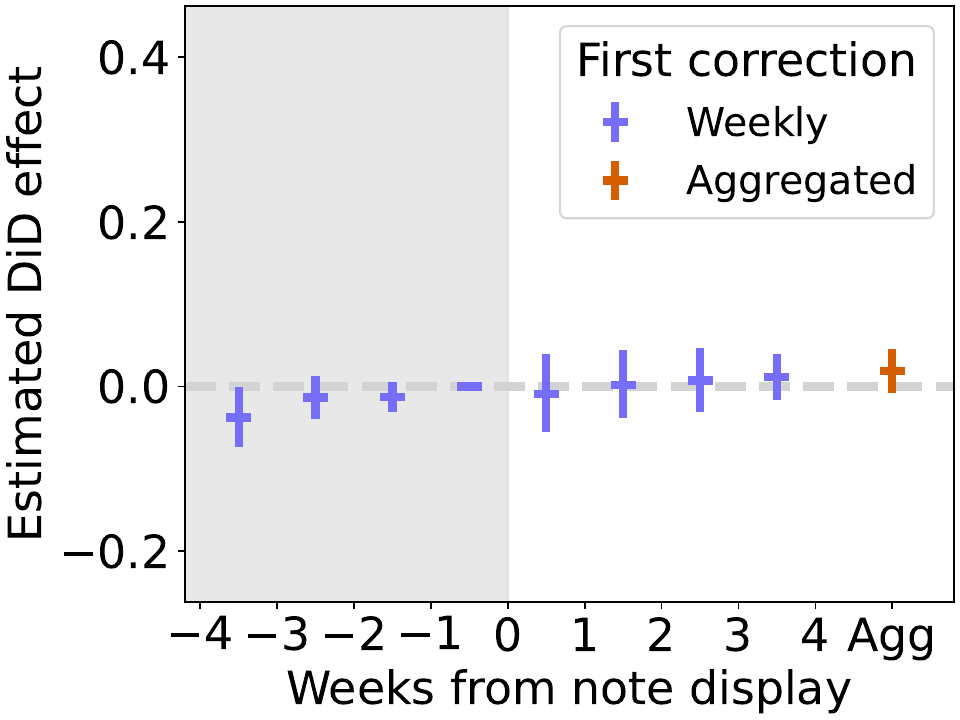}}
    \caption{\textbf{Limited overall changes in original-post activity following community corrections.} 
    \textbf{(a)}~Mean daily original-post counts for treatment and control groups over four weeks before and after the (pseudo-)correction event (\num{1169392} observations from \num{17696} accounts).  
    \textbf{(b)}~Weekly and aggregated DiD estimates of the effects of the first correction (\num{167056} observations from \num{17696} accounts).
    In (a) and (b), treatment events are restricted to accounts' first corrections, excluding those with subsequent corrections in the four-week post-treatment period. Control events include all pseudo-corrections without prior corrections (\Cref{supp:main_estimation}).
    \textbf{(c)}~Aggregated DiD estimates of the effects of subsequent focal corrections, stratified by the number of prior corrections, excluding subsequent corrections during follow-up and using control events with the same prior correction count (\Cref{supp:subsequent_corrections}).
    \textbf{(d)}~Weekly and aggregated DiD estimates of the effects of the first correction, including events with subsequent corrections during follow-up (176,552 observations from 18,627 accounts). Control events remain free of correction exposure during the corresponding observation window (see \Cref{supp:cumulative_corrections}).
    Shaded bands in (a) and error bars in (b)--(d) indicate 95\% confidence intervals (CIs); DiD standard errors are clustered at the account level.}
    \label{fig:did_main}
\end{figure}

We first examine daily original-post activity over four weeks before and after the first correction, excluding events followed by additional corrections during the post-correction observation window. As shown in \Cref{fig:did_main}a, the treatment and control groups exhibit similar posting trajectories before and after the first correction. Notably, the increase preceding note display may partly reflect how correction events are defined: a note can only be displayed after the author has published the associated post. Aligning observations around note display therefore selects periods following a posting event. The daily DiD estimates nevertheless provide no statistically significant evidence of differential pre-correction trends or an overall change in posting activity following correction (see details in \Cref{supp:main_estimation}). In the main analysis, we aggregate posting activity at the weekly level to reduce day-to-day fluctuations and account for weekly posting patterns (\Cref{fig:did_main}b). Using Week~$-1$ as the reference period, the DiD estimates for both the pre-correction period (Weeks~$-4$ to~$-2$) and the post-correction period (Weeks~$1$ to~$4$) are statistically indistinguishable from zero (all $p>0.100$; \Cref{supp:main_estimation}). The aggregated four-week effect is also close to zero and statistically insignificant (effect $=-0.001$, $z=-0.057$, $p=0.954$; 95\%~CI: $[-0.027, 0.026]$). Together, the daily and weekly analyses provide no statistically significant evidence of an overall change in original-post activity following the first correction.

We conduct several robustness and sensitivity analyses to assess the stability of these findings. The results remain consistent when applying synthetic difference-in-differences (SDiD) estimation~\cite{arkhangelsky2021synthetic} (\Cref{supp:main_estimation}), allowing for deviations from parallel trends using HonestDiD~\cite{rambachan2023honest} (\Cref{supp:honest_did}), and using alternative count regression specifications, including mixed-effects negative binomial, fixed-effects negative binomial, and conditional fixed-effects Poisson models (\Cref{supp:alternative_models}). We also obtain consistent results when controlling for note helpfulness scores (\Cref{supp:note_score_control}). We additionally examine whether posting responses differ across subsequent correction events, stratified by the number of corrections an account had previously received. The aggregated DiD estimates for the second through fifth corrections, as well as for the sixth and subsequent corrections combined, are all statistically indistinguishable from zero (all $p>0.100$; see \Cref{fig:did_main}c and \Cref{supp:subsequent_corrections}). These findings indicate that we do not detect significant changes in posting activity following subsequent corrections, consistent with the limited overall behavioral response observed after the first correction.

Finally, we assess whether retaining first-correction events followed by additional corrections during the four-week post-treatment period alters our findings. In this specification, we allow subsequent corrections in the treatment group while retaining only control events without correction exposure during the corresponding observation window. At the weekly level, the DiD estimate for Week~$-4$ is slightly negative and statistically significant (effect $= -0.038$, $z=-1.984$, $p=0.047$; 95\%~CI: $[-0.074, -0.000]$; \Cref{fig:did_main}d). However, the joint test of the three pre-correction coefficients is not statistically significant ($\chi^2(3)=5.25$, $p=0.154$). None of the four post-correction weekly estimates is statistically significant (all $p>0.050$), consistent with the findings from the more restrictive sample. The results are also consistent under SDiD estimation (\Cref{supp:cumulative_corrections}). We therefore retain all eligible first observed corrections, including those followed by additional corrections, in our subsequent analyses of posting activity and content production.

\clearpage

\clearpage
\subsection*{Heterogeneous posting responses by pre-correction misleadingness}

\begin{figure}[ht]
    \centering
    \captionsetup[subfloat]{font={bf, small}, skip=0pt, singlelinecheck=false, labelformat=simple, position=top}
    \subfloat[]{\includegraphics[width = .64\textwidth]{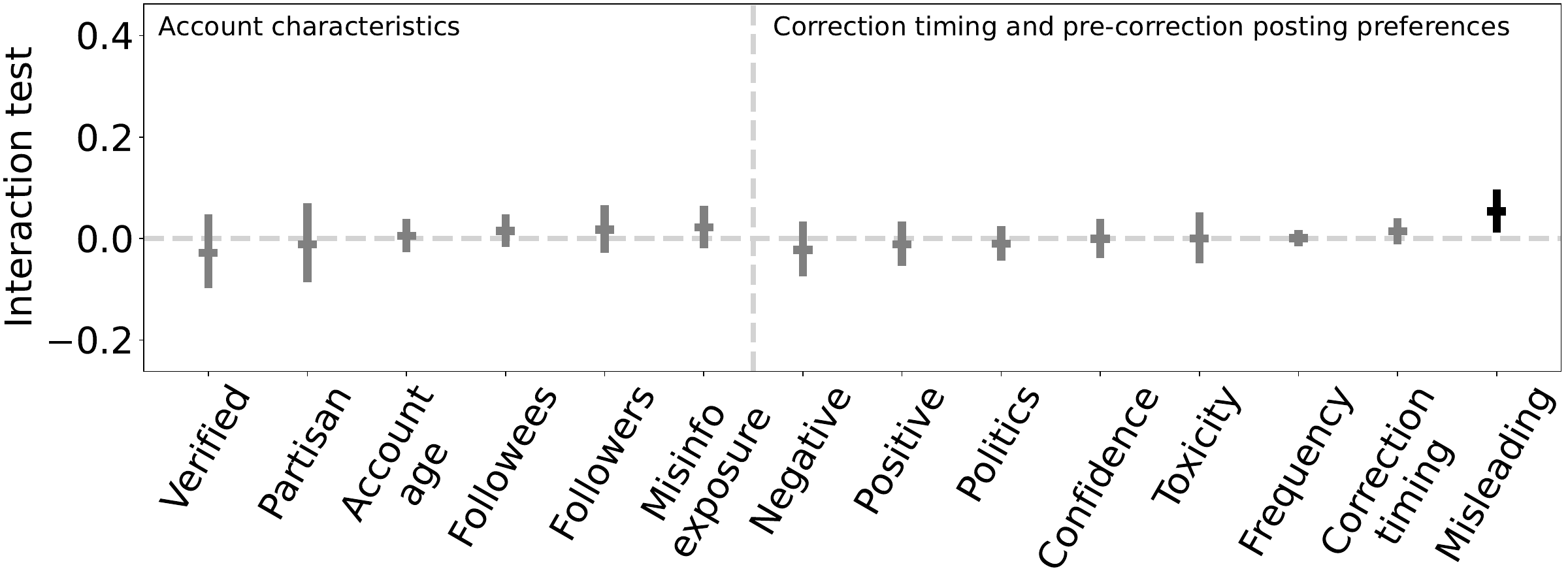}}
    \hfill
    \subfloat[]{\includegraphics[width = .32\textwidth]{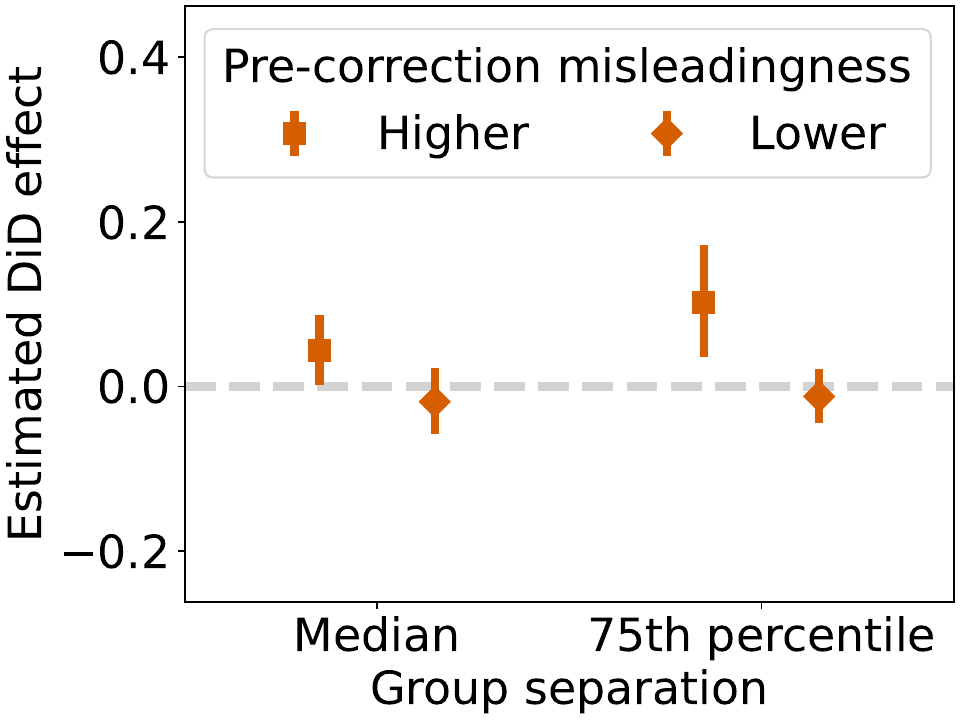}}
    \caption{\textbf{Accounts with high pre-correction GPT-estimated misleadingness increased original-post activity following their first correction.} 
    \textbf{(a)}~Joint DiD interaction estimates for author characteristics, correction timing, and pre-correction posting preferences. The estimation is based on 69,552 count observations from 7,316 accounts (see \Cref{supp:did_interaction_features}). The sample is restricted to accounts with original posts available for estimating misleadingness during the four-day pre-correction period.
    \textbf{(b)}~Subgroup analysis of accounts with higher versus lower pre-correction GPT-estimated misleadingness, using the median and 75th percentile thresholds, respectively.
    Error bars represent 95\% CIs, with standard errors clustered at the account level.}
    \label{fig:did_heterogeneity}
\end{figure}

Although we find no statistically significant overall change in original-post activity following community corrections, this average may mask heterogeneous responses across accounts with different pre-correction characteristics. We therefore examine whether posting responses vary with account characteristics, correction timing, and posting patterns measured before the first correction (see \nameref{sec:methods}). Specifically, we consider account characteristics (followers, followees, verified status, account age, political leaning, and misinformation exposure), correction timing (days from the roll-out of the Community Notes program to correction), and pre-correction posting characteristics (posting frequency, positive and negative sentiments, toxicity, GPT-estimated misleadingness, opinion confidence, and political content). Pre-correction misleadingness is measured using original posts published during the four days preceding the correction event, whereas the other posting characteristics are averaged over the four-week pre-correction period. For both treatment and control accounts, these measures exclude the focal post associated with the correction event. We then conduct a joint interaction analysis to assess whether these characteristics are associated with different posting responses while accounting for the other observed moderators. \Cref{fig:did_heterogeneity}a shows that pre-correction misleadingness is positively associated with subsequent posting responses (effect $=0.053$, $z=2.526$, $p=0.012$; 95\%~CI: $[0.012, 0.097]$), whereas none of the remaining characteristics exhibits a statistically significant interaction (all $p>0.100$; see \Cref{supp:did_interaction_features}). This suggests that accounts with higher pre-correction misleadingness exhibit a stronger increase in posting activity following their first correction. Specifically, holding the other characteristics constant, a one-standard-deviation increase in pre-correction misleadingness is associated with a 5.3\% larger post-correction change in expected original-post activity relative to control events.

To further characterize this heterogeneity, we separate accounts into higher- and lower-misleadingness groups based on their average pre-correction GPT-estimated misleadingness scores. Using the median as the threshold, accounts with higher pre-correction misleadingness publish 4.4\% more posts following their first correction (effect $= 0.044$, $z=2.074$, $p=0.038$; 95\%~CI: $[0.002, 0.087]$). In contrast, accounts with lower pre-correction misleadingness show no statistically significant change (effect $= -0.019$, $z=-0.896$, $p=0.370$; 95\%~CI: $[-0.058, 0.023]$). Using the 75th percentile as a more restrictive threshold, accounts in the highest pre-correction misleadingness quartile exhibit a more pronounced change and post 10.2\% more following their first correction (effect $= 0.102$, $z=3.083$, $p=0.002$; 95\%~CI: $[0.036, 0.172]$), while the remaining accounts show no statistically significant change (effect $= -0.012$, $z=-0.716$, $p=0.474$; 95\%~CI: $[-0.045, 0.021]$). Given the more pronounced response among accounts in the highest misleadingness quartile, we again use the joint interaction model to examine whether this heterogeneity persists after accounting for other observed characteristics. Specifically, we replace the continuous pre-correction misleadingness measure with a binary indicator for its highest quartile and find a positive and statistically significant interaction with correction exposure (effect $=0.137$, $z=3.161$, $p=0.002$; 95\%~CI: $[0.050, 0.232]$). This interaction remains statistically significant after both Holm and Benjamini-Hochberg adjustments across all 14 interaction terms (adjusted $p=0.022$ for both), while the other interaction estimates remain statistically indistinguishable from zero (see \Cref{supp:did_interaction_features}).
Additionally, we use the SDiD estimator to evaluate the sensitivity of these subgroup estimates. The estimated effect for accounts with above-median pre-correction misleadingness is positive but not statistically significant (ATT $=1.320$ original posts per week, $t=1.77$, $p=0.077$; 95\%~CI: $[-0.145, 2.785]$). For accounts in the highest pre-correction misleadingness quartile, the SDiD estimate is positive and statistically significant (ATT $=2.505$ original posts per week, $t=2.49$, $p=0.013$; 95\%~CI: $[0.535, 4.475]$; see \Cref{supp:did_interaction_features}). Together, these findings suggest that the increase in original-post activity following community corrections is concentrated among accounts with high pre-correction misleadingness.

As a complementary heterogeneity analysis, we examine whether posting responses differ according to accounts' prior NMR-note history, measured by the number of distinct posts receiving community notes that remained in the Needs More Ratings (NMR) status before the first displayed correction (see \Cref{supp:he_prior_nmr}). Accounts with NMR notes on at least two distinct posts during the preceding four weeks (\ie, repeated prior NMR notes) exhibit a statistically significant increase in posting activity following correction (effect $= 0.090$, $z=3.675$, $p<0.001$; 95\%~CI: $[0.041, 0.141]$), whereas accounts in the complementary group show no significant change. The interaction between correction exposure and receiving repeated prior NMR notes is also statistically significant (effect $= 0.096$, $z=3.211$, $p=0.001$; 95\%~CI: $[0.036, 0.159]$). Moreover, both prior NMR-note history and pre-correction misleadingness remain significant moderators when jointly included with other observed characteristics (see \Cref{fig:did_he_nmr}, \Cref{supp:he_prior_nmr}). Because prior NMR-note history may reflect contributor attention, post visibility, and account activity rather than misleadingness alone, we interpret these results as complementary exploratory evidence of heterogeneous posting responses.

We additionally examine whether posting responses vary with correction quality, audience moral outrage, and the deletion status of corrected posts~\cite{chuai2026community,chuai2025community,brady2021social}. Correction quality is measured using the note helpfulness scores~\cite{wojcik2022birdwatch}, while audience outrage is assessed from fact-checking replies in a short window ($6$ hours) surrounding note display. Across these analyses, we find no statistically significant differences in subsequent posting responses between higher- and lower-quality corrections (effect $=0.013$, $p=0.435$), between corrections associated with increasing versus non-increasing audience outrage (effect $=-0.018$, $p=0.500$), or between corrections associated with subsequently deleted versus retained posts (effect $=-0.028$, $p=0.453$; see \Cref{supp:quality_outrage_deletion}). These findings provide no statistically significant evidence that these factors distinguish subsequent posting responses.

\clearpage

\clearpage
\subsection*{No detectable reduction in misleading-post production}

We next investigate whether community corrections are associated with changes in the misleadingness of subsequently published original posts. Specifically, we examine both the number of GPT-classified misleading posts and their proportion among all original posts published during the same period, distinguishing changes in misleading-content production from changes in content composition. We additionally assess other content characteristics, including positive and negative sentiment, toxicity, opinion confidence, and political content, to provide a broader evaluation of authors' subsequent posting behavior (see \nameref{sec:methods}). To further improve comparability in pre-correction posting characteristics, we conduct an additional round of one-to-one propensity score matching of first corrections and their matched pseudo-correction counterparts (see details in \Cref{supp:overall_content}).

Across the matched sample, we find no statistically significant changes in either the production or proportion of misleading posts following the first correction. The estimated change in the number of misleading posts is positive but not statistically significant (effect $=0.038$, $z=1.653$, $p=0.098$; 95\%~CI: $[-0.007, 0.084]$), while the change in the proportion of misleading posts also remains statistically indistinguishable from zero (effect $=-0.006$, $t=-1.105$, $p=0.269$; 95\%~CI: $[-0.017, 0.005]$). We likewise find no statistically significant changes across the other examined content characteristics ($p>0.100$ for all count outcomes; see \Cref{tab:did_content_production_composition}, \Cref{supp:overall_content}). Here, the estimates for (non-)misleading posts are restricted to four days before and after correction, and other content estimates cover four weeks around correction. These results provide no statistically significant evidence of an overall reduction in misleading-content production or improvement in content composition within the follow-up period.

\begin{figure}[ht]
    \centering
    \captionsetup[subfloat]{font={bf, small}, skip=0pt, singlelinecheck=false, labelformat=simple, position=top}
    \includegraphics[width = \textwidth]{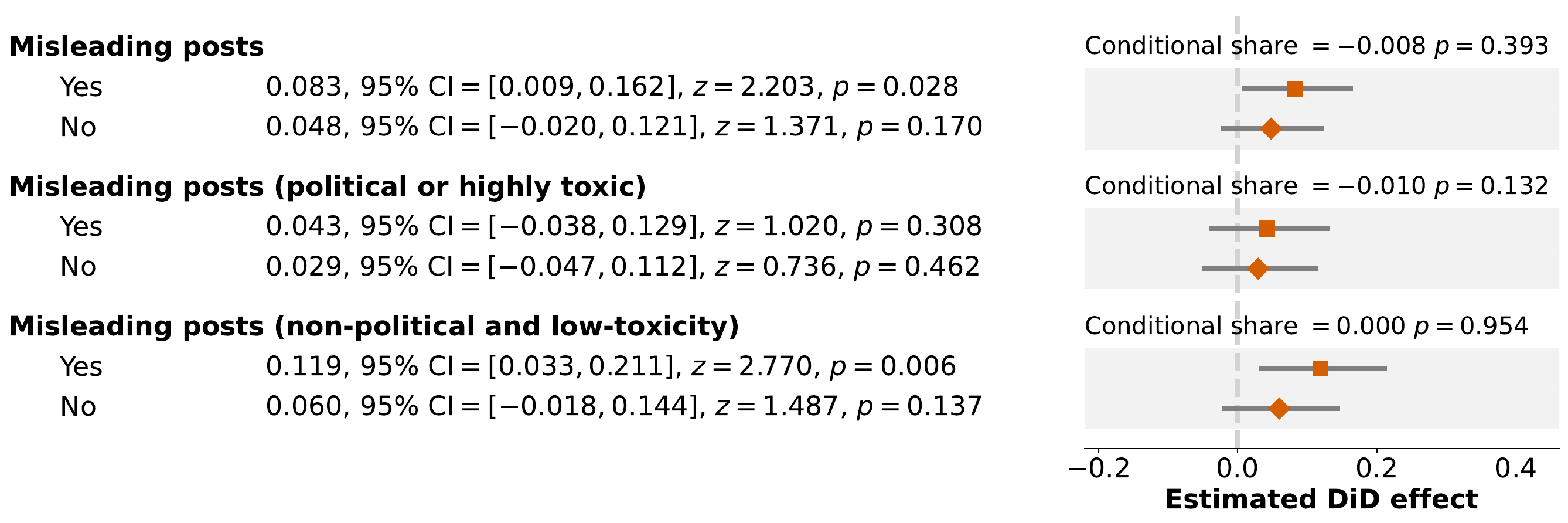}
    \caption{\textbf{Aggregated DiD estimates of changes in GPT-classified misleading-post production and composition following the first correction.} The analysis is restricted to accounts in the highest quartile of pre-correction misleadingness across three content categories: all misleading posts, misleading posts that are political or highly toxic, and misleading posts that are non-political and low-toxicity. ``Yes'' and ``No'' denote posts classified as misleading and non-misleading, respectively. Conditional share indicates changes in the proportion of category-specific posts among all original posts. Error bars represent 95\%~CIs. Full estimation results are reported in \Cref{supp:content_he_high_mis}.}
    \label{fig:content_he_mis}
\end{figure}

Given the increased posting activity observed among accounts with high pre-correction misleadingness, we explore whether this subgroup also produces more misleading posts following correction. Focusing on accounts in the highest quartile of pre-correction GPT-estimated misleadingness ($n=\num{1672}$), we additionally assess the misleadingness of \num{496505} original posts from these accounts and extend the observation window to four weeks before and after correction. We find that the number of GPT-classified misleading posts increases by 8.3\% during the four-week post-correction period (effect $=0.083$, $z=2.203$, $p=0.028$; 95\%~CI: $[0.009, 0.162]$; \Cref{fig:content_he_mis}). In contrast, the estimated change in the number of posts not classified as misleading is positive but statistically insignificant (effect $=0.048$, $z=1.371$, $p=0.170$; 95\%~CI: $[-0.020, 0.121]$). The estimated change in the proportion of misleading posts is also statistically insignificant (effect $=-0.008$, $t=-0.855$, $p=0.393$; 95\%~CI: $[-0.027, 0.011]$). For the other content dimensions, the estimated changes in the number of positive-sentiment, negative-sentiment, highly toxic, high-confidence, and political posts are not statistically significant (all $p>0.050$; \Cref{fig:content_he}, \Cref{supp:content_he_high_mis}). Across all six content dimensions, none of the estimated changes in content ratios is statistically significant (all $p>0.100$; \Cref{fig:content_he}). Our results indicate that accounts with higher baseline misleadingness are associated with more misleading-content production in absolute terms following correction, without a statistically detectable increase in the relative prevalence of misleading content.

Notably, we also observe statistically significant positive estimates for posts with low-toxicity content (effect $=0.070$, $z=2.059$, $p=0.039$; 95\%~CI: $[0.003, 0.141]$) and non-political content (effect $=0.078$, $z=1.996$, $p=0.046$; 95\%~CI: $[0.001, 0.160]$; see \Cref{fig:content_he}, \Cref{supp:content_he_high_mis}). Given the multiple outcomes examined, we interpret these nominally significant estimates as exploratory. Because content characteristics are not mutually exclusive, we further examine their overlap with misleadingness to better understand the increased production of misleading posts. Specifically, we distinguish between misleading posts that are political or highly toxic and those that are non-political and low in toxicity. As shown in \Cref{fig:content_he_mis}, we find no statistically significant change in the number of political or highly toxic misleading posts following correction (effect $=0.043$, $z=1.020$, $p=0.308$; 95\%~CI: $[-0.038, 0.129]$), whereas the change in the number of non-political and low-toxicity misleading posts is statistically significant, with an increase of 11.9\% (effect $=0.119$, $z=2.770$, $p=0.006$; 95\%~CI: $[0.033, 0.211]$). Changes in the content ratios across both groups remain not statistically significant (all $p>0.100$).

To account for multiple comparisons, we additionally adjust $p$-values across the six comparisons of misleading-content production in \Cref{fig:content_he_mis}, using both Holm and Benjamini-Hochberg procedures. After adjustment, the increase in overall misleading-post production is no longer statistically significant ($p_{\mathrm{Holm}}=0.140$, $p_{\mathrm{BH}}=0.084$). However, the increase in non-political and low-toxicity misleading posts following correction remains statistically significant (adjusted $p=0.036$ for both procedures; see \Cref{tab:did_content_high_misleadingness_combined}, \Cref{supp:content_he_high_mis}). This finding provides evidence of an increase within that category, but does not establish that its response differs from that of political or highly toxic misleading posts. We therefore retain the overall 8.3\% point estimate to summarize the estimated change in misleading-post production by accounts with pre-correction misleadingness in the highest quartile, while interpreting this aggregate finding as suggestive given the multiple-comparison adjustments. Additionally, the SDiD estimates show statistically significant increases for overall misleading posts (ATT $=1.338$ original posts per week, $t=2.41$, $p=0.016$; 95\%~CI: $[0.252, 2.425]$) and non-political and low-toxicity misleading posts (ATT $=0.882$ original posts per week, $t=2.52$, $p=0.012$; 95\%~CI: $[0.197, 1.566]$). In contrast, the change in political or highly toxic misleading posts is not statistically significant (ATT $=0.445$ original posts per week, $t=1.54$, $p=0.124$; 95\%~CI: $[-0.123, 1.014]$; see \Cref{fig:sdid_posts_weekly_content_mis_combined}, \Cref{supp:content_he_high_mis}). These estimates provide complementary evidence for the observed posting patterns.

As a complementary exploratory analysis, we examine changes in post content among accounts with repeated prior NMR notes, and find that those accounts also produce more misleading posts (effect $=0.092$, $z=2.041$, $p=0.041$; 95\%~CI: $[0.003, 0.187]$) and non-political posts following correction (effect $=0.086$, $z=2.061$, $p=0.039$; 95\%~CI: $[0.004, 0.175]$; see \Cref{tab:did_content_high_nmr}, \Cref{supp:content_he_high_nmr}). 

\clearpage

\clearpage
\section*{Discussion}

Community-based fact-checking is increasingly used to reduce the visibility and circulation of misleading content on social media platforms. Our study shifts the unit of analysis from the corrected post to the subsequent behavior of its author. While previous research has demonstrated that community notes can substantially reduce the spread of individual misleading posts after note display and facilitate the deletion of annotated posts~\cite{chuai2026community,slaughter2025community}, we find no statistically significant overall changes in authors' posting activity or misleading-content production following correction. Our findings suggest that the effectiveness of community-based fact-checking does not necessarily extend to the author level, and highlight an important distinction between reducing misinformation diffusion and reducing its subsequent production, which may require different intervention mechanisms.

The increased posting activity among accounts with higher pre-correction misleadingness further complicates expectations about corrective learning and reputational deterrence. Rather than showing improvements following correction, accounts in the highest quartile of pre-correction misleadingness produce 10.2\% more original posts, with an increase of 8.3\% in misleading-content production. One possible explanation is that authors discount corrections or experience them as reputational or identity threats, potentially leading to defensive responses. Previous experimental evidence shows that correcting misinformation sharers can produce counterproductive changes in subsequent sharing behavior, without a statistically significant effect on original posts~\cite{mosleh2021perverse}. Our study extends this discussion to original-post production using a large-scale observational dataset from a naturally deployed community-based fact-checking system. Importantly, we find no statistically significant overall change in posting activity, while increased activity and misleading-post production are particularly evident among accounts in the highest quartile of pre-correction misleadingness. Such heterogeneous responses may help explain why previous research focusing on average treatment effects did not detect changes in original-post activity~\cite{mosleh2021perverse}. Our findings are also relevant to the broader evidence that a relatively small minority of users accounts for a disproportionate share of misinformation circulation online~\cite{baribi2024supersharers}, suggesting that limited behavioral changes among highly misleading accounts may have implications beyond their relative prevalence. Furthermore, the increase in misleading-post production is not accompanied by a statistically detectable change in its relative proportion. Yet, even without measurable deterioration in content composition, greater posting activity can contribute to a larger volume of misleading content.

Our complementary analyses provide further insight into the heterogeneous responses to community corrections. Accounts with repeated prior NMR notes also increase their posting activity following correction. This association remains significant after accounting for pre-correction misleadingness and other observed characteristics, suggesting that prior community scrutiny captures additional dimensions beyond the average misleadingness of an account's content. While pre-correction misleadingness reflects an account's tendency to produce misleading content, prior NMR-note history may also reflect its visibility to active Community Notes contributors, potentially through audiences that include or attract such contributors. These findings raise the possibility that accounts subject to greater community scrutiny respond differently to public corrections, although the mechanisms underlying this association remain unclear. Moreover, we examine whether posting responses vary across characteristics related to corrections, including correction quality, audience moral outrage, and the deletion of corrected posts, but find no statistically significant differences. Although these null findings do not rule out their potential influence, they suggest that the measured characteristics of individual corrections and their immediate audience responses do not clearly explain the observed heterogeneity in posting activity.

Another notable finding is that the increase in misleading-post production among accounts with high pre-correction misleadingness following correction is concentrated in posts classified as neither political nor highly toxic. On the one hand, this highlights that misleading information extends beyond politically oriented or overtly toxic content. Evaluations focusing primarily on political misinformation or harmful language may therefore overlook other forms of misleading-content production. Moreover, given the relatively low average toxicity of posts before correction (see \Cref{tab:data_overview}, \Cref{supp:data_overview}), the observed increase may partly reflect the baseline prevalence of low-toxicity content rather than a disproportionate expansion toward this category. On the other hand, previous research shows that political content receives substantial attention from Community Notes contributors and accounts for a large share of publicly displayed corrections, compared with other topics such as economics, health, and science~\cite{allen2022birds,chuai2026community}. Because accounts with high pre-correction misleadingness post more political content before correction (see \Cref{tab:logit_account_characteristics}, \Cref{supp:did_interaction_features}), the increase in non-political misleading posts from these accounts raises the possibility that authors adapt their posting practices following correction, potentially publishing more content that attracts less community fact-checking attention. Related research has documented how social media users adapt their content to perceived moderation practices, including strategies to circumvent content restrictions~\cite{steen2023you,gillett2024just}. However, our findings do not establish whether the observed increase reflects baseline content composition or strategic adaptation, and the mechanisms underlying this pattern warrant further investigation.

Our results have important implications for both the evaluation and design of community-based fact-checking systems. Conventional measures of correction effectiveness, such as reductions in engagement and diffusion following note display, should be complemented by author-level behavioral outcomes. A correction may successfully limit the circulation of an annotated post without reducing the subsequent production of misleading content by its author. Platforms could therefore explore complementary interventions targeting subsequent posting behavior, particularly among accounts that frequently produce misleading content. Prior NMR-note history may also provide a complementary signal for identifying accounts whose subsequent behavior warrants further examination. Any differentiated intervention, however, would require transparent appeal mechanisms and safeguards against erroneous or coordinated corrections~\cite{chuai2026consensus,selvam2026gaming}. More broadly, correcting content and changing authors' behavior should be treated as related but distinct objectives of platform governance. Our exploratory audience engagement analysis also finds reductions in reposts and replies per subsequent post during the first week after correction (see \Cref{supp:post_engagement}), suggesting that changes in audience responses represent another dimension of downstream effects worth investigating.

Several limitations motivate future research. First, although our estimates are consistent with the parallel-trends assumption and remain robust across SDiD estimation, alternative model specifications, and analyses stratified by correction order, our quasi-experimental design cannot exclude unobserved time-varying confounding. Field experiments around note display could provide stronger causal evidence.
Second, our analysis focuses on the Community Notes system on \X. Although Meta has since introduced Community Notes and YouTube and TikTok are testing related systems~\cite{meta2025testing,youtube2024testing,tiktok2025testing}, differences in note selection, ranking, presentation, and associated enforcement may produce different author responses. Cross-platform research is therefore needed to assess generalizability.
Third, content proportions are defined only for periods with original posts, so the corresponding estimates describe changes in content composition conditional on posting. Our sample is also restricted to accounts accessible at the time of data collection, excluding those that had become suspended, deleted, or protected. Although we distinguish zero posting activity from unavailable count histories, we cannot reconstruct the posting behavior or disappearance rates of inaccessible accounts. Moreover, supplementary records allow us to recover correction events involving deleted posts from accessible accounts, but do not address potential selection bias associated with inaccessible accounts. Nevertheless, previous research finds that the proportion of posts associated with suspended or protected accounts is relatively low (generally below 5\%) around the note display threshold, and does not identify a statistically significant effect of note display on this outcome~\cite{chuai2026community}.
Fourth, post deletion may affect both sample inclusion and the availability of content for analysis. Although deletion is relatively uncommon and stable among posts without displayed notes, note display increases the likelihood that authors delete the annotated posts~\cite{chuai2026community}. Supplementary historical records allow us to recover some correction events involving deleted posts from accessible accounts. We find no statistically significant differences in posting responses between corrections associated with deleted and non-deleted posts. This check partially addresses the exclusion of deleted corrected posts, but does not establish whether correction also changes the deletion of other posts during the observation window.
Fifth, the Community Notes system continues to evolve, and changes to contributor eligibility, ranking algorithms, or display practices may alter both who receives corrections and how authors respond; future studies could evaluate behavioral effects around such design changes. 
Finally, we cannot directly observe whether authors notice, accept, or react defensively to notes. Combining longitudinal behavior with surveys or field experiments could test the roles of learning, reputational pressure, reactance, and habituation. 

Overall, our findings show that success in correcting individual misleading posts does not imply corresponding improvements in the behavior of their producers. Community notes can be effective at limiting individual misinformation cascades without uniformly changing the behavior of the accounts that generate them. Recognizing this distinction is essential for designing community-based fact-checking systems that not only correct misleading posts at scale but also promote more durable improvements in the surrounding information ecosystem.

\clearpage
\section*{Methods}
\label{sec:methods}

\subsection*{Data and study design}
This study examines how users' posting behaviors change after their posts are identified as misleading and corrected by community notes. To this end, we leverage a large-scale dataset sourced from prior work on the efficacy of community notes in reducing the spread of corrected misleading posts~\cite{chuai2026community}. The dataset contains all available posts corrected by publicly displayed community notes at the time of data collection (treatment group) and matched counterpart posts (control group) over a period of more than 20 months from the roll-out of the Community Notes program on \X on October 6, 2022 to June 11, 2024. Specifically, a correction event in the treatment group refers to a community note reaching helpful status and becoming publicly displayed on a post published by the corresponding account. In contrast, control events refer to posts that received proposed community notes but did not reach helpful status and therefore remained undisplayed (see \Cref{supp:data_overview}). For each control post, we assign a pseudo-correction time by adding the interval between its matched treated post's publication and first note display to the control post's publication time. This aligns each matched pair at the same elapsed time since publication. Based on this dataset, we collect longitudinal posting activity data (\ie, post counts) and post-level content data from author accounts before and after the (pseudo-)correction time. Given that the community notes predominantly fact-check original posts (3.8\% fact-checked posts in our sample are replies) our main analysis focuses on the volume and content of authors' subsequent original posts.

Using the full-archive count endpoint of \X's API, we collect daily counts of the original posts from author accounts in the treatment and control groups. The observation period spans from 28 days (4 weeks) before to 28 days (4 weeks) after the display of the community note for each post. This results in \num{3244360} daily count observations and \num{463480} weekly count observations across \num{57935} community-annotated posts from \num{19854} accounts. 
Additionally, we collect all original post content from the accounts in the treatment and control groups during the same post-specific observation windows used for the count data collection. In total, we successfully retrieve \num{11909591} original posts for the analysis of content features (for a data summary, see \Cref{tab:data_overview}, \Cref{supp:data_overview}). The treatment and control groups are constructed at the focal-correction level based on whether the posts have received displayed community notes. Consequently, an account can contribute both treatment and control events if some of its posts received displayed notes while others did not; and \num{3768} accounts appear in both groups in the full dataset. To limit treatment contamination, we exclude control events whenever any other post from the same account received a displayed note during the assigned observation window. The main first-correction analysis further restricts both groups to events with no previously displayed notes. Control events therefore correspond to accounts that had not yet received a displayed note at the assigned correction time and remained unexposed throughout the four-week follow-up period.

We assess the balance of the treatment and control groups on various factors across account characteristics and pre-correction posting preferences. Specifically, the account-level features include the number of followers (Followers), the number of followees (Followees), verified status (Verified), and account age in days (Account age). Followers and Followees were extracted from account profiles during data collection~\cite{chuai2026community}. Prior research shows that community corrections do not lead to meaningful changes in the number of followers of corrected users~\cite{bobek2026community}, supporting the use of extracted follower counts as a proxy for pre-correction audience size. We also estimate authors' political leaning from the ideological positions of the political elites they follow and their exposure to misinformation from the falsity scores associated with those elites~\cite{mosleh2022measuring}. Finally, we measure accounts' posting characteristics based on the posts created before community corrections. These posting preferences include the weekly average of the number of original posts (\ie, posting frequency) and the weekly averages of the following features (for a summary, see \Cref{supp:data_overview}):
\begin{itemize}[leftmargin=*,noitemsep,topsep=0pt]
    \item \textit{GPT-estimated misleadingness.} Large language models have demonstrated potential for identifying misleading posts and have been increasingly integrated into community-based fact-checking system to assist in drafting community notes~\cite{de2025supernotes,renault2026fact,miller2026community}. Following a previous framework for large-scale misleadingness assessment~\cite{chuai2026request}, we use GPT-5.4 mini to assign each original post a continuous score ranging from 0 (not misleading) to 1 (extremely misleading). Given the high computational demands of scoring the full corpus, we restrict this analysis to original posts published from Day~$-4$ through Day~$4$ around each actual or assigned note-display time. This procedure yields misleadingness scores for \num{2567990} unique posts. We classify individual posts as misleading using a score threshold of 0.86 (validated in \Cref{supp:gpt_misleading_val}).
    \item \textit{Positive and negative sentiment.} We apply a RoBERTa-based sentiment classifier developed for social-media text to estimate positive and negative sentiment scores for each original post~\cite{camacho2022tweetnlp,loureiro2022timelms}. We then average the respective post-level scores to characterize an account's tendency to publish positively or negatively expressed content. We classify individual posts as positive or negative using a score threshold of 0.5.
    \item \textit{Toxicity.} We use the Detoxify classifier with an AUC score of 0.921 to assign each original post a continuous toxicity score, with higher values indicating a greater presence of insulting, threatening, or otherwise toxic language~\cite{hanu2020detoxify}. We classify individual posts as toxic using a score threshold of 0.5.
    \item \textit{Opinion confidence.} We employ a validated certainty lexicon to measure the degree of confidence expressed in each original post~\cite{rocklage2023beyond}. Higher scores indicate language conveying greater certainty or conviction, whereas lower scores indicate more tentative or qualified language. We identify individual posts with high opinion confidence using a score threshold of 6.5.
    \item \textit{Political content.} We use a topic classifier fine-tuned from the pretrained TwHIN-BERT-large model to identify whether each original post concerns politics. The classifier achieves a macro-$F_1$ score of 0.816~\cite{chuai2026community}. We measure an account's preference for political content as the proportion of its pre-correction original posts classified as political.
\end{itemize}
Treatment and control groups are comparable across most measured characteristics (see \Cref{tab:covariate_balance}, \Cref{supp:data_overview}). For the heterogeneity analysis, we divide accounts into high and low subgroups at the median or 75th percentile of each continuous characteristic or posting-preference measure. Accounts with positive partisan scores are classified as right-leaning, whereas those with negative scores are classified as left-leaning. Verified status is analyzed using its corresponding categorical groups. For the content analysis, we classify individual posts using the thresholds specified by the corresponding classifiers, with the GPT-estimated misleadingness threshold calibrated against external PolitiFact labels.

\subsection*{Difference-in-differences analyses}
In this study, we use a large-scale observational dataset and adopt a DiD design to examine the changes in accounts' posting activity after being corrected by community notes~\cite{angrist2009mostly,callaway2021difference,correia2020fast}. Our DiD approach includes both weekly leads-and-lags and aggregated pre-post specifications. The leads-and-lags model estimates the temporal weekly dynamics of accounts' posting activity before and after the community corrections, allowing us to assess the parallel trend assumption and track the temporal evolution of posting activity after community corrections. The aggregated pre-post model captures the overall changes in accounts' posting activity following the corrections.

First, we examine the changes in the number of original posts from the corrected accounts over time before and after the display of community notes on the corresponding annotated posts. In the main analysis, we estimate the model using Poisson pseudo-maximum likelihood (PPML) to account for the count nature of the dependent variable~\cite{correia2020fast}. The leads-and-lags DiD specification is:
\begin{equation}
\begin{aligned}
\log(E[Y_{i,t} \mid \bm{x}_{i,t}]) = \,
& \alpha_{i} + \lambda_{t} + \textbf{b}_{1}^{\intercal}(D_{i}\times\textbf{Before}_{t})
+ \textbf{b}_{2}^{\intercal}(D_{i}\times\textbf{After}_{t}),
\end{aligned}
\end{equation}
where $Y_{i,t}$ is the number of original posts published by the account associated with the correction event $i$ during week $t$. $\alpha_{i}$ and $\lambda_{t}$ are correction-event and week fixed effects. $D_{i}$ is a dummy variable indicating whether correction event $i$ from the specific account belongs to the treatment group ($=1$) or control group ($=0$). The vectors $\textbf{Before}_{t}$ and $\textbf{After}_{t}$ denote week dummies relative to the display of community notes. The main effect of $D_{i}$ is absorbed by $\alpha_{i}$, and the week main effects are absorbed by $\lambda_{t}$. $\textbf{Before}_{t}$ covers Weeks $-4$ to $-2$, with Week $-1$ serving as the reference period, while $\textbf{After}_{t}$ spans Weeks 1 to 4. Additionally, $\bm{x}_{i,t}$ denotes the vector of regressors, comprising correction-event and relative-week indicators and treatment-by-week interactions. The DiD interaction coefficients $\textbf{b}_{1}$ capture pre-treatment (lead) effects, used to assess the parallel trends assumption, whereas $\textbf{b}_{2}$ captures post-treatment (lag) effects, measuring the weekly impact of community corrections on the accounts' subsequent posting activity. Standard errors are clustered at the account level to account for dependence across observations contributed by the same author. Correction events with zero original-post counts throughout the observation window do not contribute to estimation in the fixed-effects PPML model. As a robustness check, we therefore also estimate a mixed-effects negative binomial model with correction-event random effects, which permits the inclusion of these all-zero event histories. We further assess sensitivity using alternative fixed-effects count models (see \Cref{supp:alternative_models}). In addition to the leads-and-lags DiD model, we estimate an aggregated pre-post DiD model with a single treatment-by-post-period interaction, retaining correction-event and relative-week fixed effects:
\begin{equation}
\begin{aligned}
    log(E[Y_{i,t}|\bm{x}_{i,t}]) = \, 
    \alpha_{i} + \lambda_{t} + \beta(D_{i} \times \text{After}_{t}),
\end{aligned}
\end{equation}
where $\text{After}_{t}$ is a post-treatment indicator that equals 1 after the display of community notes and 0 before. The estimate of the DiD interaction coefficient $\beta$ captures the aggregated effect of community corrections on weekly original-post activity over the four-week post-treatment period. The main effects of $D_i$ and $\text{After}_{t}$ are absorbed by the event and week fixed effects. To better understand the effect size of community corrections in the post count analysis, we exponentiate the coefficient estimates of the DiD terms in the models:
\begin{equation}
\begin{aligned}
    \text{effect} = e^{\beta} - 1,
\end{aligned}
\end{equation}
where $\beta$ is the coefficient estimate for the specific DiD interaction term. The transformed DiD estimate indicates the proportional change in expected weekly original-post counts following correction, relative to the counterfactual counts inferred from changes in the control group under the DiD assumptions.

To examine changes in content composition, we estimate the following linear fixed-effects DiD model separately for each content category:
\begin{equation}
\begin{aligned}
R_{i,t} = \alpha_i + \lambda_t
+ \beta(D_i \times \mathrm{After}_t)
+ \varepsilon_{i,t},
\end{aligned}
\end{equation}
where $R_{i,t}$ is the number of original posts classified into the corresponding content category divided by the total original-post count for correction event $i$ during week $t$. $\varepsilon_{i,t}$ denotes the error term, and standard errors are clustered at the account level. We estimate the model using observations with positive original-post counts and proportions between zero and one. The coefficient $\beta$ captures the DiD estimate of the change in content proportion. Because proportions are undefined for periods without original posts, these estimates describe content composition conditional on posting. Here, $\beta$ is reported directly, without the transformation used for the count models.
For daily analyses, including overall misleading-post production and composition, $t$ indexes relative days rather than weeks, and the corresponding fixed effects and period indicators are defined at the daily level.

This study was approved by Oxford Internet Institute DREC (ref.~3111054). All analyses are based on publicly available social-media data. Account and post identifiers are pseudonymized to protect users' privacy.

\section*{Data and code availability}
The analysis code and supporting data will be publicly available upon publication.

\clearpage
\printbibliography

@inproceedings{camacho2022tweetnlp,
    title = "{T}weet{NLP}: Cutting-edge natural language processing for social media",
    author = "Camacho-collados, Jose  and
      Rezaee, Kiamehr  and
      Riahi, Talayeh  and
      Ushio, Asahi  and
      Loureiro, Daniel  and
      Antypas, Dimosthenis  and
      Boisson, Joanne  and
      Espinosa Anke, Luis  and
      Liu, Fangyu  and
      Mart{\'\i}nez C{\'a}mara, Eugenio",
    booktitle = "Proceedings of the 2022 Conference on Empirical Methods in Natural Language Processing: System Demonstrations",
    pages = "38--49",
    year = "2022"
}

@inproceedings{loureiro2022timelms,
    title = "{T}ime{LM}s: Diachronic language models from {T}witter",
    author = "Loureiro, Daniel  and
      Barbieri, Francesco  and
      Neves, Leonardo  and
      Espinosa Anke, Luis  and
      Camacho-collados, Jose",
    booktitle = "Proceedings of the 60th Annual Meeting of the Association for Computational Linguistics: System Demonstrations",
    pages = "251--260",
    year = "2022"
}

@article{ecker2024misinformation,
  title={Misinformation poses a bigger threat to democracy than you might think},
  author={Ecker, Ullrich and Roozenbeek, Jon and van der Linden, Sander and Tay, Li Qian and Cook, John and Oreskes, Naomi and Lewandowsky, Stephan},
  journal={Nature},
  volume={630},
  number={8015},
  pages={29--32},
  year={2024},
}

@article{drolsbach2024community,
  title={Community notes increase trust in fact-checking on social media},
  author={Drolsbach, Chiara Patricia and Solovev, Kirill and Pr{\"o}llochs, Nicolas},
  journal={PNAS Nexus},
  volume={3},
  number={7},
  pages={pgae217},
  year={2024}
}

@inproceedings{mosleh2021perverse,
  title={Perverse downstream consequences of debunking: Being corrected by another user for posting false political news increases subsequent sharing of low quality, partisan, and toxic content in a Twitter field experiment},
  author={Mosleh, Mohsen and Martel, Cameron and Eckles, Dean and Rand, David},
  booktitle={Proceedings of the 2021 CHI Conference on Human Factors in Computing Systems},
  pages={1--13},
  year={2021}
}

@inproceedings{allen2022birds,
  title={Birds of a feather don’t fact-check each other: Partisanship and the evaluation of news in Twitter’s Birdwatch crowdsourced fact-checking program},
  author={Allen, Jennifer and Martel, Cameron and Rand, David G},
  booktitle={Proceedings of the 2022 CHI Conference on Human Factors in Computing Systems},
  pages={1--19},
  year={2022}
}

@article{brady2021social,
    title={How social learning amplifies moral outrage expression in online social networks},
    author={Brady, William J and McLoughlin, Killian and Doan, Tuan N and Crockett, Molly J},
    journal={Science Advances},
    volume={7},
    number={33},
    pages={eabe5641},
    year={2021}
}

@article{chuai2024roll,
    title = {Did the roll-out of {Community} {Notes} reduce engagement with misinformation on {X/Twitter}?},
    author = {Chuai, Yuwei and Tian, Haoye and Pröllochs, Nicolas and Lenzini, Gabriele},
    journal={Proceedings of the ACM on Human-Computer Interaction},
    volume={8},
    number={CSCW2},
    pages={1--52},
    year = {2024},
}

@inproceedings{chuai2025community,
  title={Community fact-checks trigger moral outrage in replies to misleading posts on social media},
  author={Chuai, Yuwei and Sergeeva, Anastasia and Lenzini, Gabriele and Pr{\"o}llochs, Nicolas},
  booktitle={Proceedings of the 2025 CHI Conference on Human Factors in Computing Systems},
  pages={1--23},
  year={2025}
}

@inproceedings{horta2023automated,
    title = {Automated content moderation increases adherence to community guidelines},
    author = {Horta Ribeiro, Manoel and Cheng, Justin and West, Robert},
    booktitle={Proceedings of the ACM Web Conference 2023},
    pages={2666--2676},
    year = {2023},
}

@article{grinberg2019fake,
  title={Fake news on {Twitter} during the 2016 {US} presidential election},
  author={Grinberg, Nir and Joseph, Kenneth and Friedland, Lisa and Swire-Thompson, Briony and Lazer, David},
  journal={Science},
  volume={363},
  number={6425},
  pages={374--378},
  year={2019}
}

@article{baribi2024supersharers,
  title={Supersharers of fake news on {Twitter}},
  author={Baribi-Bartov, Sahar and Swire-Thompson, Briony and Grinberg, Nir},
  journal={Science},
  volume={384},
  number={6699},
  pages={979--982},
  year={2024},
  publisher={American Association for the Advancement of Science}
}

@article{pennycook2019fighting,
  title={Fighting misinformation on social media using crowdsourced judgments of news source quality},
  author={Pennycook, Gordon and Rand, David G},
  journal={Proceedings of the National Academy of Sciences},
  volume={116},
  number={7},
  pages={2521--2526},
  year={2019}
}

@article{allen2021scaling,
  title={Scaling up fact-checking using the wisdom of crowds},
  author={Allen, Jennifer and Arechar, Antonio A and Pennycook, Gordon and Rand, David G},
  journal={Science Advances},
  volume={7},
  number={36},
  pages={eabf4393},
  year={2021}
}

@inproceedings{chuai2025political,
  title={Is fact-checking politically neutral? {Asymmetries} in how {U.S.} fact-checking organizations pick up false statements mentioning political elites},
  author={Chuai, Yuwei and Zhao, Jichang and Pr{\"o}llochs, Nicolas and Lenzini, Gabriele},
  booktitle={Proceedings of the International AAAI Conference on Web and Social Media},
  volume={19},
  pages={403--429},
  year={2025}
}

@misc{youtube2024testing,
    title = {Testing new ways to offer viewers more context and information on videos},
    howpublished = {\url{https://blog.youtube/news-and-events/new-ways-to-offer-viewers-more-context/}},
    author = {YouTube},
    year = {2024},
}

@misc{meta2025testing,
    title = {Testing begins for {Community Notes} on {Facebook}, {Instagram} and {Threads}},
    howpublished = {\url{https://about.fb.com/news/2025/03/testing-begins-community-notes-facebook-instagram-threads/}},
    author = {{Meta}},
    year = {2025}
}

@misc{tiktok2025testing,
    title = {Testing a new feature to enhance content on TikTok},
    howpublished = {\url{https://newsroom.tiktok.com/en-us/footnotes}},
    author = {Presser, Adam},
    year = {2025},
}

@article{wojcik2022birdwatch,
  title={Birdwatch: Crowd wisdom and bridging algorithms can inform understanding and reduce the spread of misinformation},
  author={Wojcik, Stefan and Hilgard, Sophie and Judd, Nick and Mocanu, Delia and Ragain, Stephen and Hunzaker, MB and Coleman, Keith and Baxter, Jay},
  journal={ArXiv},
  year={2022}
}

@book{angrist2009mostly,
  title = {Mostly harmless econometrics: An empiricist's companion},
  author = {Angrist, Joshua D and Pischke, J{\"o}rn-Steffen},
  year = {2009},
  publisher = {Princeton University Press}
}

@article{mosleh2022measuring,
  title={Measuring exposure to misinformation from political elites on Twitter},
  author={Mosleh, Mohsen and Rand, David G},
  journal={Nature Communications},
  volume={13},
  number={1},
  pages={7144},
  year={2022}
}

@misc{cn2026introduction,
  author={{X}},
  title={Introduction of Community Notes},
  year={2026},
  howpublished = {\url{https://communitynotes.x.com/guide/en/about/introduction}},
  note={[Accessed: 2026-10-20]}
}

@inproceedings{de2025supernotes,
  title={Supernotes: Driving consensus in crowd-sourced fact-checking},
  author={De, Soham and Bakker, Michiel A and Baxter, Jay and Saveski, Martin},
  booktitle={Proceedings of the ACM on Web Conference 2025},
  pages={3751--3761},
  year={2025}
}

@article{chuai2026community,
  title={Community-based fact-checking reduces the spread of misleading posts on X (formerly Twitter)},
  author={Chuai, Yuwei and Pilarski, Moritz and Renault, Thomas and Restrepo-Amariles, David and Troussel-Cl{\'e}ment, Aurore and Lenzini, Gabriele and Pr{\"o}llochs, Nicolas},
  journal={Nature Communications},
  volume={17},
  pages={4070},
  year={2026}
}

@misc{hanu2020detoxify,
  title={Detoxify},
  author={Hanu, Laura and {Unitary team}},
  howpublished={https://github.com/unitaryai/detoxify},
  year={2020}
}

@article{rocklage2023beyond,
  title={Beyond sentiment: the value and measurement of consumer certainty in language},
  author={Rocklage, Matthew D and He, Sharlene and Rucker, Derek D and Nordgren, Loran F},
  journal={Journal of Marketing Research},
  volume={60},
  number={5},
  pages={870--888},
  year={2023}
}

@article{rambachan2023honest,
  title={A more credible approach to parallel trends},
  author={Rambachan, Ashesh and Roth, Jonathan},
  journal={Review of Economic Studies},
  volume={90},
  number={5},
  pages={2555--2591},
  year={2023},
  publisher={Oxford University Press US}
}

@article{callaway2021difference,
  title={Difference-in-differences with multiple time periods},
  author={Callaway, Brantly and Sant’Anna, Pedro HC},
  journal={Journal of Econometrics},
  volume={225},
  number={2},
  pages={200--230},
  year={2021},
  publisher={Elsevier}
}

@inproceedings{chuai2026request,
  title={Request a note: How the request function shapes {X}'s {Community Notes} system},
  author={Chuai, Yuwei and Zhang, Shuning and Wang, Ziming and Yi, Xin and Mosleh, Mohsen and Lenzini, Gabriele},
  booktitle={Proceedings of the 2026 CHI Conference on Human Factors in Computing Systems},
  pages={1--22},
  year={2026}
}

@article{nyhan2015effect,
  title={The effect of fact-checking on elites: A field experiment on US state legislators},
  author={Nyhan, Brendan and Reifler, Jason},
  journal={American Journal of Political Science},
  volume={59},
  number={3},
  pages={628--640},
  year={2015},
  publisher={Wiley Online Library}
}

@article{guess2019less,
  title={Less than you think: Prevalence and predictors of fake news dissemination on {Facebook}},
  author={Guess, Andrew and Nagler, Jonathan and Tucker, Joshua},
  journal={Science Advances},
  volume={5},
  number={1},
  pages={eaau4586},
  year={2019},
  publisher={American Association for the Advancement of Science}
}

@article{martel2024crowds,
  title={Crowds can effectively identify misinformation at scale},
  author={Martel, Cameron and Allen, Jennifer and Pennycook, Gordon and Rand, David G},
  journal={Perspectives on Psychological Science},
  volume={19},
  number={2},
  pages={477--488},
  year={2024},
  publisher={Sage Publications Sage CA: Los Angeles, CA}
}

@article{slaughter2025community,
  title={Community notes reduce engagement with and diffusion of false information online},
  author={Slaughter, Isaac and Peytavin, Axel and Ugander, Johan and Saveski, Martin},
  journal={Proceedings of the National Academy of Sciences},
  volume={122},
  number={38},
  pages={e2503413122},
  year={2025},
  publisher={National Academy of Sciences}
}

@article{yasseri2023can,
  title={Can crowdsourcing rescue the social marketplace of ideas?},
  author={Yasseri, Taha and Menczer, Filippo},
  journal={Communications of the ACM},
  volume={66},
  number={9},
  pages={42--45},
  year={2023},
  publisher={ACM New York, NY, USA}
}

@inproceedings{chuai2026consensus,
  title={Consensus stability of community notes on {X}},
  author={Chuai, Yuwei and Lenzini, Gabriele and Pr{\"o}llochs, Nicolas},
  booktitle={Proceedings of the ACM Web Conference 2026},
  pages={8885--8896},
  year={2026}
}

@article{bachmann2023studying,
  title={Studying the downstream effects of fact-checking on social media: Experiments on correction formats, belief accuracy, and media trust},
  author={Bachmann, Ingrid and Valenzuela, Sebasti{\'a}n},
  journal={Social Media + Society},
  volume={9},
  number={2},
  pages={20563051231179694},
  year={2023},
  publisher={SAGE Publications Sage UK: London, England}
}

@article{hoes2024prominent,
  title={Prominent misinformation interventions reduce misperceptions but increase scepticism},
  author={Hoes, Emma and Aitken, Brian and Zhang, Jingwen and Gackowski, Tomasz and Wojcieszak, Magdalena},
  journal={Nature Human Behaviour},
  volume={8},
  number={8},
  pages={1545--1553},
  year={2024},
  publisher={Nature Publishing Group UK London}
}

@article{jhaver2019does,
  title={Does transparency in moderation really matter? User behavior after content removal explanations on {Reddit}},
  author={Jhaver, Shagun and Bruckman, Amy and Gilbert, Eric},
  journal={Proceedings of the ACM on Human-Computer Interaction},
  volume={3},
  number={CSCW},
  pages={1--27},
  year={2019},
  publisher={ACM New York, NY, USA}
}

@article{zhang2026social,
  title={Social media moderation and content generation: Evidence from user bans},
  author={Zhang, Xiaohui and Wei, Zaiyan and Du, Qianzhou and Zhang, Zhongju},
  journal={MIS Quarterly},
  volume={50},
  number={1},
  pages={211--242},
  year={2026},
  publisher={Management Information Systems Research Center, University of Minnesota}
}

@article{correia2020fast,
  title={Fast Poisson estimation with high-dimensional fixed effects},
  author={Correia, Sergio and Guimar{\~a}es, Paulo and Zylkin, Tom},
  journal={The Stata Journal},
  volume={20},
  number={1},
  pages={95--115},
  year={2020},
  publisher={SAGE Publications, Sage CA: Los Angeles, CA}
}

@article{arkhangelsky2021synthetic,
  title={Synthetic difference-in-differences},
  author={Arkhangelsky, Dmitry and Athey, Susan and Hirshberg, David A and Imbens, Guido W and Wager, Stefan},
  journal={American Economic Review},
  volume={111},
  number={12},
  pages={4088--4118},
  year={2021},
  publisher={American Economic Association 2014 Broadway, Suite 305, Nashville, TN 37203}
}

@article{bouchaud2026community,
  title={{Community Notes} undermoderate polarizing content by design creating risks in electoral processes},
  author={Bouchaud, Paul and Ramaciotti, Pedro},
  journal={Science Advances},
  volume={12},
  number={27},
  pages={eaee6932},
  year={2026},
  publisher={American Association for the Advancement of Science}
}

@inproceedings{bobek2026community,
  title={Community fact-checks do not break follower loyalty},
  author={Bobek, Michelle and Pr{\"o}llochs, Nicolas},
  booktitle={Proceedings of the ACM Web Conference 2026},
  pages={9113--9124},
  year={2026}
}

@article{steen2023you,
  title={You can (not) say what you want: Using algospeak to contest and evade algorithmic content moderation on TikTok},
  author={Steen, Ella and Yurechko, Kathryn and Klug, Daniel},
  journal={Social Media+ Society},
  volume={9},
  number={3},
  pages={20563051231194586},
  year={2023},
  publisher={SAGE Publications, Sage UK: London, England}
}

@article{gillett2024just,
  title={‘Just a little hack’: Investigating cultures of content moderation circumvention by Facebook users},
  author={Gillett, Rosalie and Gray, Joanne E and Valdovinos Kaye, D Bondy},
  journal={New Media \& Society},
  volume={26},
  number={11},
  pages={6183--6204},
  year={2024},
  publisher={SAGE Publications, Sage UK: London, England}
}

@article{renault2026fact, 
    title={Fact-Checking Using AI}, 
    url={osf.io/preprints/psyarxiv/85quw_v5}, 
    journal={PsyArXiv}, 
    author={Renault, Thomas and Mosleh, Mohsen and Lin, Hause and Pennycook, Gordon and Rand, David G}, 
    year={2026}
}

@article{miller2026community,
  title={Community-driven API and AI writer design for openly scaling Community Notes},
  author={Miller, Brad and Baxter, Jay and Chao, Jiansong and Coleman, Keith and Hilgard, Sophie and Ortiz, Daniel},
  journal={ArXiv},
  year={2026}
}

@article{selvam2026gaming,
  title={Gaming consensus: Coordinated manipulation in crowdsourced fact-checking},
  author={Selvam, Nikil Roashan and Baxter, Jay and Hilgard, Sophie and Miller, Brad and Coleman, Keith and Vitercik, Ellen and Koyejo, Sanmi},
  journal={arXiv },
  year={2026}
}

@article{van2024broader,
  title={A broader view of misinformation reveals potential for intervention},
  author={Van der Linden, Sander and Kyrychenko, Yara},
  journal={Science},
  volume={384},
  number={6699},
  pages={959--960},
  year={2024},
  publisher={American Association for the Advancement of Science}
}

@article{zhang2026needs,
  title={Needs Your Help: Understanding platform-directed rating participation in Community Notes on X},
  author={Zhang, Shuning and Wen, Changxi and Wang, Jiuchang and Shi, Dai and Lenzini, Gabriele and Chuai, Yuwei and Yi, Xin},
  journal={ArXiv},
  year={2026}
}

\clearpage
\section*{Funding}

 This study was supported by research grants from the Luxembourg National Research Fund (Y.C., G.L.; C25/IS-SAS/19599536), and the German Research Foundation (N.P.; DFG grant 492310022). The funders had no role in study design, data collection and analysis, decision to publish, or preparation of the manuscript.

\section*{Author contributions}
Y.C., T.R., N.P., G.L., and M.M. conceived and designed the research. Y.C. and M.M. collected the data; Y.C. analyzed the data. Y.C., T.R., N.P., G.L., and M.M. wrote the manuscript. All authors approved the manuscript.\\
\textbf{Corresponding author:} Correspondence to Mohsen Mosleh (mohsen.mosleh@oii.ox.ac.uk).

\section*{Competing interests} 
M.M. has previously received funding from Meta and Google to conduct research related to reducing the spread of misinformation online and identifying inauthentic accounts.

\clearpage
\appendix
\crefalias{section}{appendix}
\crefalias{subsection}{appendix}

\begin{center}
    \singlespacing
    \Large \textbf{Supplementary Materials} \\
    \textbf{``\Title''}

    \vspace{2em}
    \normalsize
    Yuwei Chuai,
    Thomas Renault,
    Nicolas Pröllochs,
    Gabriele Lenzini,
    Mohsen Mosleh
\end{center}

\renewcommand{\thetable}{S\arabic{table}}
\renewcommand{\thefigure}{S\arabic{figure}}
\renewcommand\thesection{S\arabic{section}}
\setcounter{figure}{0}
\setcounter{table}{0}
\setcounter{section}{0}

\clearpage
\tableofcontents

\clearpage
\section{Data overview}
\label{supp:data_overview}

\X's Community Notes system was officially rolled out on October 6, 2022. Contributors can propose fact-checking notes on potentially misleading posts, and other contributors rate proposed notes. The note ranking algorithm then evaluates the helpfulness of community notes based on the submitted ratings and assigns each note one of three statuses: Needs More Ratings (NMR), Currently Rated Helpful (CRH), or Currently Rated Not Helpful (CRNH). NMR indicates that a note has not yet received sufficient evidence of cross-perspective agreement to be classified as helpful or not helpful. CRH indicates that contributors who have historically disagreed in their ratings broadly agree that the note is helpful, whereas CRNH indicates sufficient cross-perspective agreement that the note is not helpful. Only notes assigned CRH status are publicly displayed alongside the corresponding posts, providing correction information to readers. Notes classified as NMR or CRNH remain hidden.

\Cref{tab:data_overview} provides an overview of the dataset used in this study. 
We construct a set of pseudo-correction events as a baseline to evaluate the effect of actual corrections. Specifically, we draw on a previous study in which community fact-checked misleading posts were matched one-to-one with counterpart posts on author features (verified status, followers, followees, and account age), content characteristics (topics and sentiments), and fact-checking time relative to the roll-out of the Community Notes program~\cite{chuai2026community}.
The original dataset includes all fact-checked misleading posts that were written in English and remained available at the time of data collection, spanning more than 20 months following the roll-out of the Community Notes system. The matched comparison posts were also annotated by community notes, but their notes had not achieved displayed status. Each matched control post is assigned a pseudo-correction time by adding the interval between its paired treated post's publication and first note display to the control post's publication time, yielding \num{36077} pairs. We then use the full-archive count endpoint of \X's API to collect daily post counts of the corrected accounts between 28 days before and 28 days after each (pseudo-)correction event. After excluding corrections with no available count history, the final dataset comprises \num{57935} correction events (source posts) contributed by \num{19854} accounts, with \num{29049} misleading posts from \num{10799} accounts and \num{28886} matched posts with pseudo-corrections from \num{12823} accounts.

\begin{table}[ht]
    \centering
    \footnotesize
    \caption{\textbf{Data overview.} Reported are counts (numbers of source posts, accounts, daily/weekly count observations, and post count), means with standard deviations in parentheses (note frequency, followers, followees, account age, partisan score, misinfo exposure score, post frequency, sentiments, toxicity, and opinion confidence), or percentages (verified status and content related to politics). Treatment and control totals overlap because some accounts contribute events to both groups; the overall total counts posts once across the union of included accounts.}
    \begin{tabular}{l*3{S}}
    \toprule
    &{Total}&{With corrections}&{With pseudo-corrections}\\
    \midrule
    \# Source posts&{57,935}&{29,049}&{28,886}\\
    \# Accounts&{19,854}&{10,799}&{12,823}\\
    \# Average corrections per account&{/}&{3.346 (13.253)}&{/}\\
    Correction timing from the roll-out of CN (days)&{/}&{413.282 (141.364)}&{/}\\
    \\
    \multicolumn{4}{l}{\underline{Account characteristics}}\\
    \quad \# Followers&{340,063.134}&{419,420.332}&{417,440.310}\\
    \quad &{(2,941,502.846)}&{(3,412,354.861)}&{(3,235,505.482)}\\
    \quad \# Followees&{4,197.305}&{4,644.691}&{4,508.083}\\
    \quad &{(16,937.675)}&{(17,533.695)}&{(19,119.972)}\\
    \quad Verified&{56.6\%}&{59.4\%}&{61.4\%}\\
    \quad Account age to (pseudo-) correction (days)&{2,813.649}&{2,813.752}&{2,813.546}\\
    \quad &{(1,848.232)}&{(1,830.251)}&{(1,866.171)}\\
    \quad Partisan score&{-0.015}&{0.012}&{-0.005}\\
    \quad &{(0.805)}&{(0.806)}&{(0.800)}\\
    \quad Misinfo exposure score&{0.564}&{0.571}&{0.564}\\
    \quad &{(0.141)}&{(0.142)}&{(0.140)}\\
    \multicolumn{4}{l}{\underline{Count collection}}\\
    \quad \# Daily count observations: original posts&{3,244,360}&{1,626,744}&{1,617,616}\\
    \quad \# Weekly count observations: original posts&{463,480}&{232,392}&{231,088}\\
    \\
    \multicolumn{4}{l}{\underline{Post collection}}\\
    \quad \# Original posts from included accounts&{11,909,591}&{8,846,114}&{9,677,317}\\
    \\
    \multicolumn{4}{l}{\underline{Posting preferences before (pseudo-) correction}}\\
    \quad Post frequency (weekly)&{49.973}&{48.397}&{51.160}\\
    \quad &{(138.976)}&{(134.279)}&{(142.404)}\\
    \quad Misleadingness (daily)&{0.469}&{0.473}&{0.466}\\
    \quad &{(0.233)}&{(0.227)}&{(0.236)}\\
    \quad Positive sentiment (weekly)&{0.262}&{0.259}&{0.264}\\
    \quad &{(0.171)}&{(0.160)}&{(0.178)}\\
    \quad Negative sentiment (weekly)&{0.320}&{0.321}&{0.319}\\
    \quad &{(0.178)}&{(0.173)}&{(0.182)}\\
    \quad Toxicity (weekly)&{0.073}&{0.079}&{0.068}\\
    \quad &{(0.090)}&{(0.095)}&{(0.087)}\\
    \quad Confidence (weekly)&{6.267}&{6.253}&{6.277}\\
    \quad &{(0.468)}&{(0.473)}&{(0.463)}\\
    \quad Politics (weekly)&{21.4\%}&{19.8\%}&{22.6\%}\\
    \bottomrule
    \end{tabular}
    \label{tab:data_overview}
\end{table}

\clearpage
\subsection{GPT-estimated misleadingness}
\label{supp:gpt_misleading_val}

We assess the misleadingness of collected posts using a GPT-based misleadingness assessment framework proposed in prior work~\cite{chuai2026request}, implemented with GPT-5.4~mini (version: gpt-5.4-mini-2026-03-17) and Prompt~1 without web search. Here, we use the mini model for reasons of cost and scale, as we need to assess millions of posts. Specifically, we apply the framework to original posts published by the authors in our analysis during the four days before and four days after each (pseudo-)correction event. Of the \num{2568004} posts submitted for assessment, \num{2567990} received valid scores. We validate the performance of GPT-5.4~mini in misleadingness assessment against three sources of evidence: (i) \num{2159} claims fact-checked by PolitiFact from September 29, 2022 to July 30, 2024, covering a period similar to that of our study, (ii) \num{200} posts manually annotated by seventeen student assistants in prior work~\cite{chuai2026request}, and (iii) a comparison with associated community notes statuses.

\begin{figure}[ht]
    \centering
    \captionsetup[subfloat]{font={bf, small}, skip=0pt, singlelinecheck=false, labelformat=simple, position=top}
    \subfloat[]{\includegraphics[width = .32\textwidth]{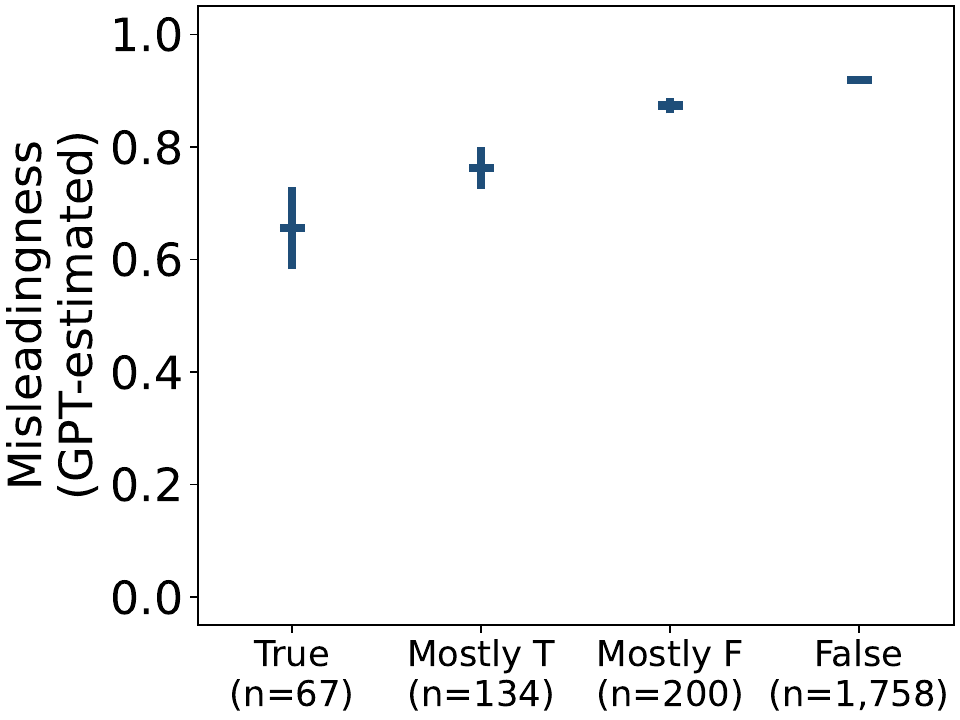}}
    \hfill
    \subfloat[]{\includegraphics[width = .32\textwidth]{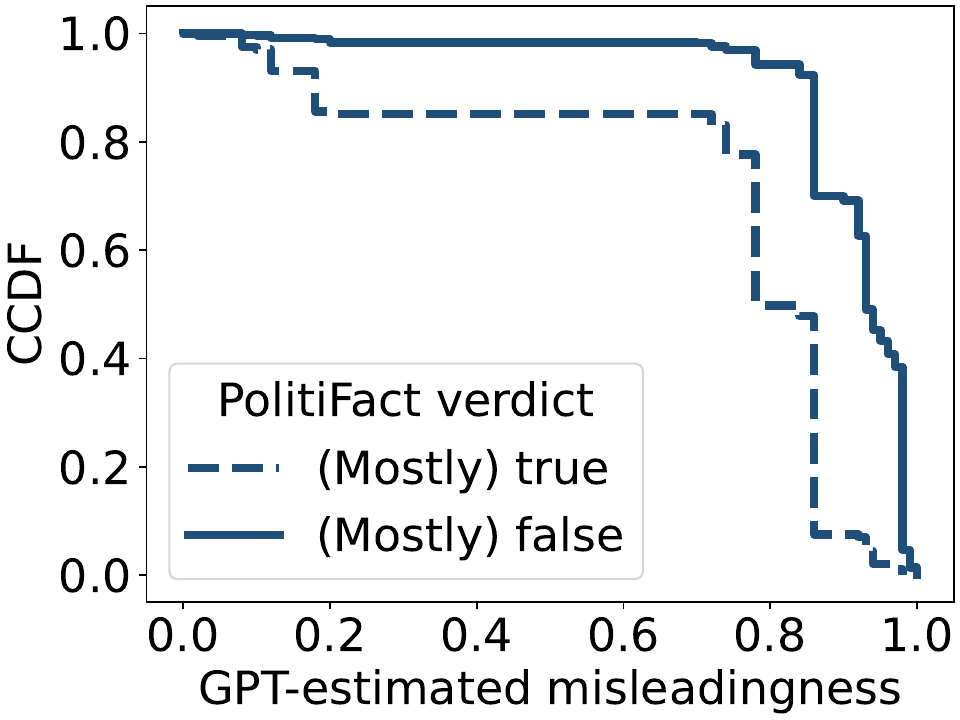}}
    \hfill
    \subfloat[]{\includegraphics[width = .32\textwidth]{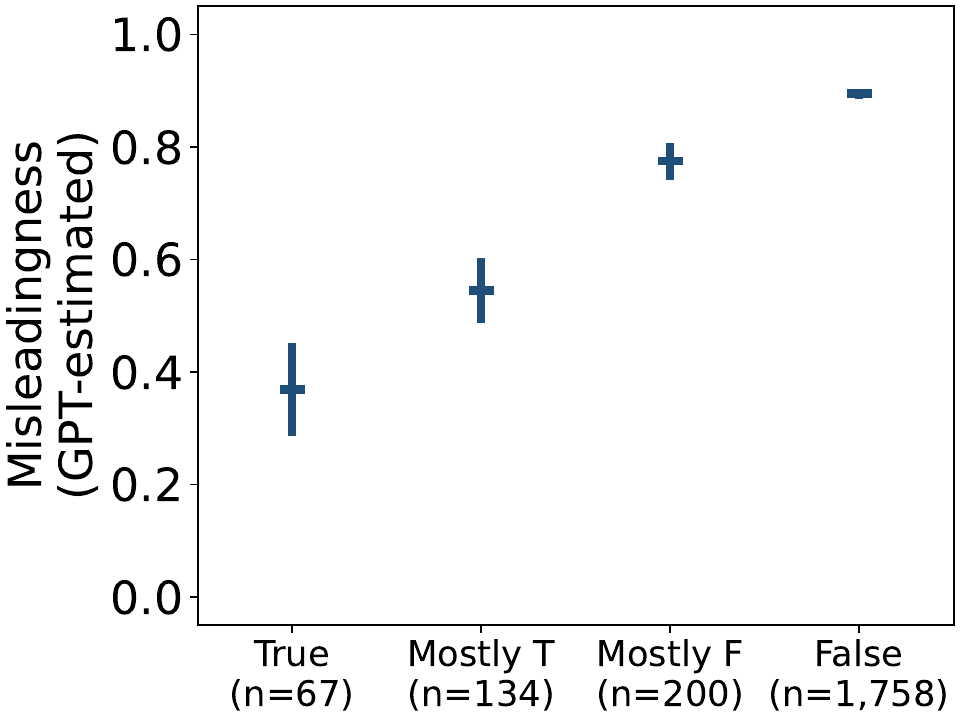}}
    \\
    \subfloat[]{\includegraphics[width = .32\textwidth]{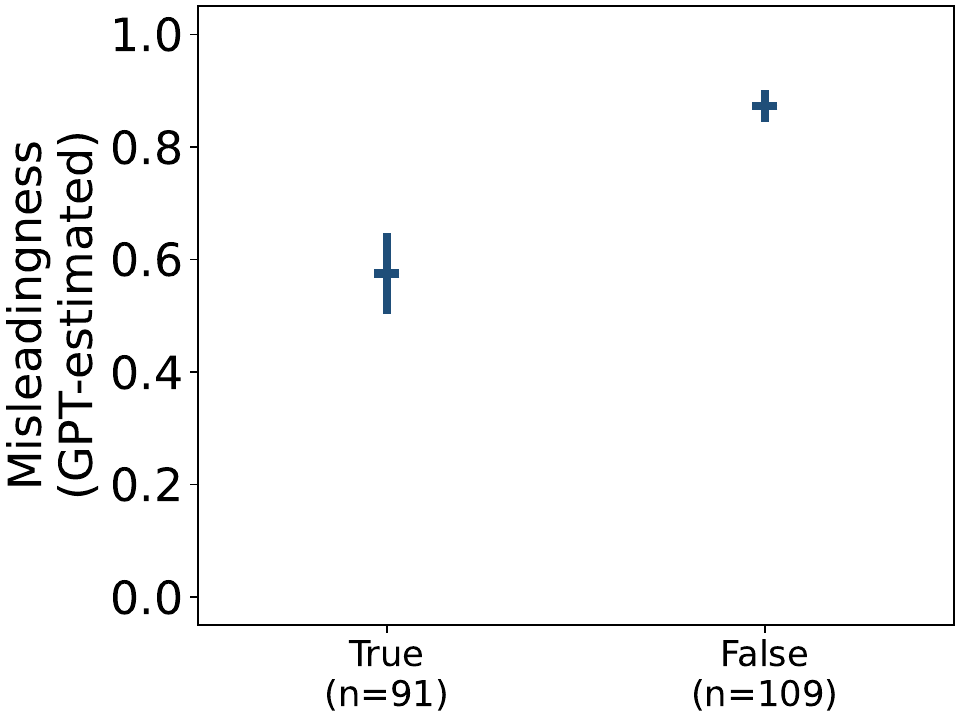}}
    \hfill
    \subfloat[]{\includegraphics[width = .32\textwidth]{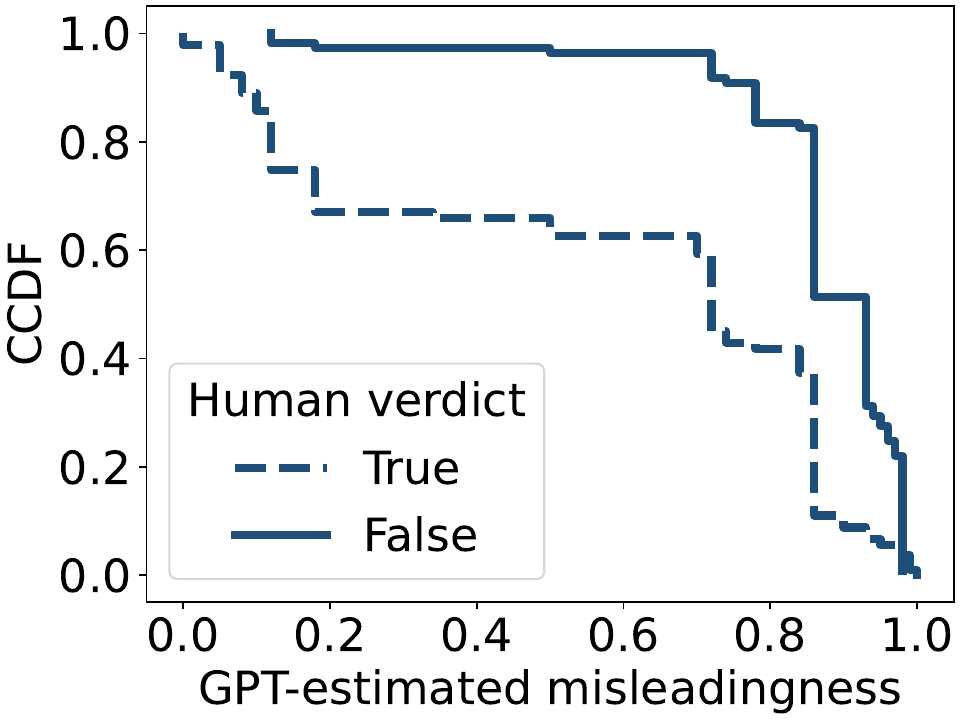}}
    \hfill
    \subfloat[]{\includegraphics[width = .32\textwidth]{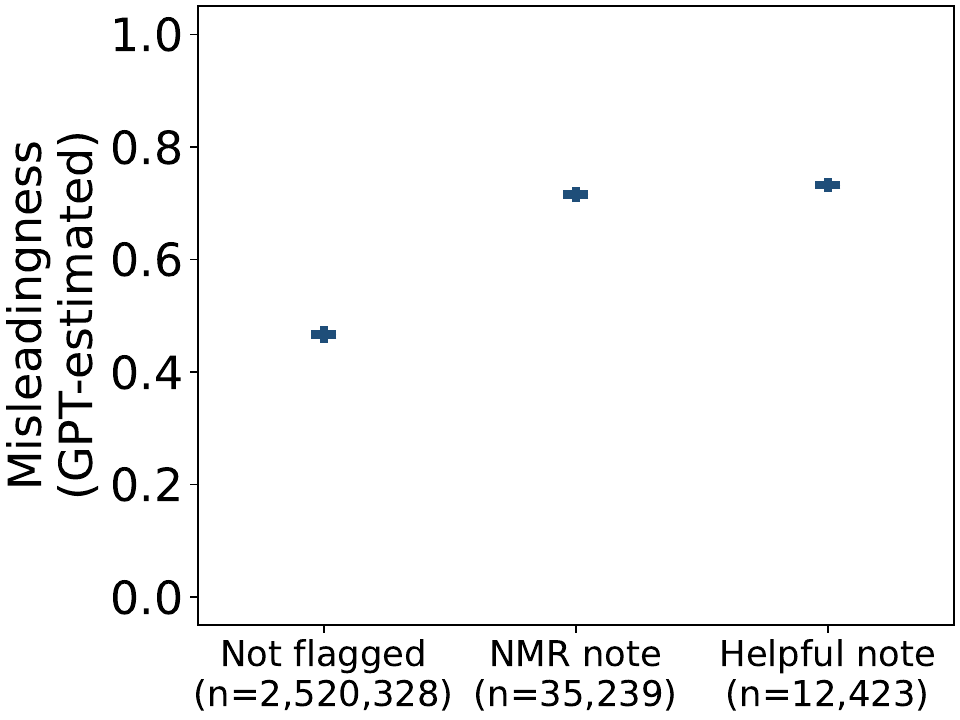}}
    \caption{\textbf{Validation of GPT-estimated misleadingness against PolitiFact verdicts, human-annotated labels, and community note statuses.} \textbf{(a)}~Mean misleadingness scores across four PolitiFact verdict categories ($n=\num{2159}$ claims).
    \textbf{(b)}~Complementary cumulative distribution functions (CCDFs) for true/mostly-true and false/mostly-false PolitiFact claims.
    \textbf{(c)}~Mean misleadingness scores estimated by the full-sized GPT-5.4 model across four PolitiFact verdict categories ($n=\num{2159}$ claims), providing a model-sensitivity comparison with (a).
    \textbf{(d)}~Mean misleadingness scores for posts classified as true ($n=91$) or false ($n=109$) by human annotators.
    \textbf{(e)}~CCDFs for the true and false groups shown in (d).
    \textbf{(f)}~Mean misleadingness scores for posts with no matched note, NMR notes, or displayed helpful notes. Displayed helpful notes take precedence when a post matches both NMR and helpful notes.
    Error bars indicate 95\% confidence intervals calculated using Student's $t$ distribution in (a), (c), and (d) and account-clustered standard errors with $t$ critical values in (f).}
    \label{fig:gpt_mis_val}
\end{figure}

\vspace{1.5em}
\noindent \textbf{PolitiFact verdicts.}\\
The PolitiFact data include \num{1758} false, \num{200} mostly-false, \num{134} mostly-true, and \num{67} true claims. We first evaluate GPT-estimated misleadingness across the four verdicts (\Cref{fig:gpt_mis_val}a), and find that the average GPT-estimated misleadingness scores increase monotonically from true (mean $=0.656$, 95\%~CI: $[0.583, 0.730]$) to mostly-true (mean $=0.763$, 95\%~CI: $[0.726, 0.800]$), mostly-false (mean $=0.873$, 95\%~CI: $[0.860, 0.887]$), and false verdicts (mean $=0.918$, 95\%~CI: $[0.913, 0.924]$). A joint Wald $F$-test using heteroskedasticity-robust standard errors rejects equality of the group means ($F(3, 2155)=48.450$, $p<0.001$). Additionally, we group false and mostly-false verdicts into a false category, and true and mostly-true verdicts into a true category (\Cref{fig:gpt_mis_val}b). GPT-estimated misleadingness for true claims (mean of 0.727) is significantly lower than the estimated misleadingness for false claims (mean of 0.914; $KS = 0.625$, $p < 0.001$). Using a misleading threshold of 0.86, which lies just below the confidence interval of the mostly-false verdict (\Cref{fig:gpt_mis_val}a), GPT-5.4~mini identifies true and false claims with a macro-$F_1$ score of 0.698.
Because the classes are heavily imbalanced toward false claims, we also report class-specific recall: 0.923 for (mostly) false claims but 0.522 for (mostly) true claims. This suggests that, at the selected threshold, the GPT-based classifier shows high sensitivity to false claims but lower specificity for true claims. Additionally, we resubmitted the assessment requests using the full-sized GPT-5.4 model (version: gpt-5.4-2026-03-05), and the estimated scores across the four verdicts are shown in \Cref{fig:gpt_mis_val}c. The estimated misleadingness scores exhibit a consistent increasing trend across the four PolitiFact verdict categories, from true (mean $=0.369$, 95\%~CI: $[0.285, 0.452]$) and mostly true (mean $=0.545$, 95\%~CI: $[0.487, 0.603]$) to mostly false (mean $=0.774$, 95\%~CI: $[0.742, 0.807]$) and false (mean $=0.895$, 95\%~CI: $[0.886, 0.904]$). In particular, posts classified as mostly false or false receive substantially higher misleadingness scores than those classified as true or mostly true ($F(3, 2155)=108.933$, $p<0.001$). The full-sized model produces lower scores for true and mostly true posts while maintaining relatively high scores for mostly false and false posts, indicating greater differentiation between the two ends of the veracity spectrum. Similarly, using the threshold defined using the bottom value ($=0.742$) of confidence interval for mostly false claims, the macro-$F_1$ score is 0.725, with recall of 0.908 for (mostly) false and 0.657 for (mostly) true claims, respectively. Although GPT-5.4 demonstrates slightly better classification performance than GPT-5.4 mini, we retain the latter for large-scale processing due to its greater computational efficiency and lower cost.

\vspace{1.5em}
\noindent \textbf{Human assessments.}\\
In addition to the validation against PolitiFact data, we evaluate GPT-5.4~mini using 200 human-annotated posts, comprising 109 false and 91 true posts, sourced from the prior work that validated the assessment framework using GPT-4.1~\cite{chuai2026request}. The mean GPT-estimated misleadingness is 0.575 (95\%~CI: $[0.503, 0.647]$) for true posts and 0.873 (95\%~CI: $[0.845, 0.901]$) for false posts (\Cref{fig:gpt_mis_val}d). The difference between true and false posts is statistically significant (false $-$ true $=0.298$, 95\%~CI: $[0.221, 0.375]$, $p<0.001$). The corresponding score distributions also differ statistically ($KS = 0.480$, $p<0.001$; \Cref{fig:gpt_mis_val}e). When using the misleading threshold of 0.86 derived from the PolitiFact validation, GPT-5.4~mini achieves a macro-$F_1$ score of 0.728, which is close to 0.800 using GPT-4.1 on the same sample. These results demonstrate that GPT-5.4~mini effectively distinguishes between human-annotated true and false posts.

\vspace{1.5em}
\noindent \textbf{Community note statuses.}\\
We compare these scores across posts with displayed helpful notes ($n=\num{12423}$), NMR notes ($n=\num{35239}$), and no match in either status ($n=\num{2520328}$; displayed helpful notes take precedence when a post has both helpful and NMR notes; see \Cref{fig:gpt_mis_val}f). Mean misleadingness is 0.732 for posts with displayed helpful notes (95\%~CI: $[0.720, 0.744]$), 0.716 for posts with NMR notes (95\%~CI: $[0.702, 0.730]$), and 0.467 for unmatched posts (95\%~CI: $[0.452, 0.481]$). Confidence intervals are calculated using account-clustered standard errors. The account-clustered joint Wald $F$-test rejects equality of group means ($F(2,15540)=657.271$, $p<0.001$). Both note-associated groups score higher than unmatched posts, and displayed-note posts score slightly higher than NMR-note posts (difference $=0.016$, 95\%~CI: $[0.005, 0.027]$, $p=0.004$). These associations provide complementary evidence supporting correspondence between GPT-estimated scores and note statuses. Additionally, the small mean difference between NMR-note and displayed-note posts is also consistent with prior research showing that cross-perspective agreement requirements can prevent notes on polarizing content from being displayed~\cite{bouchaud2026community}. Thus, the absence of a displayed note should not be interpreted as evidence that the underlying post is accurate.

Together, these validation analyses support using GPT-estimated scores to characterize relative differences in misleadingness across groups. We therefore interpret the continuous score as a proxy and threshold-based outcomes as model-classified content rather than verified misinformation.

\begin{tcolorbox}[colback=blue!5, colframe=blue!40, title=\textbf{Prompt~1. Instruction for misleadingness measurement}]
\setlength{\parskip}{0.8em}   
\setlength{\parindent}{0pt}   
\linespread{1}\selectfont     
You are a professional fact-checker to assess whether the given tweet contains misinformation. Misinformation refers to false or misleading information that is contradicted by empirical evidence or inconsistent with common or expert consensus, without imposing assumptions on the intent or the format of the content.

Indicators of misinformation or misleadingness:\\
- It contains a factual error\\
- It contains a digitally altered photo or video\\
- It contains outdated information that may be misleading\\
- It is a misrepresentation or missing important context\\
- It presents an unverified claim as a fact\\
- It is a joke or satire that might be misinterpreted as a fact

Return ONLY a float number from 0 to 1 to indicate the misleadingness of the tweet, where 0 = not misleading, 1 = extremely misleading.\\
Do not include any explanation or text.

\end{tcolorbox}

\clearpage

\clearpage
\subsection{Balance assessment}
We assess the balance between correction (treatment) and pseudo-correction (control) groups across covariates using standardized mean differences (SMDs), with an absolute SMD below 0.10 indicating acceptable balance. As shown in \Cref{tab:covariate_balance}, 13 of the 14 examined covariates meet this criterion. The groups are well balanced in correction timing, basic account characteristics (followers, followees, verified status, and account age), and most pre-correction posting characteristics, including posting frequency, sentiment, toxicity, misleadingness, opinion confidence, and political content. Only misinformation exposure score exhibits a modest difference between the two groups (Abs.~SMD $=0.108$). Overall, these results indicate that the treatment and control groups are comparable across most observed characteristics.

\begin{table}[ht]
\footnotesize
\centering
\caption{\textbf{Covariate balance between treatment and control groups.} Balance is assessed at the correction-event level using standardized mean differences (SMDs), with an absolute SMD below 0.10 indicating acceptable balance. Sample sizes represent nonmissing observations for each covariate and may vary due to missing data. Individual accounts may contribute multiple correction events.}
\label{tab:covariate_balance}
\begin{tabular}{llrrSSl}
\toprule
Covariate & Type & N treated & N control & {SMD} & {Abs. SMD} & Balanced \\
\midrule
\multicolumn{7}{l}{\underline{Account characteristics}}\\
\# Followers & continuous & 29,049 & 28,886 & -0.014 & 0.014 & Yes \\
\# Followees & continuous & 29,049 & 28,886 & 0.016 & 0.016 & Yes \\
Verified & binary & 29,049 & 28,886 & 0.030 & 0.030 & Yes \\
Account age & continuous & 29,049 & 28,886 & 0.000 & 0.000 & Yes \\
Partisan score & continuous & 16,748 & 17,227 & 0.092 & 0.092 & Yes \\
Misinfo exposure score & continuous & 16,748 & 17,227 & 0.108 & 0.108 & No \\
\\
\multicolumn{7}{l}{\underline{Correction timing and pre-correction posting preferences}}\\
Correction timing & continuous & 29,049 & 28,886 & 0.027 & 0.027 & Yes \\
Frequency (weekly) & continuous & 29,049 & 28,886 & 0.069 & 0.069 & Yes \\
Misleadingness (daily) & continuous & 16,593 & 17,597 & 0.082 & 0.082 & Yes \\
Positive sentiment (weekly) & continuous & 17,421 & 18,514 & -0.066 & 0.066 & Yes \\
Negative sentiment (weekly) & continuous & 17,421 & 18,514 & 0.017 & 0.017 & Yes \\
Toxicity (weekly) & continuous & 17,421 & 18,514 & 0.076 & 0.076 & Yes \\
Confidence (weekly) & continuous & 16,913 & 17,931 & -0.028 & 0.028 & Yes \\
Politics (weekly) & continuous & 17,421 & 18,514 & -0.039 & 0.039 & Yes \\
\bottomrule
\end{tabular}
\end{table}

\clearpage
\subsection{Correction frequency and historical coverage across accounts}
\label{supp:note_frequency}
Our dataset includes correction events for which both the associated posts and their accounts remained accessible at the time of data collection. Consequently, deleted posts and accounts that had been suspended or protected were excluded. Given that account suspension and protection occur at the account level and that displayed community notes have no statistically significant effects on these outcomes, but can significantly increase the likelihood of post deletion~\cite{chuai2026community}, we additionally consider deleted posts from accounts included in our dataset to obtain a more comprehensive account of their correction histories. However, the current \X API endpoints do not provide author information for deleted posts, preventing us from directly linking these posts to their authors. To remedy this, we supplement our dataset with historical correction records from the Community Notes Leaderboard (\url{https://community-notes-leaderboard.com/}), which regularly collects information on displayed community notes and their associated posts and authors on a daily basis. These records enable us to recover additional correction histories involving deleted posts and link them to the accounts in our dataset. We downloaded the available correction records from the Community Notes Leaderboard on September 27, 2026, and matched them to accounts included in our dataset using author identifiers. This process identifies an additional 404 deleted posts associated with accounts in our dataset, allowing us to complement their observed correction histories.

\begin{figure}[ht]
    \centering
    \captionsetup[subfloat]{font={bf, small}, skip=0pt, singlelinecheck=false, labelformat=simple, position=top}
    \subfloat[]{\includegraphics[width = .32\textwidth]{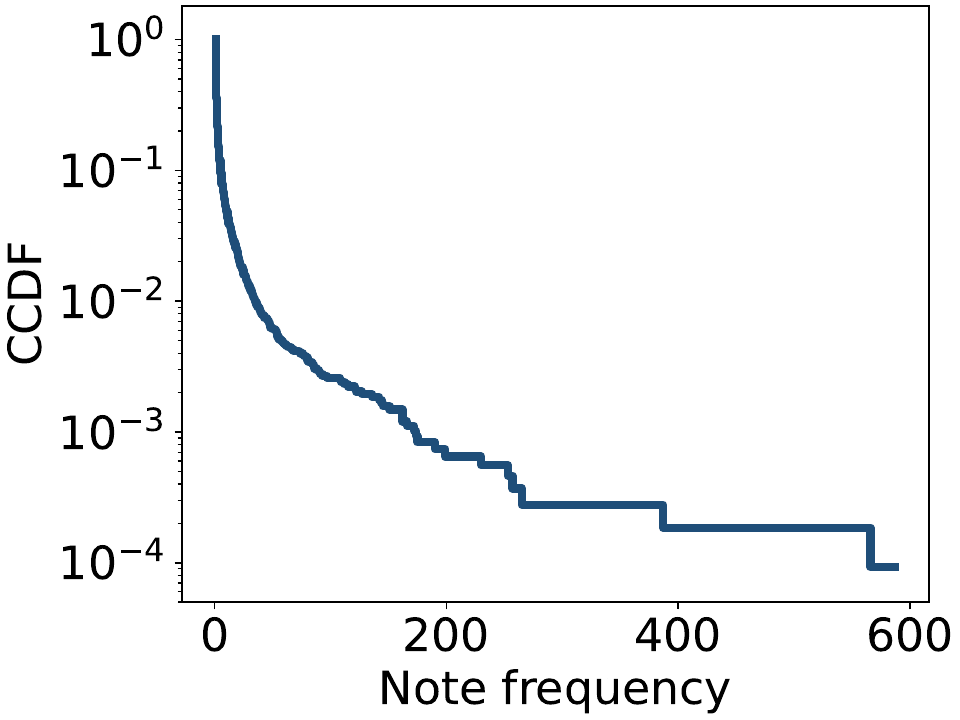}}
    \hfill
    \subfloat[]{\includegraphics[width = .32\textwidth]{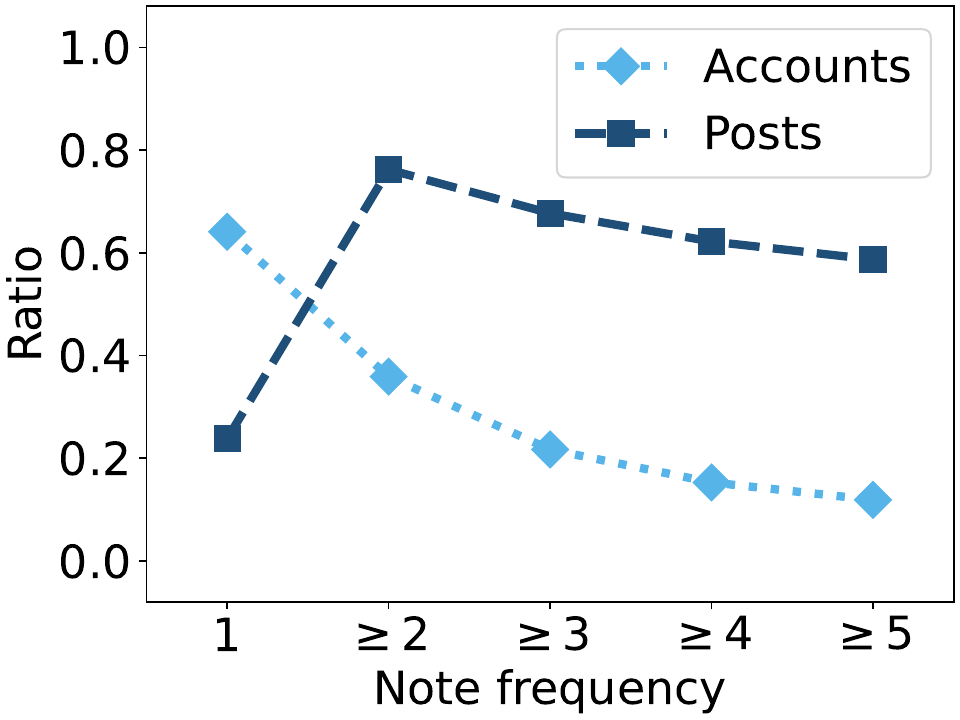}}
    \hfill
    \subfloat[]{\includegraphics[width = .32\textwidth]{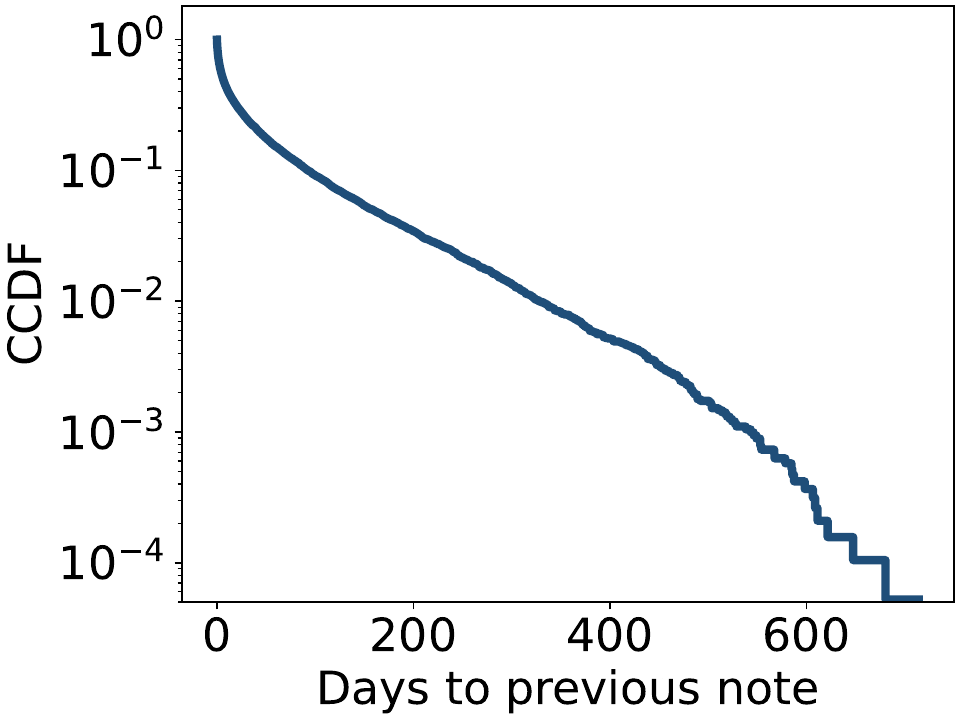}}
    \caption{\textbf{Distribution of note frequency (the number of corrections).} \textbf{(a)}~The CCDF (Complementary Cumulative Distribution Function) of note corrections received by each corrected account. \textbf{(b)}~The distribution of corrections by note frequency across accounts and their corrected posts. \textbf{(c)}~The CCDF of days to previous note for accounts that received two or more note corrections.}
    \label{fig:note_ccdf}
\end{figure}

Among all accounts whose misleading posts were corrected, the majority received only one community note. \Cref{fig:note_ccdf}a shows the distribution of note frequency for corrected accounts: \num{6922} (64.1\%) accounts received only one note, while \num{3877} (35.9\%), \num{2341} (21.7\%), \num{1649} (15.3\%) and \num{1287} (11.9\%) accounts received $\geq 2$, $\geq 3$, $\geq 4$, $\geq 5$ community notes, respectively. However, the majority of fact-checked misleading posts, \num{22127} in total (76.2\%), originated from accounts that received $\geq 2$ community notes, suggesting that a small subset of repeatedly corrected accounts accounts for a disproportionate share of the misinformation flagged on \X (\Cref{fig:note_ccdf}b). Additionally, for those accounts with $\geq 2$ community note corrections (\Cref{fig:note_ccdf}c), the average interval between consecutive notes is 31.988 days.

\clearpage

\clearpage
\subsection{Statistics of prior corrections for focal correction events}
\label{supp:prior_corrections}

Because posting activity was collected at the correction-event level, individual accounts may contribute multiple focal correction events to the treatment group and may also contribute matched posts with pseudo-correction events to the control group. Consequently, correction events are not necessarily independent at the account level, and accounts may have experienced previous corrections before a given focal (pseudo-)correction event. Here, we examine the number of prior corrections preceding each focal event in the treatment and control groups. 

As shown in \Cref{tab:prior_correction}, 34.2\% of treatment events and 45.7\% of control events are associated with accounts that have not previously received a correction (\ie, displayed community note). The proportions of events associated with one to four prior corrections are considerably smaller in both groups. However, when considering events with five or more prior corrections collectively, these account for 42.0\% of treatment events and 31.6\% of control events. This indicates that repeated correction exposure is common among accounts contributing events to both groups, particularly the treatment group. To minimize potential confounding from previous corrections, we focus our main analysis on focal events with no prior corrections, thereby examining posting activity around the first observed correction. In additional analyses, we extend the investigation to multiple correction events associated with one account, examining posting activity conditional on prior corrections. This approach allows us to assess whether the estimated response to a focal correction varies with an account's correction history, while distinguishing the effects associated with first corrections from those associated with subsequent corrections.

\begin{table}[ht]
    \centering
    \caption{\textbf{Distribution of prior corrections across focal (pseudo-)correction events.}}
    \setlength{\tabcolsep}{20pt}
    \begin{tabular}{lcc}
    \toprule
    \# Prior corrections & Control, n (\%) & Treatment, n (\%) \\
    \midrule
    0 & 13,211 (45.7\%) & 9,948 (34.2\%) \\
    1 & 2,921 (10.1\%) & 2,984 (10.3\%) \\
    2 & 1,611 (5.6\%) & 1,739 (6.0\%) \\
    3 & 1,180 (4.1\%) & 1,208 (4.2\%) \\
    4 & 837 (2.9\%) & 971 (3.3\%) \\
    $\geq$ 5 & 9,126 (31.6\%) & 12,199 (42.0\%) \\
    \bottomrule
    \end{tabular}
    \label{tab:prior_correction}
\end{table}

\clearpage
\subsection{Correction recurrence at fixed horizons}
\label{supp:correction_recurrence}

\begin{figure}[ht]
    \centering
    \captionsetup[subfloat]{font={bf, small}, skip=0pt, singlelinecheck=false, labelformat=simple, position=top}
    \subfloat[]{\includegraphics[width = .47\textwidth]{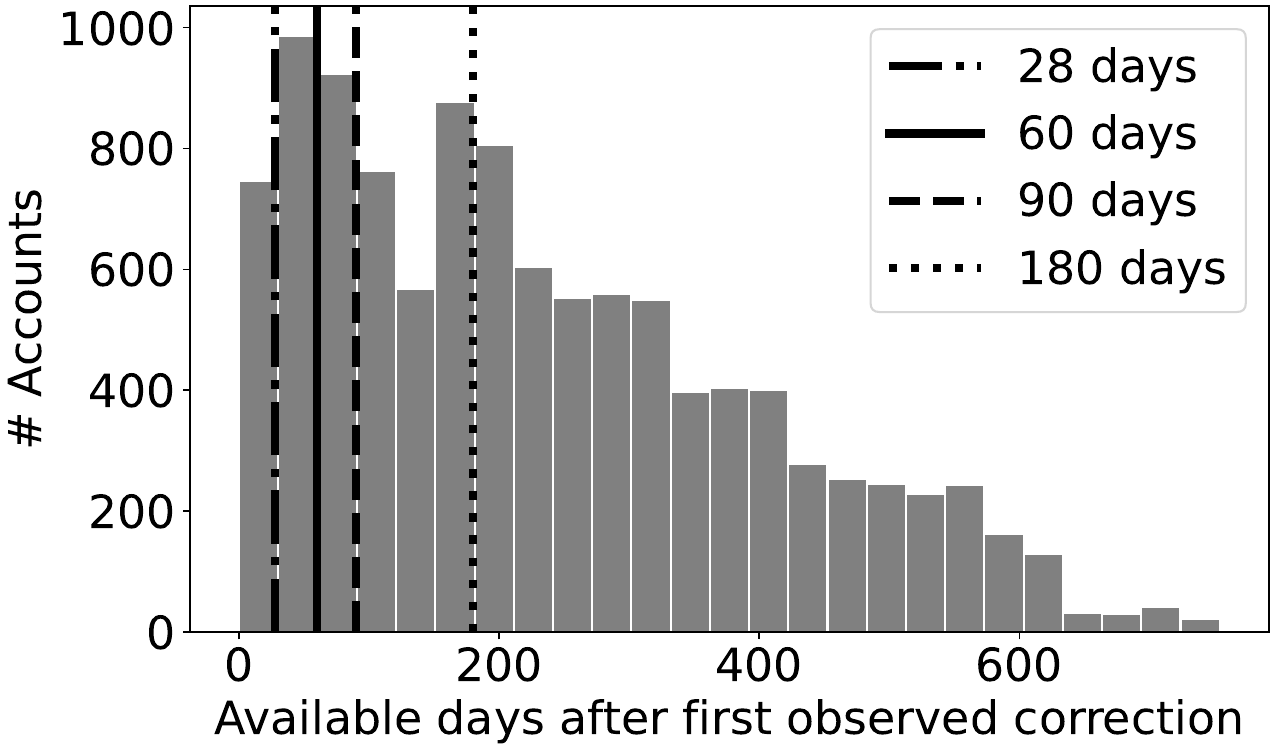}}
    \hfill
    \subfloat[]{\includegraphics[width = .47\textwidth]{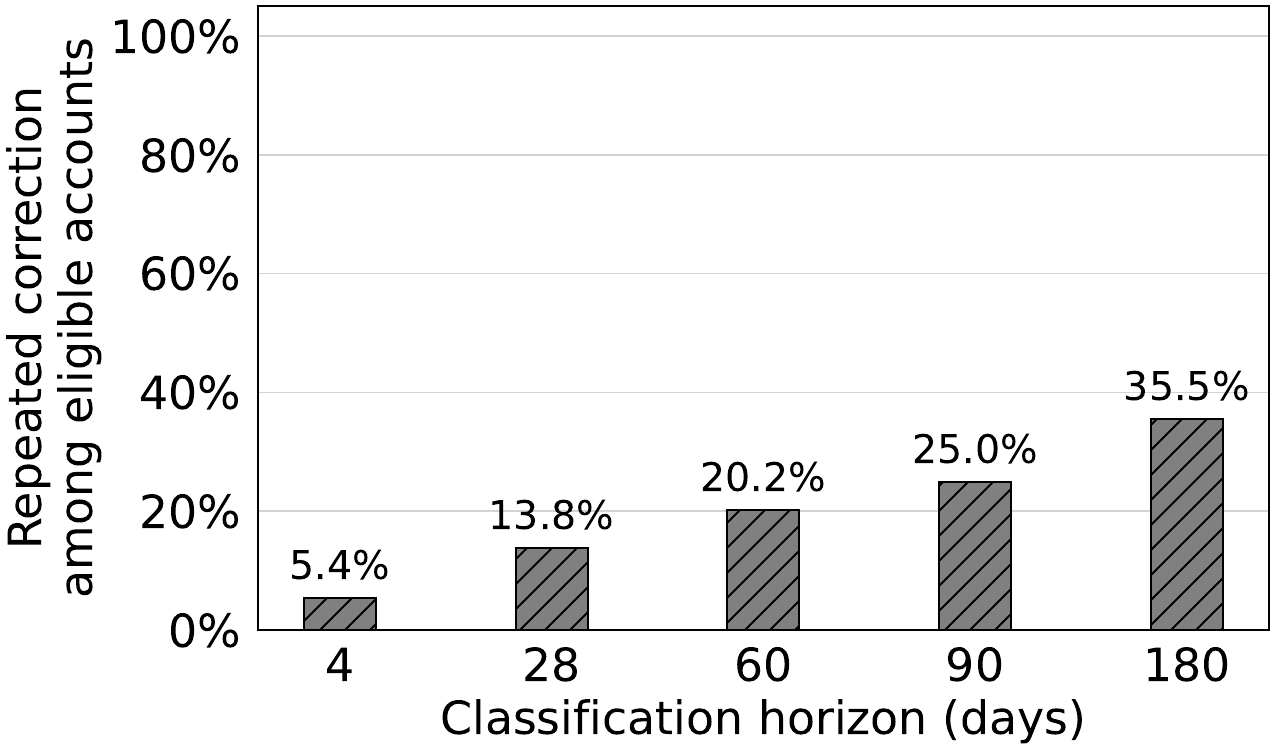}}
    
    \subfloat[]{\includegraphics[width = .47\textwidth]{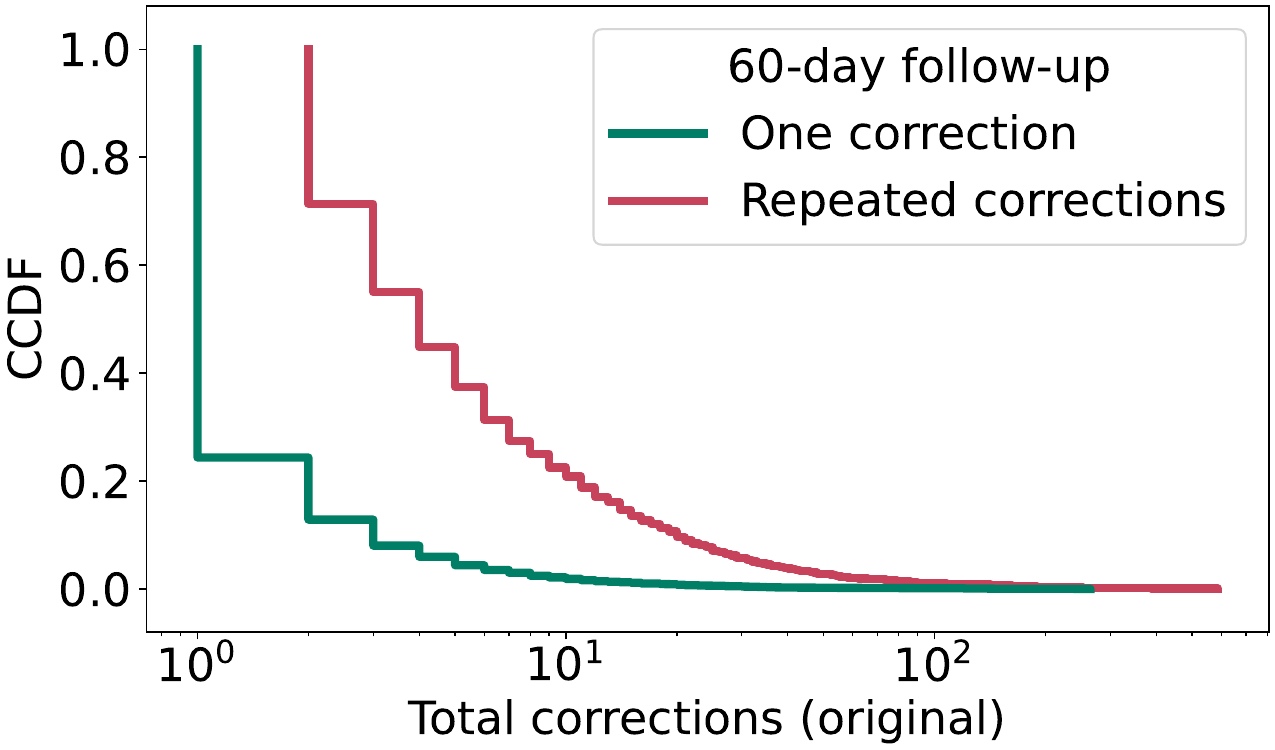}}
    \hfill
    \subfloat[]{\includegraphics[width = .47\textwidth]{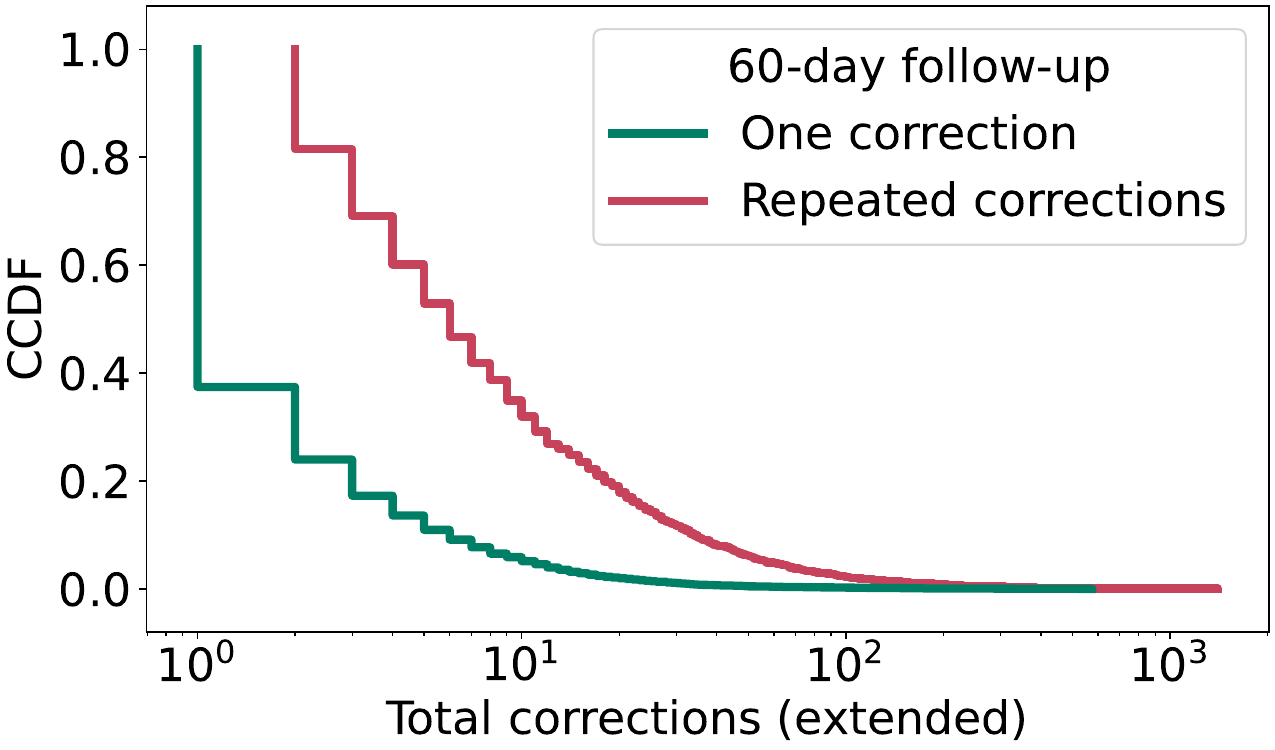}}
    \caption{\textbf{Repeated corrections at fixed follow-up horizons and the complementary cumulative distribution functions (CCDFs) for total corrections received by accounts during original and extended histories.} 
    \textbf{(a)}~The distribution of available follow-up days after the first observed correction for each account. 
    \textbf{(b)}~The proportions of accounts with observed repeated corrections among eligible accounts across different follow-up horizons. 
    \textbf{(c)}~The CCDFs of total corrections for one-correction and repeated-correction groups during 60 follow-up days since the first correction, based on original correction history in this study. 
    \textbf{(d)}~The CCDFs of total corrections for one-correction and repeated-correction groups during 60 follow-up days since the first correction, based on extended correction history collected from the Community Notes Leaderboard (\url{https://community-notes-leaderboard.com/}) on September 27, 2026.}
    \label{fig:fixed_horizons}
\end{figure}

We examine correction recurrence following accounts' first corrections. The available follow-up period is defined as the number of days between an account's first correction and the cutoff date of the correction history (\ie, June 11, 2024). As shown in \Cref{fig:fixed_horizons}a, the available follow-up periods vary substantially, ranging from 0.352 to 754.003 days, with an average of 234.570 days. This motivates the use of fixed observation horizons to ensure comparable opportunities for observing subsequent corrections. Here, we consider five follow-up horizons of 4, 28, 60, 90, and 180 days, which correspond to \num{10728} (99.3\%), \num{10110} (93.6\%), \num{9092} (84.2\%), \num{8163} (75.6\%), and \num{5961} (55.2\%) eligible accounts, respectively. The proportion of accounts receiving repeated corrections increases with the follow-up horizon, from 5.4\% within 4 days to 13.8\% within 28 days, 20.2\% within 60 days, 25.0\% within 90 days, and 35.5\% within 180 days (\Cref{fig:fixed_horizons}b).

We further compare the distributions of total corrections received by accounts classified as having one versus repeated corrections within the 60-day follow-up period. We select this horizon to balance sample coverage with a sufficiently long observation period beyond the 28-day post-treatment window in the main analysis. As shown in \Cref{fig:fixed_horizons}c, accounts experiencing repeated corrections within 60 days (mean $=10.283$) tend to have higher total correction counts in the original observation history than accounts receiving only one correction within the same period (mean $=2.062$; $KS = 0.756$, $p < 0.001$). This pattern remains evident when using the extended correction history collected from the Community Notes Leaderboard (\Cref{fig:fixed_horizons}d): mean $= 16.891$ for repeatedly corrected accounts and mean $=3.389$ for accounts with one correction ($KS = 0.626$, $p < 0.001$). Together, these results indicate that correction recurrence increases with longer follow-up periods, and is associated with greater overall correction exposure across both the original and extended observation histories. 

\clearpage

\clearpage
\subsection{Prior NMR-note history at fixed horizons}
\label{supp:prior_nmr_recurrence}

We further analyze corrected accounts' prior NMR-note histories, measured by the number of distinct posts associated with NMR notes that indicate misleadingness. Although NMR notes are not publicly displayed, they indicate that the corresponding posts have attracted attention from Community Notes contributors. We define repeated prior NMR notes as having at least two distinct posts associated with NMR notes within a fixed pre-correction observation window. We examine five horizons (4, 28, 60, 90, and 180 days), restricting each analysis to accounts with sufficient observation history before their first correction.

\begin{figure}[ht]
    \centering
    \captionsetup[subfloat]{font={bf, small}, skip=0pt, singlelinecheck=false, labelformat=simple, position=top}
    \includegraphics[width = .47\textwidth]{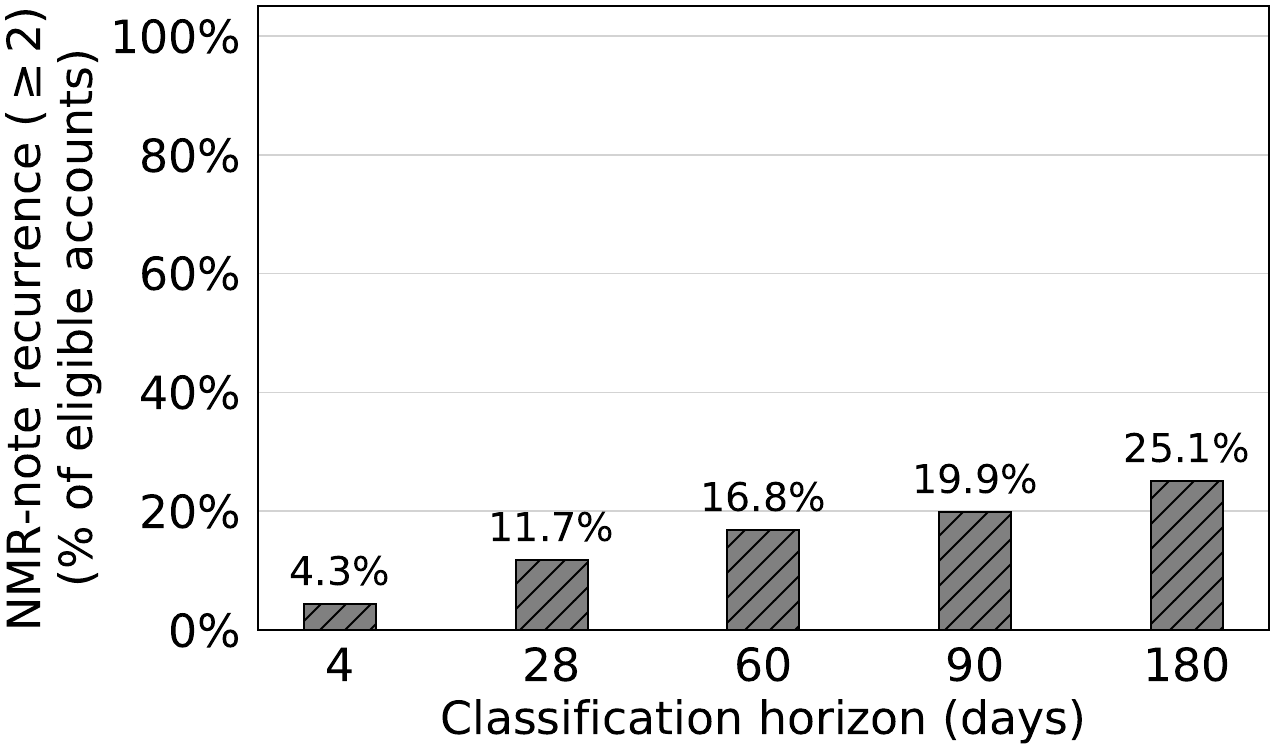}
    \caption{\textbf{Prevalence of accounts receiving repeated NMR notes across pre-correction observation horizons.} Percentage of eligible corrected accounts with at least two distinct posts associated with NMR notes within 4, 28, 60, 90, and 180 days before their first correction. The numbers of eligible accounts are \num{10558}, \num{10458}, \num{10224}, \num{10007}, and \num{9369}, respectively. Eligibility requires sufficient pre-correction observation history for the corresponding horizon.}
    \label{fig:prior_nmr_recurrence}
\end{figure}

As shown in \Cref{fig:prior_nmr_recurrence}, the proportion of eligible accounts exhibiting recurrent NMR-note histories increases with the observation horizon, from 4.3\% within 4 days (\num{459} of \num{10558} accounts) to 11.7\% within 28 days (\num{1227} of \num{10458}), 16.8\% within 60 days (\num{1719} of \num{10224}), 19.9\% within 90 days (\num{1991} of \num{10007}), and 25.1\% within 180 days (\num{2351} of \num{9369}). These results indicate that receiving repeated prior NMR notes is concentrated among a subset of corrected accounts, with NMR note recurrence becoming more prevalent over longer observation windows. We use the 28-day horizon for the primary prior-NMR-based heterogeneity analysis to align with the four-week pre-correction observation period, and examine longer horizons as sensitivity analyses.

\clearpage
\section{Changes in posting activity}
\label{supp:overall_changes}

\subsection{Estimation results for main analysis}
\label{supp:main_estimation}

\begin{figure}[ht]
    \centering
    \captionsetup[subfloat]{font={bf, small}, skip=0pt, singlelinecheck=false, labelformat=simple, position=top}
    \includegraphics[width= \linewidth]{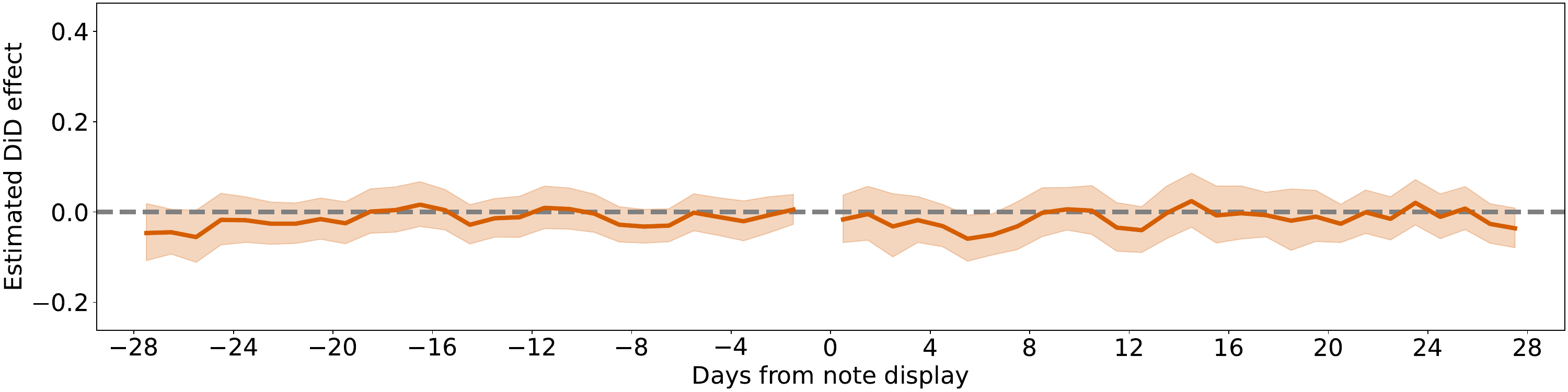}
    \caption{\textbf{Daily DiD estimates for the effects of the first correction on the original-post activity.} The estimation is based on 1,169,392 daily count observations across 20,882 correction events contributed by 17,696 accounts, using Day~$-1$ as the reference period. Treatment events are restricted to accounts' first corrections, excluding those with subsequent corrections in the 28-day post-treatment period. Control events include all pseudo-corrections without prior corrections. Error band represents 95\%~CIs. Full estimation results are reported in \Cref{tab:did_daily_long}.}
    \label{fig:did_daily_long}
\end{figure}

The daily leads-and-lags DiD estimates are shown in \Cref{fig:did_daily_long} and reported in \Cref{tab:did_daily_long}, corresponding to the changes in original-post counts in \Cref{fig:did_main}a. Using Day~$-1$ as the reference period, we find that none of the individual pre-correction daily DiD estimates, covering Days~$-28$ to~$-2$, is statistically distinguishable from zero at the 5\% level (all $p > 0.050$, see full estimation results in \Cref{tab:did_daily_long}). A joint Wald $\chi^2$ test also provides no statistically significant evidence of differential pre-correction changes ($\chi^2 (27) = 29.14$, $p=0.354$).
Following correction, the daily DiD estimates fluctuate around zero, with statistically significant negative effects observed at Day 6 (effect $= -0.059$, $z=-2.207$, $p=0.027$; 95\%~CI: $[-0.109, -0.007]$) and Day 7 (effect $= -0.051$, $z=-2.121$, $p=0.034$; 95\%~CI: $[-0.095, -0.004]$). However, a joint Wald $\chi^2$ test for the 28 post-correction estimates does not reject the null hypothesis that all effects are zero at the 5\% significance level ($\chi^2(28)=40.76$, $p=0.057$).
Overall, the aggregated daily effect (effect $= -0.001$, $z=-0.057$, $p=0.954$; 95\%~CI: $[-0.027, 0.026]$) is identical to the aggregated weekly effect, indicating that the estimated treatment effect is robust to the temporal aggregation of posting activity. The weekly leads-and-lags DiD estimates in \Cref{fig:did_main}b and are reported in \Cref{tab:did_weekly}.

\begin{figure}[ht]
    \centering
    \captionsetup[subfloat]{font={bf, small}, skip=0pt, singlelinecheck=false, labelformat=simple, position=top}
    \subfloat[]{\includegraphics[width= .8\linewidth]{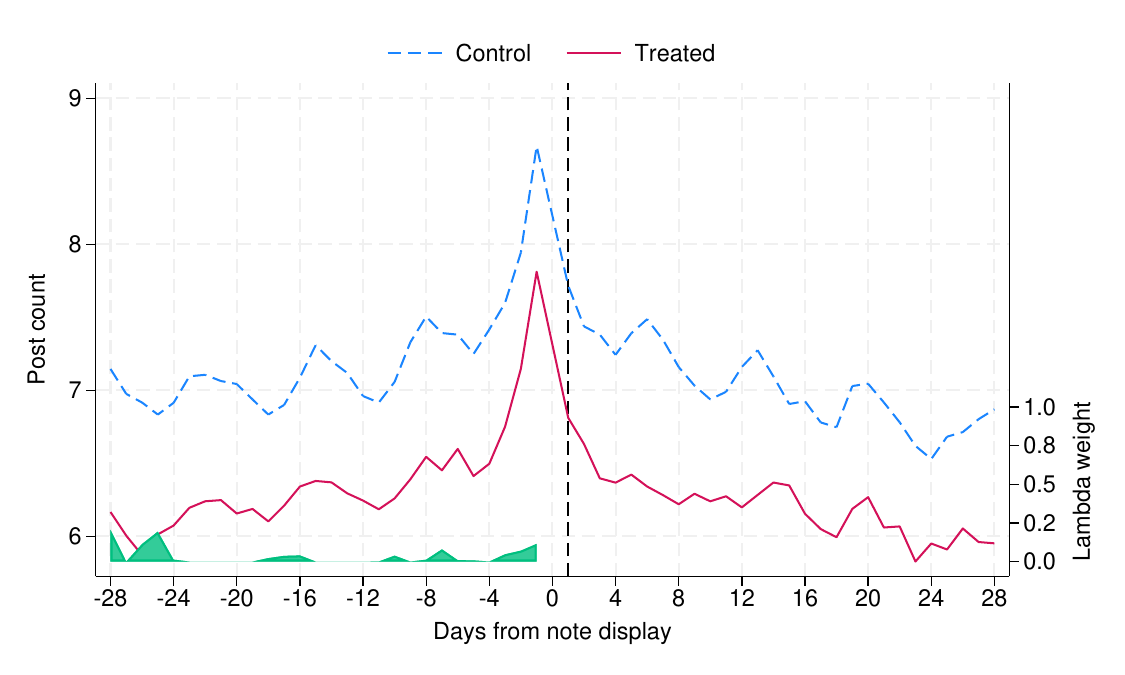}}
    \hfill
    \subfloat[]{\includegraphics[width= .8\linewidth]{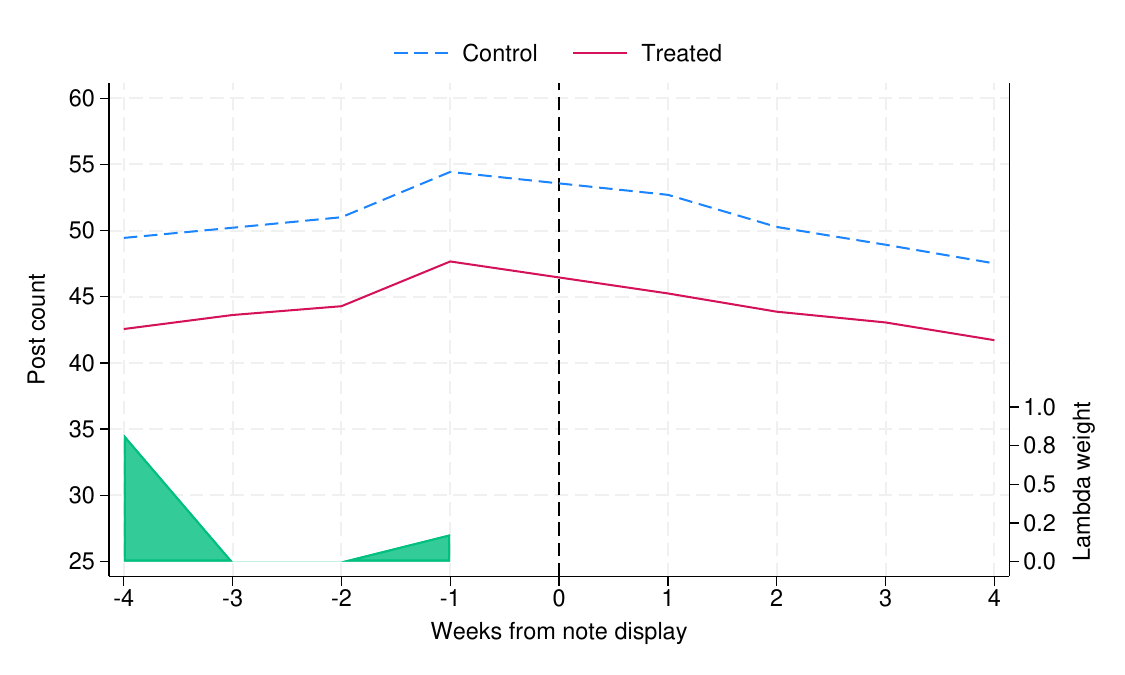}}
    \caption{\textbf{SDiD estimates of posting activity following the first correction at the daily and weekly levels.} \textbf{(a)}~Daily original-post activity. \textbf{(b)}~Weekly original-post activity. Solid red and dashed blue lines indicate the average daily or weekly original-post counts for treated and control accounts, respectively. Treatment events are restricted to accounts' first corrections, excluding those with subsequent corrections in the post-treatment period. Control events include all pseudo-corrections without prior corrections. The green area indicates the estimated pre-treatment time weights ($\lambda$). Posting activity is measured using daily or weekly original-post counts. Standard errors are estimated using 100 bootstrap replications with clustering at the account level.}
    \label{fig:sdid_posts_daily}
\end{figure}

Additionally, we employ synthetic difference-in-differences (SDiD) estimation as a robustness check, which constructs a weighted control group and assigns weights to pre-treatment periods to improve comparability between the treatment and control groups~\cite{arkhangelsky2021synthetic}. Here, we use the original post counts as the outcome to facilitate interpretation, although the estimates may be sensitive to highly active accounts due to the skewed distribution of posting activity. Therefore, we retain the Poisson DiD model as our main specification to better accommodate count outcomes and use SDiD as a complementary robustness check. As shown in \Cref{fig:sdid_posts_daily}, both groups exhibit similar posting trajectories before and after the first correction. Consistent with our main DiD findings, the estimated average treatment effects are close to zero and statistically insignificant (ATT $=0.059$, $t=0.84$, $p=0.398$; 95\%~CI: $[-0.078, 0.196]$). This provides further evidence that our findings are robust to an alternative estimation approach, with no statistically significant change in subsequent posting activity following the first correction.

\clearpage
\begin{table*}[ht]
\centering
\caption{\textbf{Leads-and-lags DiD estimates over 28 days before and after correction treatment.} Day~$-1$ serves as the reference period. The estimation is based on 1,169,392 daily count observations across 20,882 correction events contributed by 17,696 accounts. Treatment events are restricted to accounts' first corrections, excluding those with subsequent corrections in the 28-day post-treatment period. Control events include all pseudo-corrections without prior corrections. Event and day fixed effects are absorbed, and standard errors are clustered at the account level.}
\setlength{\tabcolsep}{4pt}
\begin{tabular}{@{}rrrrcrrrrc@{}}
\toprule
\multicolumn{5}{c}{Pre-correction} &
\multicolumn{5}{c}{Post-correction} \\
\cmidrule(lr){1-5}\cmidrule(lr){6-10}
Day & {Estimate} & {$z$} & {$p$} & 95\% CI &
Day & {Estimate} & {$z$} & {$p$} & 95\% CI \\
\midrule
$-28$ & $-0.047$ & $-1.415$ & $0.157$ & $[-0.107, 0.019]$ &
$1$   & $-0.016$ & $-0.610$ & $0.542$ & $[-0.067, 0.037]$ \\

$-27$ & $-0.045$ & $-1.729$ & $0.084$ & $[-0.093, 0.006]$ &
$2$   & $-0.005$ & $-0.150$ & $0.880$ & $[-0.062, 0.057]$ \\

$-26$ & $-0.056$ & $-1.837$ & $0.066$ & $[-0.111, 0.004]$ &
$3$   & $-0.032$ & $-0.884$ & $0.377$ & $[-0.099, 0.040]$ \\

$-25$ & $-0.017$ & $-0.590$ & $0.555$ & $[-0.073, 0.041]$ &
$4$   & $-0.018$ & $-0.687$ & $0.492$ & $[-0.067, 0.034]$ \\

$-24$ & $-0.018$ & $-0.706$ & $0.480$ & $[-0.067, 0.033]$ &
$5$   & $-0.031$ & $-1.290$ & $0.197$ & $[-0.077, 0.017]$ \\

$-23$ & $-0.026$ & $-1.071$ & $0.284$ & $[-0.071, 0.022]$ &
$6$   & $-0.059$ & $-2.207$ & $0.027$ & $[-0.109, -0.007]$ \\

$-22$ & $-0.026$ & $-1.120$ & $0.263$ & $[-0.069, 0.020]$ &
$7$   & $-0.051$ & $-2.121$ & $0.034$ & $[-0.095, -0.004]$ \\

$-21$ & $-0.016$ & $-0.672$ & $0.502$ & $[-0.060, 0.031]$ &
$8$   & $-0.032$ & $-1.159$ & $0.246$ & $[-0.083, 0.023]$ \\

$-20$ & $-0.025$ & $-1.043$ & $0.297$ & $[-0.070, 0.022]$ &
$9$   & $-0.002$ & $-0.067$ & $0.947$ & $[-0.054, 0.054]$ \\

$-19$ & $0.001$  & $0.039$  & $0.969$ & $[-0.047, 0.051]$ &
$10$  & $0.006$  & $0.251$  & $0.802$ & $[-0.040, 0.054]$ \\

$-18$ & $0.004$  & $0.163$  & $0.870$ & $[-0.045, 0.055]$ &
$11$  & $0.003$  & $0.107$  & $0.915$ & $[-0.050, 0.058]$ \\

$-17$ & $0.016$  & $0.649$  & $0.517$ & $[-0.032, 0.067]$ &
$12$  & $-0.035$ & $-1.240$ & $0.215$ & $[-0.087, 0.021]$ \\

$-16$ & $0.004$  & $0.182$  & $0.855$ & $[-0.040, 0.050]$ &
$13$  & $-0.040$ & $-1.527$ & $0.127$ & $[-0.090, 0.012]$ \\

$-15$ & $-0.028$ & $-1.256$ & $0.209$ & $[-0.071, 0.016]$ &
$14$  & $-0.003$ & $-0.086$ & $0.931$ & $[-0.059, 0.057]$ \\

$-14$ & $-0.014$ & $-0.635$ & $0.525$ & $[-0.056, 0.030]$ &
$15$  & $0.024$  & $0.806$  & $0.420$ & $[-0.034, 0.086]$ \\

$-13$ & $-0.012$ & $-0.496$ & $0.620$ & $[-0.056, 0.035]$ &
$16$  & $-0.008$ & $-0.234$ & $0.815$ & $[-0.069, 0.057]$ \\

$-12$ & $0.009$  & $0.384$  & $0.701$ & $[-0.037, 0.057]$ &
$17$  & $-0.003$ & $-0.091$ & $0.927$ & $[-0.060, 0.058]$ \\

$-11$ & $0.007$  & $0.283$  & $0.777$ & $[-0.038, 0.053]$ &
$18$  & $-0.007$ & $-0.278$ & $0.781$ & $[-0.055, 0.044]$ \\

$-10$ & $-0.004$ & $-0.168$ & $0.867$ & $[-0.045, 0.039]$ &
$19$  & $-0.019$ & $-0.551$ & $0.582$ & $[-0.085, 0.051]$ \\

$-9$  & $-0.028$ & $-1.396$ & $0.163$ & $[-0.066, 0.012]$ &
$20$  & $-0.010$ & $-0.357$ & $0.721$ & $[-0.065, 0.048]$ \\

$-8$  & $-0.032$ & $-1.671$ & $0.095$ & $[-0.069, 0.006]$ &
$21$  & $-0.026$ & $-1.196$ & $0.232$ & $[-0.067, 0.017]$ \\

$-7$  & $-0.030$ & $-1.593$ & $0.111$ & $[-0.066, 0.007]$ &
$22$  & $-0.001$ & $-0.026$ & $0.979$ & $[-0.048, 0.049]$ \\

$-6$  & $-0.001$ & $-0.071$ & $0.943$ & $[-0.042, 0.040]$ &
$23$  & $-0.015$ & $-0.619$ & $0.536$ & $[-0.062, 0.034]$ \\

$-5$  & $-0.011$ & $-0.518$ & $0.605$ & $[-0.052, 0.032]$ &
$24$  & $0.020$  & $0.799$  & $0.424$ & $[-0.029, 0.072]$ \\

$-4$  & $-0.021$ & $-0.903$ & $0.367$ & $[-0.064, 0.025]$ &
$25$  & $-0.011$ & $-0.424$ & $0.672$ & $[-0.059, 0.040]$ \\

$-3$  & $-0.007$ & $-0.353$ & $0.724$ & $[-0.046, 0.034]$ &
$26$  & $0.008$  & $0.315$  & $0.753$ & $[-0.039, 0.056]$ \\

$-2$  & $0.005$  & $0.314$  & $0.754$ & $[-0.027, 0.039]$ &
$27$  & $-0.026$ & $-1.169$ & $0.243$ & $[-0.069, 0.018]$ \\

      &         &          &         &                   &
$28$  & $-0.036$ & $-1.585$ & $0.113$ & $[-0.079, 0.009]$ \\
\bottomrule
\end{tabular}
\label{tab:did_daily_long}
\end{table*}

\clearpage
\begin{table*}[ht]
\centering
\caption{\textbf{Leads-and-lags DiD estimates over four weeks before and after correction treatment.} Week~$-1$ serves as the reference period. The estimation is based on 167,056 weekly count observations across 20,882 correction events contributed by 17,696 accounts. Treatment events are restricted to accounts' first corrections, excluding those with subsequent corrections in the four-week post-treatment period. Control events include all pseudo-corrections without prior corrections. Event and week fixed effects are absorbed, and standard errors are clustered at the account level.}
\setlength{\tabcolsep}{4pt}
\begin{tabular}{@{}rrrrcrrrrc@{}}
\toprule
\multicolumn{5}{c}{Pre-correction} &
\multicolumn{5}{c}{Post-correction} \\
\cmidrule(lr){1-5}\cmidrule(lr){6-10}
Week & {Estimate} & {$z$} & {$p$} & 95\% CI &
Week & {Estimate} & {$z$} & {$p$} & 95\% CI \\
\midrule
$-4$ & $-0.025$ & $-1.311$ & $0.190$ & $[-0.061, 0.013]$ &
$1$  & $-0.022$ & $-0.884$ & $0.377$ & $[-0.068, 0.027]$ \\
 
$-3$ & $0.002$  & $0.187$  & $0.852$ & $[-0.023, 0.028]$ &
$2$  & $-0.006$ & $-0.282$ & $0.778$ & $[-0.047, 0.036]$ \\
 
$-2$ & $-0.002$ & $-0.216$ & $0.829$ & $[-0.021, 0.017]$ &
$3$  & $0.002$  & $0.088$  & $0.930$ & $[-0.040, 0.045]$ \\
 
     &          &          &         &                   &
$4$  & $-0.000$ & $-0.003$ & $0.997$ & $[-0.028, 0.029]$ \\
\bottomrule
\end{tabular}
\label{tab:did_weekly}
\end{table*}

\clearpage
\subsection{HonestDiD estimation}
\label{supp:honest_did}

We adopt the HonestDiD estimator to account for potential violations of parallel-trends assumption and conduct robust inference and sensitivity analysis for our DiD estimates~\cite{rambachan2023honest}. Specifically, we implement the Conditional Least Favorable Hybrid (C-LF) method under the relative-magnitudes restriction, assuming that any deviations from parallel trends in the post-correction periods do not exceed $\overline{M}$ times the largest deviation observed in the pre-correction periods. 

As shown in \Cref{tab:honest_did_weekly}, we examine four post-correction weeks under different sensitivity parameters ($\overline{M}=0.5, 1, 1.5,$ and $2$). The original 95\% confidence intervals include zero across all four weeks, indicating no statistically significant changes in posting activity under the conventional parallel trends assumption. When allowing for potential deviations from parallel trends, the HonestDiD robust confidence intervals become progressively wider as $\overline{M}$ increases, reflecting greater uncertainty under weaker identifying assumptions. Nevertheless, the robust confidence intervals continue to include zero across all four weeks and all examined sensitivity parameters. Together, these findings indicate that our conclusion of no statistically significant changes in posting activity following the first correction remains unchanged when allowing for increasingly large deviations from parallel trends.

\begin{table}[ht]
\centering
\caption{\textbf{HonestDiD sensitivity analysis of weekly posting activity following the first observed correction.} Reported are the original 95\% confidence intervals and robust confidence intervals estimated using the C-LF method under the relative-magnitudes restriction, allowing post-treatment deviations from parallel trends of up to $\overline{M}$ times the maximum pre-treatment deviation (\Cref{fig:did_main}b).}
\label{tab:honest_did_weekly}
\setlength{\tabcolsep}{12pt}
\begin{tabular}{lcccc}
\toprule
$\overline{M}$ &
Week 1 & Week 2 & Week 3 & Week 4 \\
\midrule
Original
& $[-0.070,\ 0.026]$
& $[-0.048,\ 0.036]$
& $[-0.040,\ 0.044]$
& $[-0.028,\ 0.028]$ \\

0.5
& $[-0.087,\ 0.027]$
& $[-0.087,\ 0.045]$
& $[-0.099,\ 0.081]$
& $[-0.111,\ 0.111]$ \\

1.0
& $[-0.111,\ 0.033]$
& $[-0.135,\ 0.087]$
& $[-0.171,\ 0.153]$
& $[-0.213,\ 0.213]$ \\

1.5
& $[-0.135,\ 0.045]$
& $[-0.189,\ 0.135]$
& $[-0.249,\ 0.231]$
& $[-0.315,\ 0.315]$ \\

2.0
& $[-0.159,\ 0.063]$
& $[-0.237,\ 0.183]$
& $[-0.327,\ 0.303]$
& $[-0.423,\ 0.423]$ \\
\bottomrule
\end{tabular}
\end{table}

\clearpage
\subsection{Alternative model specifications}
\label{supp:alternative_models}

\begin{figure}[ht]
    \centering
    \captionsetup[subfloat]{font={bf, small}, skip=0pt, singlelinecheck=false, labelformat=simple, position=top}
    \subfloat[]{\includegraphics[width = .32\textwidth]{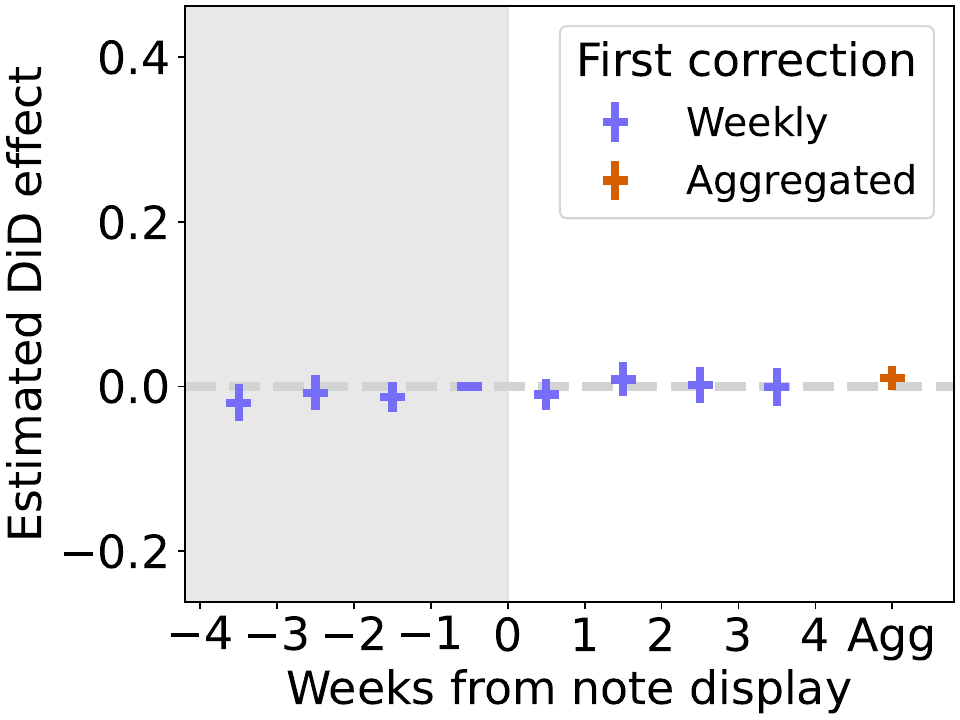}}
    \hfill
    \subfloat[]{\includegraphics[width = .32\textwidth]{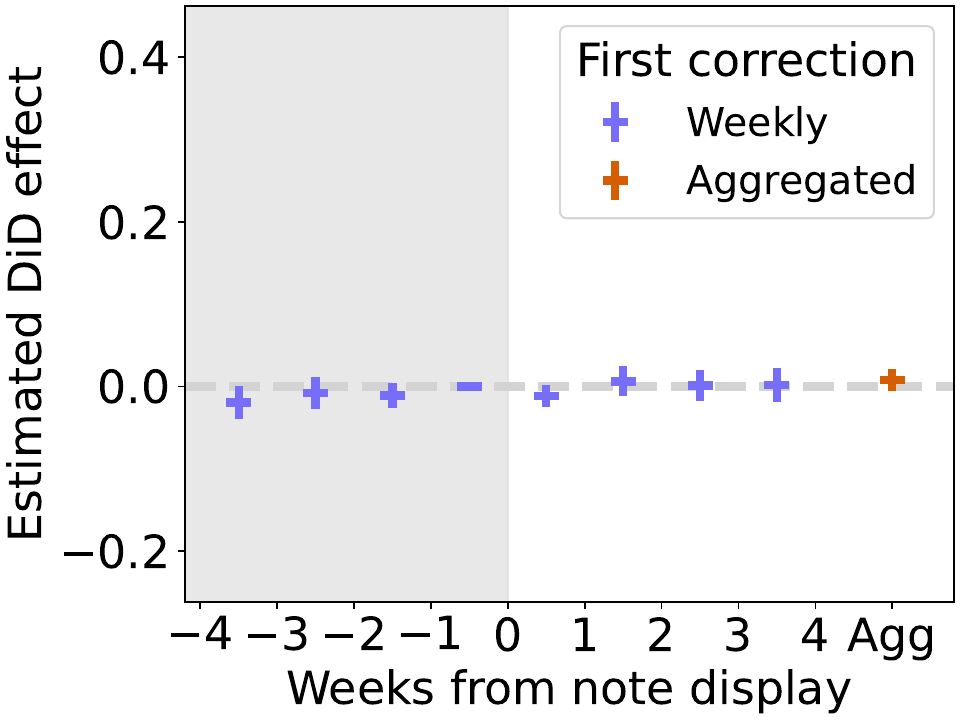}}
    \hfill
    \subfloat[]{\includegraphics[width = .32\textwidth]{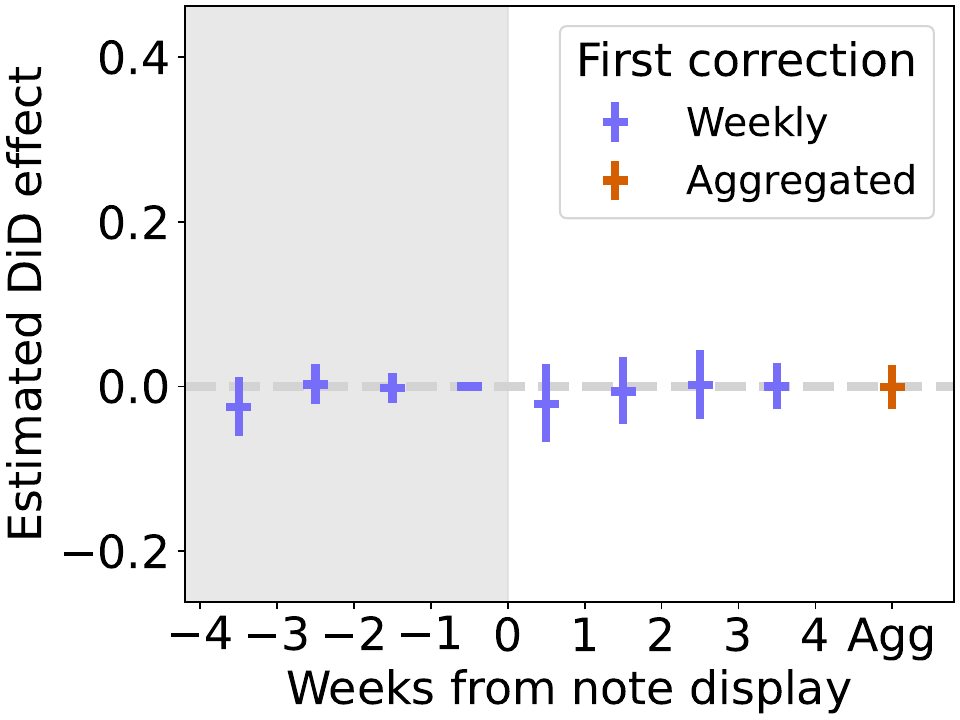}}
    \caption{\textbf{Weekly and aggregated DiD estimates of the effects of the first correction on posting activity under alternative regression specifications.} \textbf{(a)}~Mixed-effects negative binomial regression with event-level random effects and account-clustered standard errors. Full estimation results are reported in \Cref{tab:did_weekly_me}. \textbf{(b)}~Fixed-effects negative binomial regression with event-level fixed effects and bootstrapped standard errors with 100 replications. Full estimation results are reported in \Cref{tab:did_weekly_fe}. \textbf{(c)}~Conditional fixed-effects Poisson regression with event-level fixed effects and event-clustered standard errors. Full estimation results are reported in \Cref{tab:did_weekly_fe_poisson}. 
    Error bars indicate 95\% CIs.}
    \label{fig:did_weekly_alter}
\end{figure}

To assess the robustness of our main findings to alternative modeling assumptions and estimation procedures, we re-estimate the weekly DiD models using three specifications: mixed-effects negative binomial regression, fixed-effects negative binomial regression, and conditional fixed-effects Poisson regression. While our main analysis employs Poisson pseudo-maximum likelihood regression with high-dimensional fixed effects~\cite{correia2020fast}, the alternative specifications allow us to examine the sensitivity of our findings to different count-data models and approaches to accounting for unobserved heterogeneity. As shown in \Cref{fig:did_weekly_alter} and \Cref{tab:did_weekly_me,tab:did_weekly_fe,tab:did_weekly_fe_poisson}, the three specifications yield consistent results: across all models, the pre-correction estimates for Weeks~$-4$ to~$-2$ are statistically indistinguishable from zero (all $p>0.050$); and following the first correction, none of the weekly treatment effects from Week~1 to Week~4 is statistically significant (all $p>0.050$). The aggregated post-correction effects are also statistically insignificant across all three specifications: $0.010$ ($z=1.340$, $p=0.180$; 95\%~CI: $[-0.005, 0.025]$) for the mixed-effects negative binomial model, $0.008$ ($z=1.157$, $p=0.247$; 95\%~CI: $[-0.006, 0.022]$) for the fixed-effects negative binomial model, and $-0.001$ ($z=-0.058$, $p=0.954$; 95\%~CI: $[-0.027, 0.026]$) for the conditional fixed-effects Poisson model. Together, these findings indicate that the absence of statistically significant changes in posting activity following the first correction remains consistent across alternative model specifications.

\begin{table*}[ht]
\centering
\caption{\textbf{Weekly and aggregated effects of the first correction on posting activity estimated by mixed-effects negative binomial regression with event-level random effects and account-clustered standard errors.} The estimation is based on 167,992 weekly count observations across 20,999 correction events contributed by 17,812 accounts. Treatment events are restricted to accounts' first corrections, excluding those with subsequent corrections in the 4-week post-treatment period. Control events include all pseudo-corrections without prior corrections.}
\setlength{\tabcolsep}{4pt}
\begin{tabular}{@{}rrrrcrrrrc@{}}
\toprule
\multicolumn{5}{c}{Pre-correction} &
\multicolumn{5}{c}{Post-correction} \\
\cmidrule(lr){1-5}\cmidrule(lr){6-10}
Week & {Estimate} & {$z$} & {$p$} & 95\% CI &
Week & {Estimate} & {$z$} & {$p$} & 95\% CI \\
\midrule
$-4$ & $-0.020$ & $-1.748$ & $0.080$ & $[-0.042, 0.002]$ &
$1$  & $-0.010$ & $-1.050$ & $0.294$ & $[-0.028, 0.009]$ \\

$-3$ & $-0.008$ & $-0.718$ & $0.473$ & $[-0.029, 0.014]$ &
$2$  & $0.008$  & $0.796$  & $0.426$ & $[-0.012, 0.029]$ \\

$-2$ & $-0.013$ & $-1.357$ & $0.175$ & $[-0.031, 0.006]$ &
$3$  & $0.001$  & $0.130$  & $0.897$ & $[-0.020, 0.024]$ \\

     &          &          &         &                   &
$4$  & $-0.000$ & $-0.027$ & $0.978$ & $[-0.023, 0.023]$ \\
\bottomrule
\end{tabular}
\label{tab:did_weekly_me}
\end{table*}

\begin{table*}[ht]
\centering
\caption{\textbf{Weekly and aggregated effects of the first correction on posting activity estimated by fixed-effects negative binomial regression with event-level fixed effects and bootstrapped standard errors.} The estimation is based on 167,056 weekly count observations across 20,882 correction events contributed by 17,696 accounts. Treatment events are restricted to accounts' first corrections, excluding those with subsequent corrections in the 4-week post-treatment period. Control events include all pseudo-corrections without prior corrections.}
\setlength{\tabcolsep}{4pt}
\begin{tabular}{@{}rrrrcrrrrc@{}}
\toprule
\multicolumn{5}{c}{Pre-correction} &
\multicolumn{5}{c}{Post-correction} \\
\cmidrule(lr){1-5}\cmidrule(lr){6-10}
Week & {Estimate} & {$z$} & {$p$} & 95\% CI &
Week & {Estimate} & {$z$} & {$p$} & 95\% CI \\
\midrule
$-4$ & $-0.019$ & $-1.682$ & $0.093$ & $[-0.042, 0.003]$ &
$1$  & $-0.012$ & $-1.375$ & $0.169$ & $[-0.028, 0.005]$ \\

$-3$ & $-0.008$ & $-0.736$ & $0.462$ & $[-0.029, 0.014]$ &
$2$  & $0.006$  & $0.611$  & $0.541$ & $[-0.013, 0.026]$ \\

$-2$ & $-0.011$ & $-1.161$ & $0.246$ & $[-0.029, 0.008]$ &
$3$  & $0.001$  & $0.100$  & $0.921$ & $[-0.020, 0.023]$ \\

     &          &          &         &                   &
$4$  & $0.002$  & $0.144$  & $0.885$ & $[-0.020, 0.024]$ \\
\bottomrule
\end{tabular}
\label{tab:did_weekly_fe}
\end{table*}

\begin{table*}[ht]
\centering
\caption{\textbf{Weekly and aggregated effects of the first correction on posting activity estimated by conditional fixed-effects Poisson regression with event-level fixed effects and event-clustered standard errors.} The estimation is based on 167,056 weekly count observations across 20,882 correction events contributed by 17,696 accounts. Treatment events are restricted to accounts' first corrections, excluding those with subsequent corrections in the 4-week post-treatment period. Control events include all pseudo-corrections without prior corrections.}
\setlength{\tabcolsep}{4pt}
\begin{tabular}{@{}rrrrcrrrrc@{}}
\toprule
\multicolumn{5}{c}{Pre-correction} &
\multicolumn{5}{c}{Post-correction} \\
\cmidrule(lr){1-5}\cmidrule(lr){6-10}
Week & {Estimate} & {$z$} & {$p$} & 95\% CI &
Week & {Estimate} & {$z$} & {$p$} & 95\% CI \\
\midrule
$-4$ & $-0.025$ & $-1.345$ & $0.179$ & $[-0.060, 0.012]$ &
$1$  & $-0.022$ & $-0.878$ & $0.380$ & $[-0.068, 0.027]$ \\

$-3$ & $0.002$  & $0.194$  & $0.846$ & $[-0.022, 0.027]$ &
$2$  & $-0.006$ & $-0.286$ & $0.775$ & $[-0.046, 0.036]$ \\

$-2$ & $-0.002$ & $-0.219$ & $0.827$ & $[-0.020, 0.017]$ &
$3$  & $0.002$  & $0.089$  & $0.929$ & $[-0.039, 0.045]$ \\

     &          &          &         &                   &
$4$  & $-0.000$ & $-0.004$ & $0.997$ & $[-0.027, 0.028]$ \\
\bottomrule
\end{tabular}
\label{tab:did_weekly_fe_poisson}
\end{table*}

\clearpage
\subsection{Note score control}
\label{supp:note_score_control}
We additionally examine the sensitivity of our findings to note scores by restricting control events to source posts associated with community notes having a helpfulness score of at least 0.35. Under the core Community Notes scoring algorithm, notes with a score of 0.40 or higher generally qualify as helpful and are displayed on the corresponding misleading posts. We therefore adopt a threshold of 0.35 to retain control posts associated with notes that exhibit relatively high helpfulness scores but have not achieved displayed status. This restriction allows us to compare corrected posts with more closely comparable uncorrected posts in terms of their notes' helpfulness. As shown in \Cref{tab:did_weekly_note_score}, all weekly DiD estimates across the pre- and post-correction periods are not statistically significant (all $p>0.050$). The aggregated effect estimate for the post-correction period is $0.013$ ($z=0.423$, $p=0.672$; 95\%~CI: $[-0.047, 0.077]$) and also statistically indistinguishable from zero. Together, the results provide no statistically significant evidence of changes in posting activity following the first correction, consistent with our main findings.

\begin{table*}[ht]
\centering
\caption{\textbf{Weekly DiD estimation of the effects of the first correction on posting activity after note score control.} The estimation is based on 73,744 weekly count observations across 9,218 correction events contributed by 9,126 accounts. Treatment events are restricted to accounts' first corrections, excluding those with subsequent corrections in the 4-week post-treatment period. Control events include all pseudo-corrections without prior corrections. Event and week fixed effects are absorbed, and standard errors are clustered at the account level.}
\setlength{\tabcolsep}{4pt}
\begin{tabular}{@{}rrrrcrrrrc@{}}
\toprule
\multicolumn{5}{c}{Pre-correction} &
\multicolumn{5}{c}{Post-correction} \\
\cmidrule(lr){1-5}\cmidrule(lr){6-10}
Week & {Estimate} & {$z$} & {$p$} & 95\% CI &
Week & {Estimate} & {$z$} & {$p$} & 95\% CI \\
\midrule
$-4$ & $0.061$ & $1.445$ & $0.148$ & $[-0.021, 0.150]$ &
$1$  & $0.045$ & $1.340$ & $0.180$ & $[-0.020, 0.113]$ \\

$-3$ & $0.060$ & $1.523$ & $0.128$ & $[-0.016, 0.142]$ &
$2$  & $0.074$ & $1.711$ & $0.087$ & $[-0.010, 0.165]$ \\

$-2$ & $0.025$ & $0.772$ & $0.440$ & $[-0.037, 0.091]$ &
$3$  & $0.045$ & $0.910$ & $0.363$ & $[-0.050, 0.149]$ \\

     &         &         &         &                   &
$4$  & $0.030$ & $0.596$ & $0.551$ & $[-0.066, 0.137]$ \\
\bottomrule
\end{tabular}
\label{tab:did_weekly_note_score}
\end{table*}

\clearpage
\subsection{Posting responses following subsequent corrections}
\label{supp:subsequent_corrections}

We further examine changes in posting activity following subsequent focal corrections, stratified by accounts' prior correction histories, to assess whether posting responses differ across accounts with varying levels of prior correction exposure. Specifically, we group correction events according to the number of corrections an account had received before the focal event and estimate the corresponding DiD effects on subsequent posting activity. 

\begin{figure}[ht]
    \centering
    \captionsetup[subfloat]{font={bf, small}, skip=0pt, singlelinecheck=false, labelformat=simple, position=top}
    \subfloat[]{\includegraphics[width = .32\textwidth]{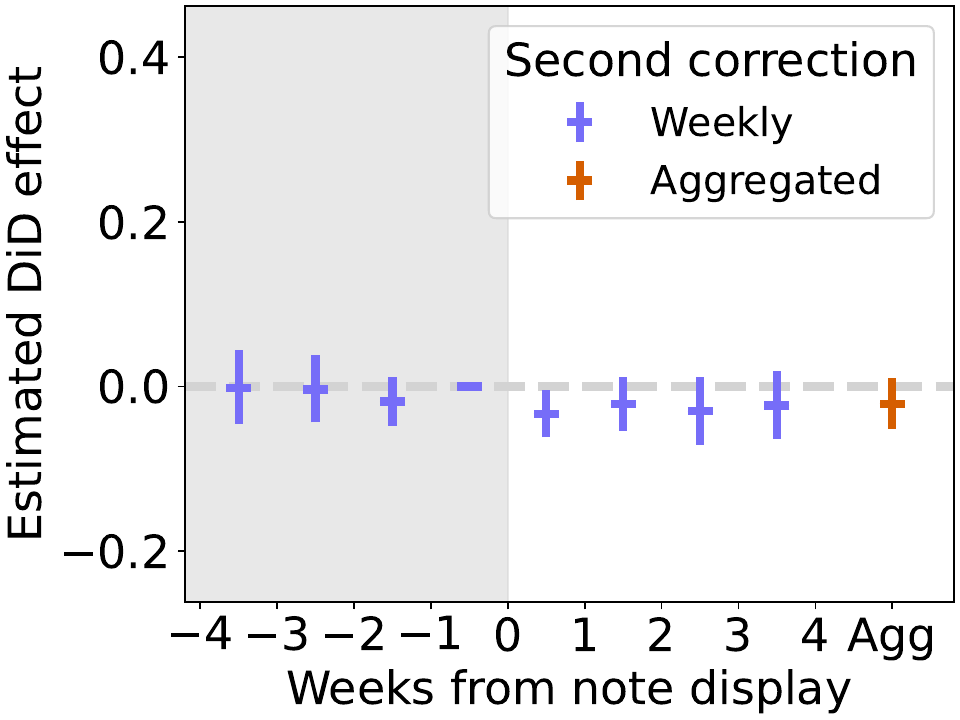}}
    \hfill
    \subfloat[]{\includegraphics[width = .32\textwidth]{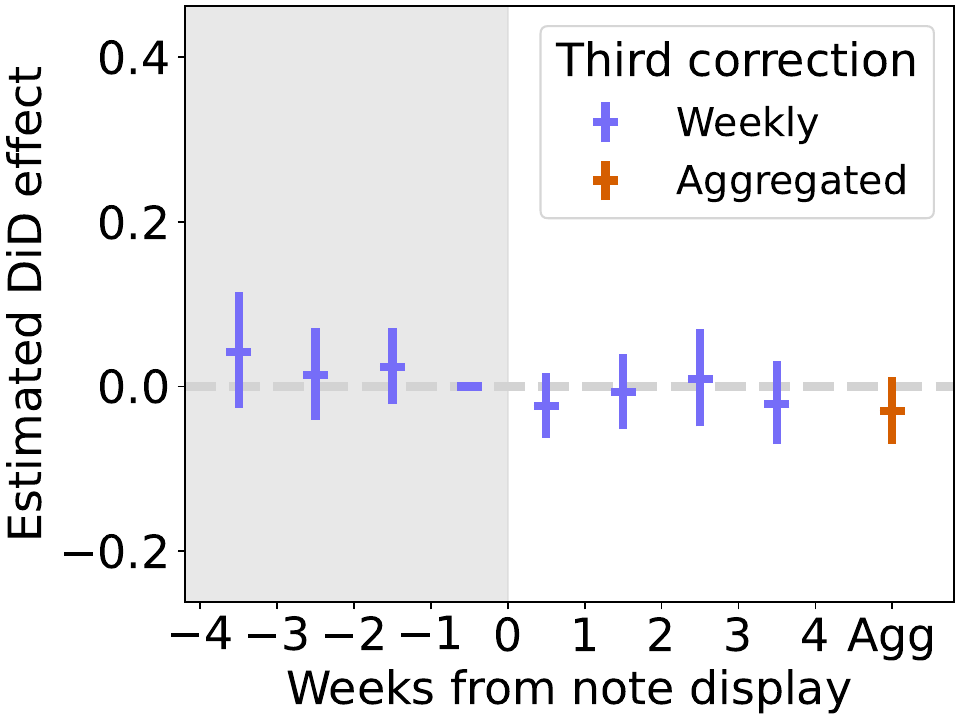}}
    \hfill
    \subfloat[]{\includegraphics[width = .32\textwidth]{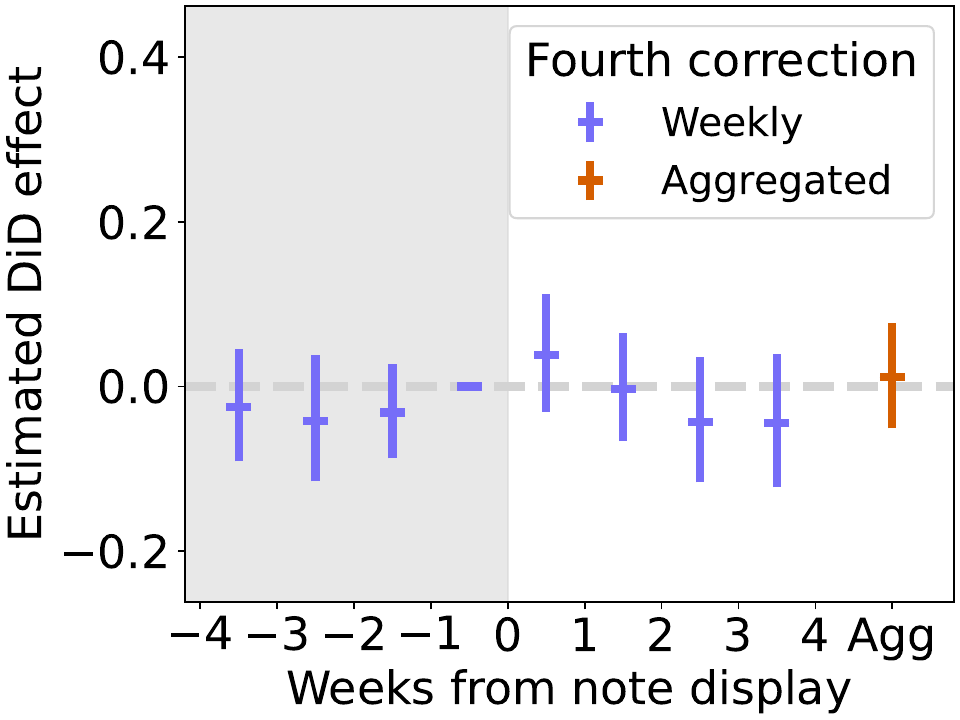}}
    \\
    \subfloat[]{\includegraphics[width = .32\textwidth]{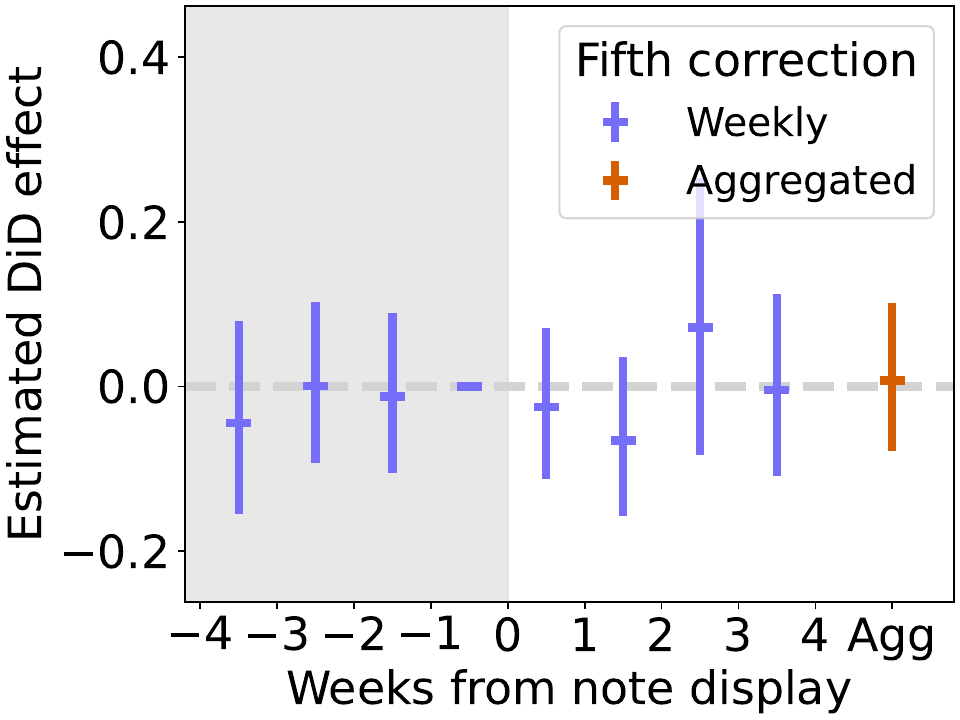}}
    \hfill
    \subfloat[]{\includegraphics[width = .32\textwidth]{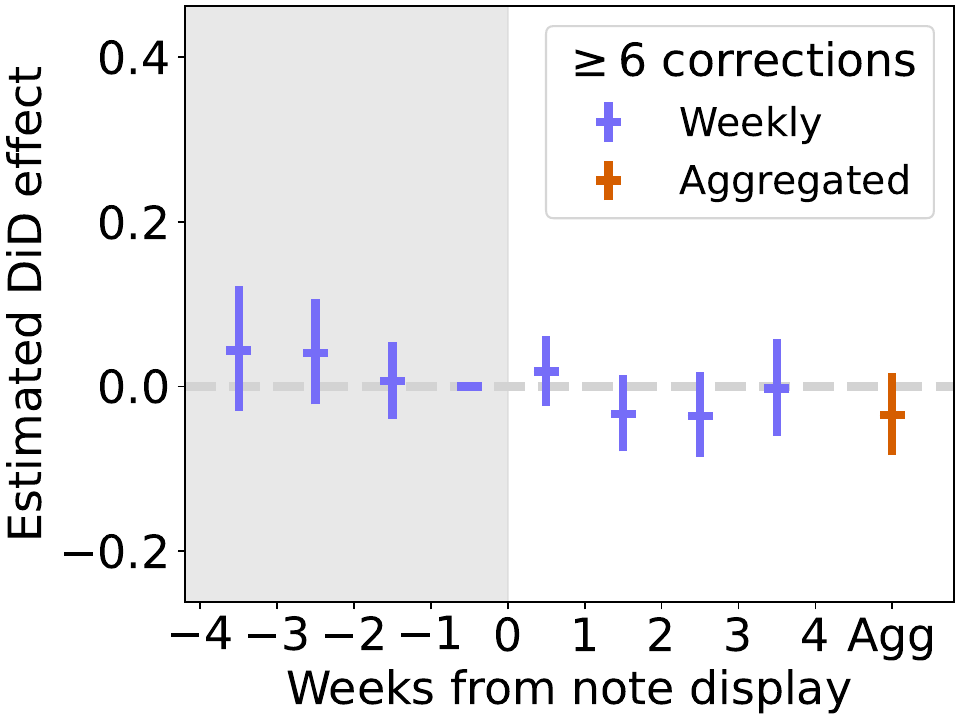}}
    \hfill
    \subfloat[]{\includegraphics[width = .32\textwidth]{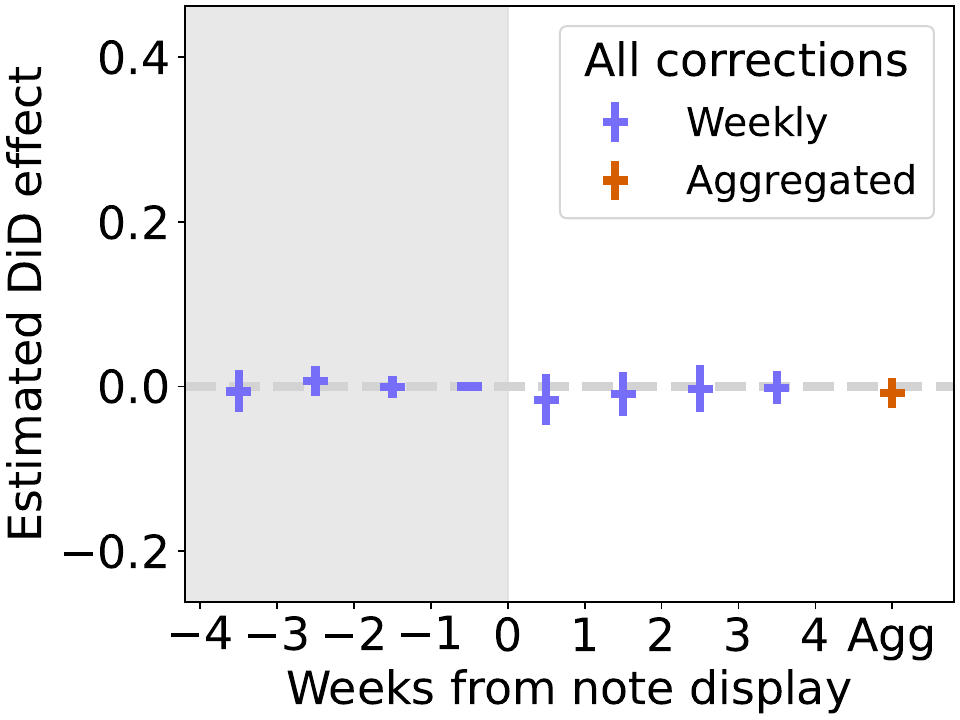}}
    \caption{\textbf{Weekly and aggregated DiD estimates of original-post activity following subsequent focal corrections.} Panels \textbf{(a)}--\textbf{(e)} show estimates following the second, third, fourth, fifth, and sixth or later observed corrections, respectively. Panel \textbf{(f)} pools all eligible correction events regardless of prior correction history. Week~$-1$ serves as the reference period. Treatment events exclude additional corrections during the four-week post-treatment period, and control events are selected according to the corresponding prior correction histories. Error bars indicate 95\% confidence intervals, with standard errors clustered at the account level. Full estimation results are reported in \Cref{tab:did_second_correction,tab:did_third_correction,tab:did_fourth_correction,tab:did_fifth_correction,tab:did_sixth_plus_correction,tab:did_all_corrections}.}
    \label{fig:did_weekly_prior}
\end{figure}

As shown in \Cref{fig:did_weekly_prior}, the aggregated DiD estimates for the second (effect $= -0.021$, $z=-1.346$, $p=0.178$; 95\%~CI: $[-0.051, 0.010]$), third (effect $= -0.030$, $z=-1.417$, $p=0.157$; 95\%~CI: $[-0.069, 0.012]$), fourth (effect $= 0.011$, $z=0.344$, $p=0.731$; 95\%~CI: $[-0.051, 0.077]$), and fifth corrections (effect $= 0.007$, $z=0.160$, $p=0.873$; 95\%~CI: $[-0.079, 0.101]$), as well as the sixth and subsequent corrections combined (effect $= -0.035$, $z=-1.336$, $p=0.182$; 95\%~CI: $[-0.083, 0.017]$), are all statistically indistinguishable from zero. The weekly analysis of the second correction shows a significant reduction in posting activity during Week~1 (effect $= -0.033$, $z=-2.247$, $p=0.025$; 95\%~CI: $[-0.061, -0.004]$), but the remaining weekly estimates and the joint test of the four post-correction coefficients are not statistically significant ($\chi^2(4)=5.19$, $p=0.268$). We additionally pool all eligible correction events, regardless of prior correction history, and find no statistically significant changes in posting activity following correction (effect $= -0.008$, $z=-0.830$, $p=0.407$; 95\%~CI: $[-0.026, 0.011]$). Together, these results provide no statistically significant evidence that subsequent corrections affect posting activity, regardless of the number of corrections an account had previously received.

\clearpage

\begin{table*}[ht]
\centering
\caption{\textbf{Leads-and-lags DiD estimates over four weeks before and after the second correction, conditional on the first correction.} Week~$-1$ serves as the reference period. The estimation is based on 24,576 weekly count observations across 3,072 correction events contributed by 2,299 accounts. Treatment events are restricted to second corrections, excluding those followed by additional corrections during the four-week post-treatment period. Control events include pseudo-corrections associated with accounts having one prior correction. Event and week fixed effects are absorbed, and standard errors are clustered at the account level.}
\setlength{\tabcolsep}{3pt}
\begin{tabular}{@{}rrrrcrrrrc@{}}
\toprule
\multicolumn{5}{c}{Pre-correction} &
\multicolumn{5}{c}{Post-correction} \\
\cmidrule(lr){1-5}\cmidrule(lr){6-10}
Week & {Estimate} & {$z$} & {$p$} & 95\% CI &
Week & {Estimate} & {$z$} & {$p$} & 95\% CI \\
\midrule
$-4$ & $-0.001$ & $-0.064$ & $0.949$ & $[-0.045, 0.044]$ &
$1$  & $-0.033$ & $-2.247$ & $0.025$ & $[-0.061, -0.004]$ \\

$-3$ & $-0.004$ & $-0.183$ & $0.855$ & $[-0.044, 0.038]$ &
$2$  & $-0.021$ & $-1.278$ & $0.201$ & $[-0.054, 0.012]$ \\

$-2$ & $-0.018$ & $-1.203$ & $0.229$ & $[-0.048, 0.012]$ &
$3$  & $-0.030$ & $-1.410$ & $0.159$ & $[-0.071, 0.012]$ \\

     &          &          &         &                   &
$4$  & $-0.023$ & $-1.078$ & $0.281$ & $[-0.063, 0.019]$ \\
\bottomrule
\end{tabular}
\label{tab:did_second_correction}
\end{table*}

\begin{table*}[ht]
\centering
\caption{\textbf{Leads-and-lags DiD estimates over four weeks before and after the third correction, conditional on the first two corrections.} Week~$-1$ serves as the reference period. The estimation is based on 10,536 weekly count observations across 1,317 correction events contributed by 926 accounts. Treatment events are restricted to third corrections, excluding those followed by additional corrections during the four-week post-treatment period. Control events include pseudo-corrections associated with accounts having two prior corrections. Event and week fixed effects are absorbed, and standard errors are clustered at the account level.}
\setlength{\tabcolsep}{4pt}
\begin{tabular}{@{}rrrrcrrrrc@{}}
\toprule
\multicolumn{5}{c}{Pre-correction} &
\multicolumn{5}{c}{Post-correction} \\
\cmidrule(lr){1-5}\cmidrule(lr){6-10}
Week & {Estimate} & {$z$} & {$p$} & 95\% CI &
Week & {Estimate} & {$z$} & {$p$} & 95\% CI \\
\midrule
$-4$ & $0.042$  & $1.198$  & $0.231$ & $[-0.026, 0.115]$ &
$1$  & $-0.024$ & $-1.159$ & $0.247$ & $[-0.063, 0.017]$ \\

$-3$ & $0.014$  & $0.488$  & $0.625$ & $[-0.040, 0.071]$ &
$2$  & $-0.007$ & $-0.300$ & $0.764$ & $[-0.052, 0.040]$ \\

$-2$ & $0.024$  & $1.023$  & $0.306$ & $[-0.021, 0.071]$ &
$3$  & $0.009$  & $0.294$  & $0.769$ & $[-0.049, 0.070]$ \\

     &          &          &         &                   &
$4$  & $-0.021$ & $-0.815$ & $0.415$ & $[-0.070, 0.030]$ \\
\bottomrule
\end{tabular}
\label{tab:did_third_correction}
\end{table*}

\begin{table*}[ht]
\centering
\caption{\textbf{Leads-and-lags DiD estimates over four weeks before and after the fourth correction, conditional on the first three corrections.} Week~$-1$ serves as the reference period. The estimation is based on 6,128 weekly count observations across 766 correction events contributed by 525 accounts. Treatment events are restricted to fourth corrections, excluding those followed by additional corrections during the four-week post-treatment period. Control events include pseudo-corrections associated with accounts having three prior corrections. Event and week fixed effects are absorbed, and standard errors are clustered at the account level.}
\setlength{\tabcolsep}{3pt}
\begin{tabular}{@{}rrrrcrrrrc@{}}
\toprule
\multicolumn{5}{c}{Pre-correction} &
\multicolumn{5}{c}{Post-correction} \\
\cmidrule(lr){1-5}\cmidrule(lr){6-10}
Week & {Estimate} & {$z$} & {$p$} & 95\% CI &
Week & {Estimate} & {$z$} & {$p$} & 95\% CI \\
\midrule
$-4$ & $-0.025$ & $-0.710$ & $0.478$ & $[-0.091, 0.046]$ &
$1$  & $0.038$  & $1.070$  & $0.284$ & $[-0.031, 0.112]$ \\

$-3$ & $-0.042$ & $-1.045$ & $0.296$ & $[-0.115, 0.038]$ &
$2$  & $-0.003$ & $-0.088$ & $0.930$ & $[-0.066, 0.065]$ \\

$-2$ & $-0.032$ & $-1.065$ & $0.287$ & $[-0.087, 0.027]$ &
$3$  & $-0.043$ & $-1.085$ & $0.278$ & $[-0.116, 0.036]$ \\

     &          &          &         &                   &
$4$  & $-0.045$ & $-1.054$ & $0.292$ & $[-0.122, 0.040]$ \\
\bottomrule
\end{tabular}
\label{tab:did_fourth_correction}
\end{table*}

\begin{table*}[ht]
\centering
\caption{\textbf{Leads-and-lags DiD estimates over four weeks before and after the fifth correction, conditional on the first four corrections.} Week~$-1$ serves as the reference period. The estimation is based on 3,352 weekly count observations across 419 correction events contributed by 280 accounts. Treatment events are restricted to fifth corrections, excluding those followed by additional corrections during the four-week post-treatment period. Control events include pseudo-corrections associated with accounts having four prior corrections. Event and week fixed effects are absorbed, and standard errors are clustered at the account level.}
\setlength{\tabcolsep}{3pt}
\begin{tabular}{@{}rrrrcrrrrc@{}}
\toprule
\multicolumn{5}{c}{Pre-correction} &
\multicolumn{5}{c}{Post-correction} \\
\cmidrule(lr){1-5}\cmidrule(lr){6-10}
Week & {Estimate} & {$z$} & {$p$} & 95\% CI &
Week & {Estimate} & {$z$} & {$p$} & 95\% CI \\
\midrule
$-4$ & $-0.045$ & $-0.731$ & $0.465$ & $[-0.155, 0.080]$ &
$1$  & $-0.025$ & $-0.528$ & $0.598$ & $[-0.113, 0.071]$ \\

$-3$ & $0.000$  & $0.005$  & $0.996$ & $[-0.093, 0.103]$ &
$2$  & $-0.065$ & $-1.286$ & $0.198$ & $[-0.157, 0.036]$ \\

$-2$ & $-0.012$ & $-0.249$ & $0.804$ & $[-0.105, 0.089]$ &
$3$  & $0.072$  & $0.870$  & $0.384$ & $[-0.083, 0.253]$ \\

     &          &          &         &                   &
$4$  & $-0.004$ & $-0.073$ & $0.942$ & $[-0.108, 0.112]$ \\
\bottomrule
\end{tabular}
\label{tab:did_fifth_correction}
\end{table*}

\begin{table*}[ht]
\centering
\caption{\textbf{Leads-and-lags DiD estimates over four weeks before and after the sixth and later corrections, conditional on at least five prior corrections.} Week~$-1$ serves as the reference period. The estimation is based on 11,376 weekly count observations across 1,422 correction events contributed by 555 accounts. Treatment events are restricted to sixth and later corrections, excluding those followed by additional corrections during the post-treatment period. Control events include pseudo-corrections associated with accounts having at least five prior corrections. Event and week fixed effects are absorbed, and standard errors are clustered at the account level.}
\setlength{\tabcolsep}{4pt}
\begin{tabular}{@{}rrrrcrrrrc@{}}
\toprule
\multicolumn{5}{c}{Pre-correction} &
\multicolumn{5}{c}{Post-correction} \\
\cmidrule(lr){1-5}\cmidrule(lr){6-10}
Week & {Estimate} & {$z$} & {$p$} & 95\% CI &
Week & {Estimate} & {$z$} & {$p$} & 95\% CI \\
\midrule
$-4$ & $0.044$  & $1.156$  & $0.248$ & $[-0.029, 0.122]$ &
$1$  & $0.018$  & $0.850$  & $0.396$ & $[-0.023, 0.062]$ \\

$-3$ & $0.040$  & $1.259$  & $0.208$ & $[-0.022, 0.106]$ &
$2$  & $-0.033$ & $-1.385$ & $0.166$ & $[-0.079, 0.014]$ \\

$-2$ & $0.007$  & $0.277$  & $0.782$ & $[-0.039, 0.054]$ &
$3$  & $-0.036$ & $-1.326$ & $0.185$ & $[-0.086, 0.017]$ \\

     &          &          &         &                   &
$4$  & $-0.003$ & $-0.084$ & $0.933$ & $[-0.060, 0.058]$ \\
\bottomrule
\end{tabular}
\label{tab:did_sixth_plus_correction}
\end{table*}

\begin{table*}[ht]
\centering
\caption{\textbf{Leads-and-lags DiD estimates over four weeks before and after correction pooling all corrections.} Week~$-1$ serves as the reference period. The estimation is based on 223,024 weekly count observations across 27,878 correction events contributed by 18,834 accounts. Treatment events include first and later corrections, excluding those followed by additional corrections during the four-week post-treatment period. Control events include all pseudo-corrections without corrections across the observation window. Event and week fixed effects are absorbed, and standard errors are clustered at the account level.}
\setlength{\tabcolsep}{4pt}
\begin{tabular}{@{}rrrrcrrrrc@{}}
\toprule
\multicolumn{5}{c}{Pre-correction} &
\multicolumn{5}{c}{Post-correction} \\
\cmidrule(lr){1-5}\cmidrule(lr){6-10}
Week & {Estimate} & {$z$} & {$p$} & 95\% CI &
Week & {Estimate} & {$z$} & {$p$} & 95\% CI \\
\midrule
$-4$ & $-0.006$ & $-0.471$ & $0.638$ & $[-0.031, 0.020]$ &
$1$  & $-0.016$ & $-1.008$ & $0.313$ & $[-0.047, 0.016]$ \\

$-3$ & $0.007$  & $0.750$  & $0.453$ & $[-0.011, 0.025]$ &
$2$  & $-0.009$ & $-0.681$ & $0.496$ & $[-0.036, 0.018]$ \\

$-2$ & $-0.001$ & $-0.116$ & $0.908$ & $[-0.014, 0.013]$ &
$3$  & $-0.003$ & $-0.189$ & $0.850$ & $[-0.031, 0.026]$ \\

     &          &          &         &                   &
$4$  & $-0.002$ & $-0.180$ & $0.857$ & $[-0.022, 0.019]$ \\
\bottomrule
\end{tabular}
\label{tab:did_all_corrections}
\end{table*}

\clearpage
\subsection{Sensitivity to additional corrections during the post-treatment period}
\label{supp:cumulative_corrections}
\begin{figure}[ht]
    \centering
    \captionsetup[subfloat]{font={bf, small}, skip=0pt, singlelinecheck=false, labelformat=simple, position=top}
    \includegraphics[width= .8\linewidth]{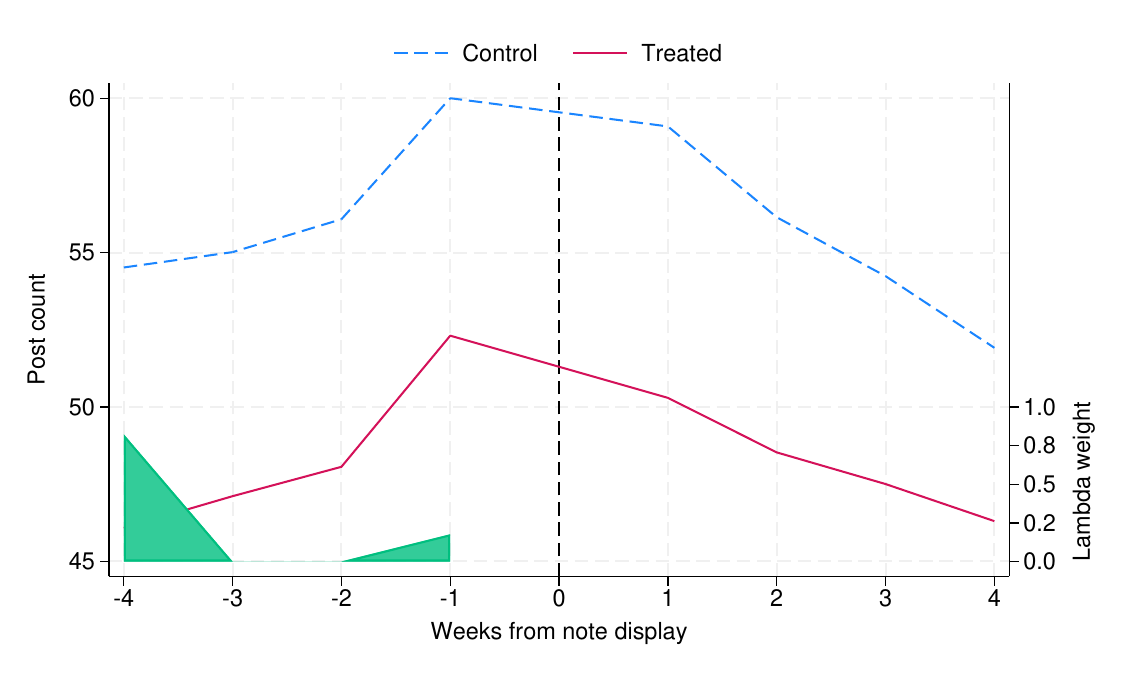}
    \caption{\textbf{SDiD estimates of weekly posting activity following the first correction.}  Treatment events are restricted to accounts' first corrections, including those with subsequent corrections in the 4-week post-treatment period. Control events include all pseudo-corrections without prior corrections. Solid red and dashed blue lines indicate the average weekly original-post counts for treated and control accounts, respectively. The green area indicates the estimated pre-treatment time weights ($\lambda$). Posting activity is measured using weekly original-post counts. Standard errors are estimated using 100 bootstrap replications with clustering at the account level.}
    \label{fig:sdid_posts_weekly_cum}
\end{figure}

We assess the sensitivity of our findings by retaining first-correction events followed by additional corrections during the four-week post-treatment period, while restricting control events to those without correction exposure during the corresponding window. \Cref{tab:did_weekly_cum} reports the weekly DiD estimates for posting activity following the first correction. Additionally, we conduct a robustness check using SDiD estimator (\Cref{fig:sdid_posts_weekly_cum}), and find that the original-post counts show no statistically significant change following the first correction (ATT $=1.116$, $t=1.57$, $p=0.116$; 95\%~CI: $[-0.275, 2.508]$). 

Additionally, we conduct a robustness analysis excluding control events associated with accounts that also appear in the treatment group. All weekly DiD estimates remain statistically insignificant (\Cref{tab:did_weekly_treated_control}), and the aggregated effect is also not statistically significant (effect $=0.024$, $z=1.407$, $p=0.160$; 95\%~CI: $[-0.009, 0.058]$).

Together, our findings remain robust and do not support that community corrections significantly change accounts' posting activity, when including additional corrections during the post-treatment period and excluding control events associated with accounts that also appear in the treatment group.

\begin{table*}[ht]
\centering
\caption{\textbf{Leads-and-lags DiD estimates over four weeks before and after correction} Week~$-1$ serves as the reference period. The estimation is based on 176,552 weekly count observations across 22,069 correction events contributed by 18,627 accounts. Treatment events are restricted to accounts' first corrections, including those with subsequent corrections in the 4-week post-treatment period. Control events include all pseudo-corrections without prior corrections. Event and week fixed effects are absorbed, and standard errors are clustered at the account level.}
\setlength{\tabcolsep}{3pt}
\begin{tabular}{@{}rrrrcrrrrc@{}}
\toprule
\multicolumn{5}{c}{Pre-correction} &
\multicolumn{5}{c}{Post-correction} \\
\cmidrule(lr){1-5}\cmidrule(lr){6-10}
Week & {Estimate} & {$z$} & {$p$} & 95\% CI &
Week & {Estimate} & {$z$} & {$p$} & 95\% CI \\
\midrule
$-4$ & $-0.038$ & $-1.984$ & $0.047$ & $[-0.074, -0.000]$ &
$1$  & $-0.009$ & $-0.371$ & $0.710$ & $[-0.055, 0.039]$ \\

$-3$ & $-0.013$ & $-1.008$ & $0.313$ & $[-0.039, 0.013]$ &
$2$  & $0.002$  & $0.093$  & $0.926$ & $[-0.038, 0.044]$ \\

$-2$ & $-0.013$ & $-1.369$ & $0.171$ & $[-0.032, 0.006]$ &
$3$  & $0.007$  & $0.363$  & $0.717$ & $[-0.031, 0.047]$ \\

     &          &          &         &                   &
$4$  & $0.011$  & $0.802$  & $0.423$ & $[-0.016, 0.039]$ \\
\bottomrule
\end{tabular}
\label{tab:did_weekly_cum}
\end{table*}

\begin{table*}[ht]
\centering
\caption{\textbf{Leads-and-lags DiD estimates over four weeks before and after correction.} Week~$-1$ serves as the reference period. The estimation is based on 162,648 weekly count observations across 20,331 correction events contributed by 18,624 accounts. Treatment events are restricted to each account's first observed correction. Control events associated with accounts that also appear in the treatment group are excluded. Event and week fixed effects are absorbed, and standard errors are clustered at the account level.}
\setlength{\tabcolsep}{3pt}
\begin{tabular}{@{}rrrrcrrrrc@{}}
\toprule
\multicolumn{5}{c}{Pre-correction} &
\multicolumn{5}{c}{Post-correction} \\
\cmidrule(lr){1-5}\cmidrule(lr){6-10}
Week & {Estimate} & {$z$} & {$p$} & 95\% CI &
Week & {Estimate} & {$z$} & {$p$} & 95\% CI \\
\midrule
$-4$ & $-0.034$ & $-1.476$ & $0.140$ & $[-0.078, 0.011]$ &
$1$  & $-0.013$ & $-0.412$ & $0.681$ & $[-0.073, 0.051]$ \\

$-3$ & $-0.009$ & $-0.630$ & $0.529$ & $[-0.037, 0.020]$ &
$2$  & $0.014$  & $0.506$  & $0.613$ & $[-0.039, 0.069]$ \\

$-2$ & $-0.009$ & $-0.889$ & $0.374$ & $[-0.029, 0.011]$ &
$3$  & $0.017$  & $0.732$  & $0.464$ & $[-0.028, 0.064]$ \\

     &          &          &         &                   &
$4$  & $0.028$  & $1.772$  & $0.076$ & $[-0.003, 0.059]$ \\
\bottomrule
\end{tabular}
\label{tab:did_weekly_treated_control}
\end{table*}

\clearpage
\section{Heterogeneity analysis of posting activity}
\label{supp:account_he}

\subsection{Heterogeneity by account profiles and posting characteristics}
\label{supp:did_interaction_features}

The results for the joint estimation of interactions between the DiD term (Treated $\times$ After) and variables capturing account characteristics, correction timing, and pre-correction posting behavior are reported in \Cref{tab:did_interaction_features}. Among these interaction terms, only the interaction with pre-correction misleadingness is positive and statistically significant, suggesting that accounts with higher baseline misleadingness exhibit a greater increase in posting activity following their first corrections, conditional on the other characteristics included in the model. The subgroup DiD estimates based on the median and 75th-percentile splits are reported in \Cref{tab:did_subgroup_features_median} and \Cref{tab:did_subgroup_features_p75}, respectively.

\begin{figure}[ht]
    \centering
    \captionsetup[subfloat]{font={bf, small}, skip=0pt, singlelinecheck=false, labelformat=simple, position=top}
    \subfloat[]{\includegraphics[width = .8\textwidth]{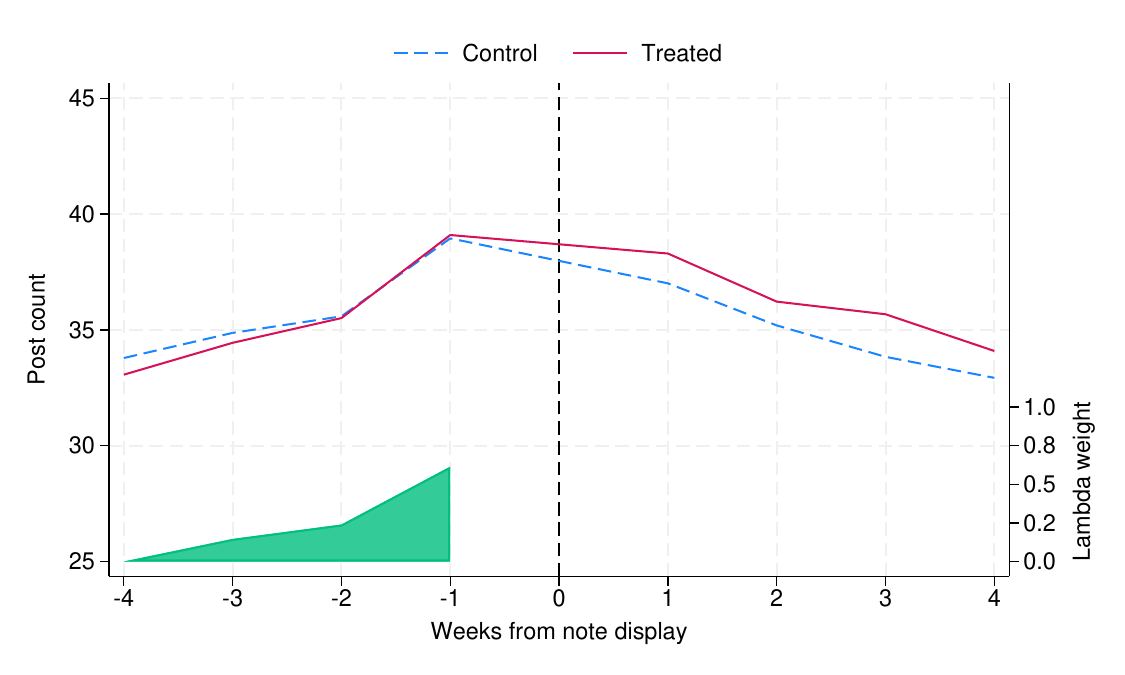}}
    \\
    \subfloat[]{\includegraphics[width = .8\textwidth]{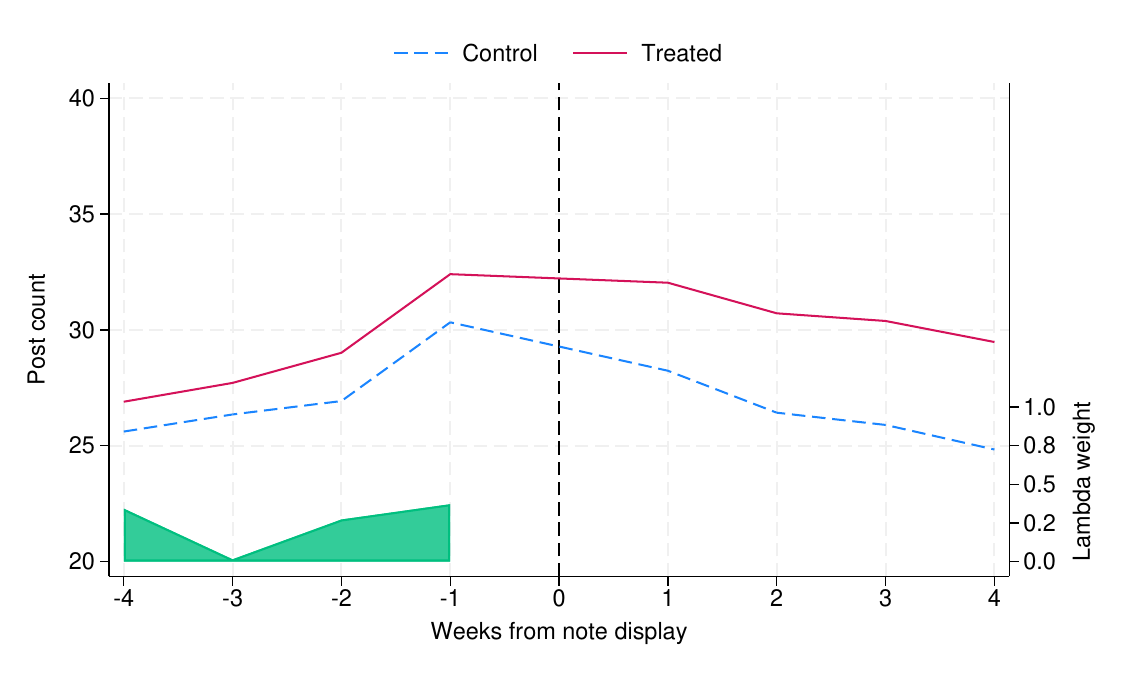}}
    \caption{\textbf{SDiD estimates of weekly posting activity following the first correction among accounts with high pre-correction misleadingness.}
    \textbf{(a)}~SDiD estimation for accounts with average pre-correction misleadingness scores at or above the median.
    \textbf{(b)}~SDiD estimation for accounts with average pre-correction misleadingness scores at or above the 75th percentile.
    Solid red and dashed blue lines indicate the average weekly original-post counts for treated and control accounts, respectively. The green area indicates the estimated pre-treatment time weights ($\lambda$). Posting activity is measured using weekly original-post counts. Standard errors are estimated using 100 bootstrap replications with clustering at the account level.}
    \label{fig:sdid_posts_weekly_mis}
\end{figure}

Additionally, we employ the SDiD estimator to examine changes in weekly posting activity among accounts with high pre-correction misleadingness, defined using the median and 75th-percentile thresholds. For accounts above the median, the estimated effect is positive but not statistically significant (ATT $=1.320$, $t=1.77$, $p=0.077$; 95\%~CI: $[-0.145, 2.785]$; \Cref{fig:sdid_posts_weekly_mis}a). In contrast, for accounts in the top quartile of pre-correction misleadingness, the estimated effect is positive and statistically significant (ATT $=2.505$, $t=2.49$, $p=0.013$; 95\%~CI: $[0.535, 4.475]$; \Cref{fig:sdid_posts_weekly_mis}b), corresponding to an estimated increase of approximately 2.5 original posts per week following their first corrections. These findings are consistent with the subgroup DiD results, suggesting that increased posting activity following corrections is concentrated among accounts with particularly high pre-correction misleadingness.

\Cref{tab:did_interaction_features_p75} reports estimation results for the joint interaction model, where the pre-correction misleadingness is represented by a binary indicator identifying accounts in the highest quartile, and \Cref{tab:logit_account_characteristics} reports the multivariable logistic regression results for characterizing these accounts. We find that accounts with higher misinformation exposure, political right-leaning, and higher pre-correction negative sentiment, opinion confidence, and political content are more likely to belong to the highest misleadingness quartile. In contrast, higher follower counts, older account age, greater posting frequency, higher positive sentiment, and higher toxicity are negatively associated with accounts in this group. These findings provide additional insights into the observed characteristics of accounts with high pre-correction misleadingness.

\begin{table}[ht]
\centering
\caption{\textbf{DiD interaction estimates across account characteristics, correction time, and pre-correction posting preferences.} All interaction terms are estimated jointly in a single model. Each row reports the transformed interaction coefficient ($e^\beta-1$) between the DiD term (Treated $\times$ After) and the indicated variable. The estimation is based on 69,552 observations across 8,694 correction events contributed by 7,316 accounts. Event and week fixed effects are absorbed, and standard errors are clustered at the account level.}
\setlength{\tabcolsep}{10pt}
\begin{tabular}{@{}lrrrc@{}}
\toprule
Variable & {Estimate} & {$z$} & {$p$} & 95\% CI \\
\midrule
\multicolumn{5}{@{}l}{\underline{Panel A: Account characteristics}} \\
Followers
  & $0.017$  & $0.732$  & $0.464$ & $[-0.029, 0.066]$ \\
Followees
  & $0.015$  & $0.926$  & $0.355$ & $[-0.016, 0.047]$ \\
Verified
  & $-0.028$ & $-0.742$ & $0.458$ & $[-0.098, 0.048]$ \\
Account age
  & $0.005$  & $0.295$  & $0.768$ & $[-0.027, 0.038]$ \\
Political leaning
  & $-0.011$ & $-0.287$ & $0.774$ & $[-0.087, 0.070]$ \\
Misinfo exposure
  & $0.022$  & $1.042$  & $0.297$ & $[-0.019, 0.064]$ \\
\\
\multicolumn{5}{@{}l}{\underline{Panel B: Correction timing and pre-correction posting preferences}} \\
Correction timing
  & $0.014$  & $1.095$  & $0.273$ & $[-0.011, 0.040]$ \\
Frequency
  & $0.001$  & $0.120$  & $0.905$ & $[-0.015, 0.017]$ \\
Misleadingness
  & $0.053$  & $2.526$  & $0.012$ & $[0.012, 0.097]$ \\
Positive
  & $-0.012$ & $-0.524$ & $0.600$ & $[-0.055, 0.033]$ \\
Negative
  & $-0.022$ & $-0.804$ & $0.421$ & $[-0.075, 0.033]$ \\
Toxicity
  & $0.000$  & $0.002$  & $0.998$ & $[-0.049, 0.052]$ \\
Confidence
  & $-0.001$ & $-0.037$ & $0.970$ & $[-0.038, 0.038]$ \\
Politics
  & $-0.010$ & $-0.583$ & $0.560$ & $[-0.044, 0.024]$ \\
\bottomrule
\end{tabular}
\label{tab:did_interaction_features}
\end{table}

\clearpage

\clearpage
\begin{table*}[ht]
\centering
\footnotesize
\caption{\textbf{Subgroup DiD estimates of the effects of the first correction on weekly original-post activity.} Each characteristic is used to divide the sample into Group~1 and Group~0, with treatment effects estimated separately for each subgroup. For continuous variables, Group~1 indicates values at or above the median and Group~0 indicates values below the median; for binary variables, the groups correspond to values of 1 and 0, respectively. Event and week fixed effects are absorbed, and standard errors are clustered at the account level.}
\setlength{\tabcolsep}{5pt}
\begin{tabular}{@{}llrrrcrrr@{}}
\toprule
Variable & Group & {Estimate} & {$z$} & {$p$} & 95\% CI & Obs. & Events & Accounts \\
\midrule
\multicolumn{9}{@{}l}{\underline{Panel A: Account characteristics}} \\
Followers
  & 1 & $-0.005$ & $-0.300$ & $0.764$ & $[-0.040, 0.031]$ & 34,784 & 4,348 & 3,305 \\
  & 0 & $0.034$ & $1.230$ & $0.219$ & $[-0.020, 0.090]$ & 34,768 & 4,346 & 4,011 \\
\addlinespace
Followees
  & 1 & $0.015$ & $0.658$ & $0.510$ & $[-0.030, 0.063]$ & 34,792 & 4,349 & 3,731 \\
  & 0 & $-0.006$ & $-0.311$ & $0.756$ & $[-0.045, 0.034]$ & 34,760 & 4,345 & 3,585 \\
\addlinespace
Verified
  & 1 & $0.005$ & $0.295$ & $0.768$ & $[-0.029, 0.040]$ & 45,232 & 5,654 & 4,497 \\
  & 0 & $0.011$ & $0.383$ & $0.701$ & $[-0.044, 0.068]$ & 24,320 & 3,040 & 2,819 \\
\addlinespace
Account age
  & 1 & $0.008$ & $0.416$ & $0.677$ & $[-0.030, 0.048]$ & 34,800 & 4,350 & 3,642 \\
  & 0 & $-0.001$ & $-0.048$ & $0.962$ & $[-0.047, 0.047]$ & 34,752 & 4,344 & 3,694 \\
\addlinespace
Political leaning
  & 1 & $0.026$ & $1.297$ & $0.195$ & $[-0.013, 0.067]$ & 30,992 & 3,874 & 3,219 \\
  & 0 & $-0.010$ & $-0.465$ & $0.642$ & $[-0.052, 0.034]$ & 38,560 & 4,820 & 4,097 \\
\addlinespace
Misinfo exposure
  & 1 & $0.022$ & $1.149$ & $0.251$ & $[-0.015, 0.062]$ & 34,928 & 4,366 & 3,653 \\
  & 0 & $-0.010$ & $-0.425$ & $0.671$ & $[-0.052, 0.035]$ & 34,624 & 4,328 & 3,663 \\
\addlinespace
\multicolumn{9}{@{}l}{\underline{Panel B: Correction timing and pre-correction posting preferences}} \\
Correction timing
  & 1 & $-0.003$ & $-0.140$ & $0.889$ & $[-0.043, 0.039]$ & 34,816 & 4,352 & 3,911 \\
  & 0 & $0.012$ & $0.522$ & $0.601$ & $[-0.032, 0.057]$ & 34,736 & 4,342 & 3,764 \\
\addlinespace
Frequency
  & 1 & $0.002$ & $0.101$ & $0.920$ & $[-0.030, 0.035]$ & 34,840 & 4,355 & 3,593 \\
  & 0 & $0.035$ & $1.612$ & $0.107$ & $[-0.007, 0.078]$ & 34,712 & 4,339 & 3,864 \\
\addlinespace
Misleadingness
  & 1 & $0.044$ & $2.074$ & $0.038$ & $[0.002, 0.087]$ & 34,784 & 4,348 & 3,716 \\
  & 0 & $-0.019$ & $-0.896$ & $0.370$ & $[-0.058, 0.023]$ & 34,768 & 4,346 & 3,852 \\
\addlinespace
Positive
  & 1 & $-0.013$ & $-0.646$ & $0.518$ & $[-0.052, 0.027]$ & 34,784 & 4,348 & 3,836 \\
  & 0 & $0.016$ & $0.757$ & $0.449$ & $[-0.024, 0.058]$ & 34,768 & 4,346 & 3,626 \\
\addlinespace
Negative
  & 1 & $0.019$ & $0.921$ & $0.357$ & $[-0.021, 0.062]$ & 34,776 & 4,347 & 3,676 \\
  & 0 & $-0.003$ & $-0.164$ & $0.870$ & $[-0.044, 0.039]$ & 34,776 & 4,347 & 3,789 \\
\addlinespace
Toxicity
  & 1 & $0.023$ & $1.125$ & $0.261$ & $[-0.017, 0.066]$ & 34,776 & 4,347 & 3,807 \\
  & 0 & $-0.006$ & $-0.271$ & $0.787$ & $[-0.045, 0.036]$ & 34,776 & 4,347 & 3,673 \\
\addlinespace
Confidence
  & 1 & $-0.005$ & $-0.244$ & $0.807$ & $[-0.047, 0.038]$ & 34,776 & 4,347 & 3,816 \\
  & 0 & $0.013$ & $0.596$ & $0.551$ & $[-0.029, 0.056]$ & 34,776 & 4,347 & 3,859 \\
\addlinespace
Politics
  & 1 & $0.006$ & $0.315$ & $0.753$ & $[-0.033, 0.048]$ & 34,776 & 4,347 & 3,569 \\
  & 0 & $0.005$ & $0.223$ & $0.823$ & $[-0.038, 0.050]$ & 34,776 & 4,347 & 3,865 \\
\bottomrule
\end{tabular}
\label{tab:did_subgroup_features_median}
\end{table*}

\clearpage
\begin{table*}[ht]
\centering
\footnotesize
\caption{\textbf{Subgroup DiD estimates of the effects of the first correction on weekly original-post activity, stratified by the 75th percentile thresholds.} The estimations are across continuous account characteristics, correction timing, and pre-correction posting preferences. For each characteristic, Group~1 includes observations at or above the 75th percentile, while Group~0 includes observations below this threshold. Treatment effects are estimated separately for each subgroup. Event and week fixed effects are absorbed, and standard errors are clustered at the account level.}
\setlength{\tabcolsep}{5pt}
\begin{tabular}{@{}llrrrcrrr@{}}
\toprule
Variable & Group & {Estimate} & {$z$} & {$p$} & 95\% CI & Obs. & Events & Accounts \\
\midrule
\multicolumn{9}{@{}l}{\underline{Panel A: Account characteristics}} \\
Followers
  & 1 & $0.001$ & $0.033$ & $0.973$ & $[-0.043, 0.046]$ & 17,392 & 2,174 & 1,547 \\
  & 0 & $0.008$ & $0.408$ & $0.683$ & $[-0.031, 0.049]$ & 52,160 & 6,520 & 5,769 \\
\addlinespace
Followees
  & 1 & $0.036$ & $1.181$ & $0.237$ & $[-0.023, 0.100]$ & 17,392 & 2,174 & 1,845 \\
  & 0 & $-0.007$ & $-0.406$ & $0.685$ & $[-0.041, 0.028]$ & 52,160 & 6,520 & 5,471 \\
\addlinespace
Account age
  & 1 & $0.005$ & $0.192$ & $0.848$ & $[-0.047, 0.060]$ & 17,416 & 2,177 & 1,804 \\
  & 0 & $0.006$ & $0.318$ & $0.750$ & $[-0.028, 0.041]$ & 52,136 & 6,517 & 5,549 \\
\addlinespace
Misinfo exposure
  & 1 & $0.035$ & $1.473$ & $0.141$ & $[-0.011, 0.084]$ & 17,416 & 2,177 & 1,839 \\
  & 0 & $-0.005$ & $-0.281$ & $0.779$ & $[-0.040, 0.031]$ & 52,136 & 6,517 & 5,477 \\
\addlinespace
\multicolumn{9}{@{}l}{\underline{Panel B: Correction timing and pre-correction posting preferences}} \\
Correction timing
  & 1 & $0.027$ & $1.023$ & $0.306$ & $[-0.024, 0.081]$ & 17,384 & 2,173 & 2,051 \\
  & 0 & $-0.001$ & $-0.058$ & $0.953$ & $[-0.036, 0.035]$ & 52,168 & 6,521 & 5,544 \\
\addlinespace
Frequency
  & 1 & $0.001$ & $0.050$ & $0.960$ & $[-0.036, 0.039]$ & 17,432 & 2,179 & 1,796 \\
  & 0 & $0.016$ & $0.816$ & $0.414$ & $[-0.022, 0.056]$ & 52,120 & 6,515 & 5,628 \\
\addlinespace
Misleadingness
  & 1 & $0.102$ & $3.083$ & $0.002$ & $[0.036, 0.172]$ & 17,400 & 2,175 & 1,887 \\
  & 0 & $-0.012$ & $-0.716$ & $0.474$ & $[-0.045, 0.021]$ & 52,152 & 6,519 & 5,648 \\
\addlinespace
Positive
  & 1 & $0.010$ & $0.402$ & $0.688$ & $[-0.038, 0.060]$ & 17,392 & 2,174 & 1,952 \\
  & 0 & $0.004$ & $0.241$ & $0.810$ & $[-0.030, 0.039]$ & 52,160 & 6,520 & 5,480 \\
\addlinespace
Negative
  & 1 & $-0.001$ & $-0.016$ & $0.987$ & $[-0.063, 0.066]$ & 17,384 & 2,173 & 1,867 \\
  & 0 & $0.006$ & $0.350$ & $0.726$ & $[-0.027, 0.040]$ & 52,168 & 6,521 & 5,613 \\
\addlinespace
Toxicity
  & 1 & $0.034$ & $1.038$ & $0.299$ & $[-0.029, 0.102]$ & 17,384 & 2,173 & 1,959 \\
  & 0 & $0.000$ & $0.013$ & $0.989$ & $[-0.033, 0.035]$ & 52,168 & 6,521 & 5,477 \\
\addlinespace
Confidence
  & 1 & $-0.003$ & $-0.073$ & $0.942$ & $[-0.075, 0.075]$ & 17,392 & 2,174 & 1,977 \\
  & 0 & $0.006$ & $0.373$ & $0.709$ & $[-0.026, 0.040]$ & 52,160 & 6,520 & 5,611 \\
\addlinespace
Politics
  & 1 & $0.000$ & $0.002$ & $0.998$ & $[-0.066, 0.071]$ & 17,392 & 2,174 & 1,795 \\
  & 0 & $0.006$ & $0.364$ & $0.716$ & $[-0.027, 0.040]$ & 52,160 & 6,520 & 5,653 \\
\bottomrule
\end{tabular}
\label{tab:did_subgroup_features_p75}
\end{table*}

\clearpage
\begin{table}[ht]
\centering
\caption{\textbf{DiD interaction estimates across account characteristics, correction timing, and pre-correction posting preferences.}
All interaction terms are estimated jointly in a single model.
Pre-correction misleadingness is represented by a binary indicator identifying accounts in the highest quartile (above the 75th percentile), while the remaining variables follow the original specification.
Each row reports the transformed interaction coefficient between the DiD term (Treated $\times$ After) and the indicated variable.
Original $p$-values and adjusted $p$-values using the Holm and Benjamini-Hochberg (FDR-BH) procedures are reported, with adjustments applied across all 14 interaction terms.
The estimation is based on 69,552 observations across 8,694 correction events contributed by 7,316 accounts.
Event and week fixed effects are absorbed, and standard errors are clustered at the account level.}
\setlength{\tabcolsep}{7pt}
\begin{tabular}{@{}lrrrrrc@{}}
\toprule
Variable & Estimate & $z$ & $p$ & $p_{\mathrm{Holm}}$ & $p_{\mathrm{BH}}$ & 95\% CI \\
\midrule
\multicolumn{7}{@{}l}{\underline{Panel A: Account characteristics}} \\
Followers
  & $0.013$ & $0.529$ & $0.597$ & $1.000$ & $0.941$ & $[-0.033, 0.060]$ \\
Followees
  & $0.016$ & $1.002$ & $0.317$ & $1.000$ & $0.941$ & $[-0.015, 0.049]$ \\
Verified
  & $-0.026$ & $-0.685$ & $0.493$ & $1.000$ & $0.941$ & $[-0.097, 0.050]$ \\
Account age
  & $0.003$ & $0.159$ & $0.874$ & $1.000$ & $0.941$ & $[-0.029, 0.036]$ \\
Political leaning
  & $-0.016$ & $-0.415$ & $0.678$ & $1.000$ & $0.941$ & $[-0.090, 0.063]$ \\
Misinfo exposure
  & $0.025$ & $1.235$ & $0.217$ & $1.000$ & $0.941$ & $[-0.015, 0.067]$ \\
\\
\multicolumn{7}{@{}l}{\underline{Panel B: Correction timing and pre-correction posting preferences}} \\
Correction timing
  & $0.014$ & $1.037$ & $0.300$ & $1.000$ & $0.941$ & $[-0.012, 0.040]$ \\
Frequency
  & $0.001$ & $0.161$ & $0.872$ & $1.000$ & $0.941$ & $[-0.015, 0.018]$ \\
Misleadingness (top quartile)
  & $0.137$ & $3.161$ & $0.002$ & $0.022$ & $0.022$ & $[0.050, 0.232]$ \\
Positive
  & $-0.012$ & $-0.518$ & $0.605$ & $1.000$ & $0.941$ & $[-0.054, 0.033]$ \\
Negative
  & $-0.015$ & $-0.555$ & $0.579$ & $1.000$ & $0.941$ & $[-0.066, 0.039]$ \\
Toxicity
  & $0.000$ & $0.003$ & $0.998$ & $1.000$ & $0.998$ & $[-0.049, 0.052]$ \\
Confidence
  & $-0.005$ & $-0.256$ & $0.798$ & $1.000$ & $0.941$ & $[-0.043, 0.034]$ \\
Politics
  & $-0.006$ & $-0.381$ & $0.704$ & $1.000$ & $0.941$ & $[-0.039, 0.027]$ \\
\bottomrule
\end{tabular}
\label{tab:did_interaction_features_p75}
\end{table}

\clearpage
\begin{table}[ht]
\centering
\caption{\textbf{Logistic regression estimates of account characteristics associated with high pre-correction misleadingness.} The dependent variable is a binary indicator equal to 1 if an account's average pre-correction GPT-estimated misleadingness score is at or above the 75th percentile, and 0 otherwise. The independent variables include account characteristics, correction timing, and pre-correction posting preferences. The estimation includes 7,317 accounts.}
\setlength{\tabcolsep}{10pt}
\begin{tabular}{@{}lrrrc@{}}
\toprule
Variable & Estimate & $z$ & $p$ & 95\% CI \\
\midrule
\multicolumn{5}{@{}l}{\underline{Panel A: Account characteristics}} \\
Followers
  & $-0.177$ & $-4.133$ & $<0.001$ & $[-0.261, -0.093]$ \\
Followees
  & $0.023$ & $0.685$ & $0.493$ & $[-0.042, 0.087]$ \\
Verified
  & $0.123$ & $1.689$ & $0.091$ & $[-0.020, 0.266]$ \\
Account age
  & $-0.174$ & $-5.172$ & $<0.001$ & $[-0.239, -0.108]$ \\
Political leaning
  & $0.421$ & $5.021$ & $<0.001$ & $[0.257, 0.586]$ \\
Misinfo exposure
  & $0.279$ & $6.465$ & $<0.001$ & $[0.194, 0.363]$ \\
\\
\multicolumn{5}{@{}l}{\underline{Panel B: Correction timing and pre-correction posting preferences}} \\
Correction timing
  & $-0.043$ & $-1.346$ & $0.178$ & $[-0.105, 0.019]$ \\
Frequency
  & $-0.827$ & $-6.701$ & $<0.001$ & $[-1.069, -0.585]$ \\
Positive
  & $-0.719$ & $-10.871$ & $<0.001$ & $[-0.849, -0.589]$ \\
Negative
  & $0.469$ & $8.457$ & $<0.001$ & $[0.361, 0.578]$ \\
Toxicity
  & $-0.145$ & $-4.024$ & $<0.001$ & $[-0.216, -0.075]$ \\
Confidence
  & $0.199$ & $5.147$ & $<0.001$ & $[0.123, 0.275]$ \\
Politics
  & $0.190$ & $5.680$ & $<0.001$ & $[0.125, 0.256]$ \\
\bottomrule
\end{tabular}
\label{tab:logit_account_characteristics}
\end{table}

\clearpage
\subsection{Heterogeneity by pre-correction NMR-note history}
\label{supp:he_prior_nmr}

We examine heterogeneity in posting responses according to accounts' prior NMR-note histories, measured during the 28 days preceding their first correction. We estimate a DiD model incorporating an interaction between the DiD term (Treated $\times$ After) and repeated prior NMR notes, a binary indicator for receiving NMR notes on at least two distinct posts during the 28-day pre-correction period ($=1$, otherwise $=0$). The estimation includes \num{175016} weekly count observations across \num{21877} correction events contributed by \num{18484} accounts with an eligible 28-day pre-correction tracking period. The baseline DiD estimate for accounts with fewer than two prior NMR notes is statistically insignificant (effect $=-0.005$, $z=-0.327$, $p=0.744$; 95\%~CI: $[-0.037, 0.027]$). In contrast, the interaction estimate is positive and statistically significant (effect $=0.096$, $z=3.211$, $p=0.001$; 95\%~CI: $[0.036, 0.159]$), indicating a significantly greater post-correction change in posting activity among accounts with at least two prior NMR notes relative to those fewer than two. These findings suggest that posting responses to the first correction vary with accounts' prior NMR-note histories.

\begin{figure}[ht]
    \centering
    \captionsetup[subfloat]{font={bf, small}, skip=0pt, singlelinecheck=false, labelformat=simple, position=top}
    \subfloat[]{\includegraphics[width = .32\textwidth]{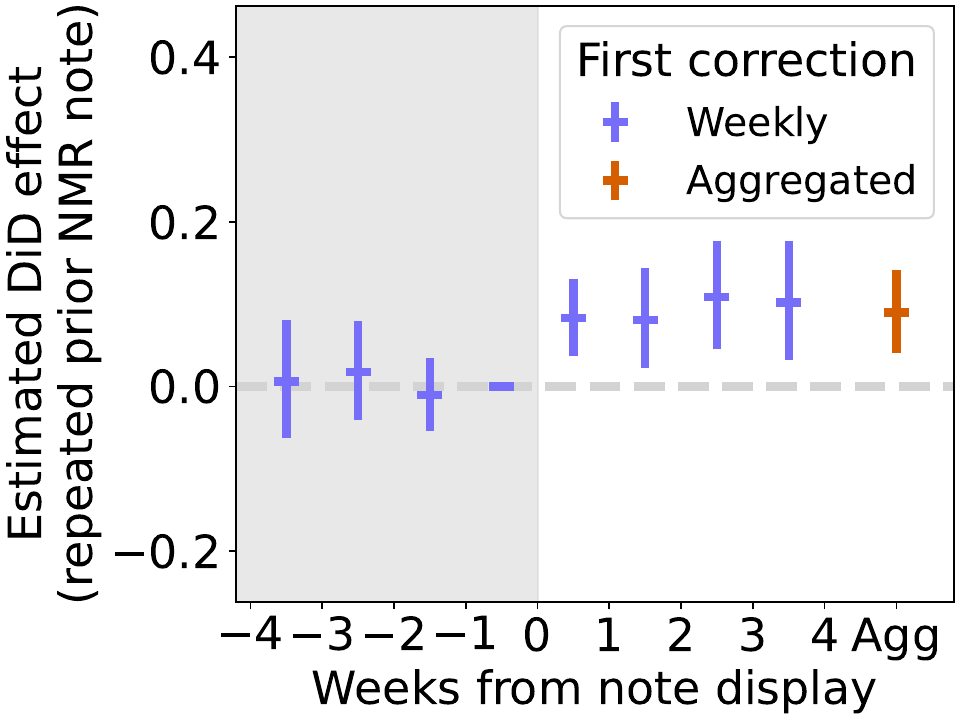}}
    \hspace{2cm}
    \subfloat[]{\includegraphics[width = .32\textwidth]{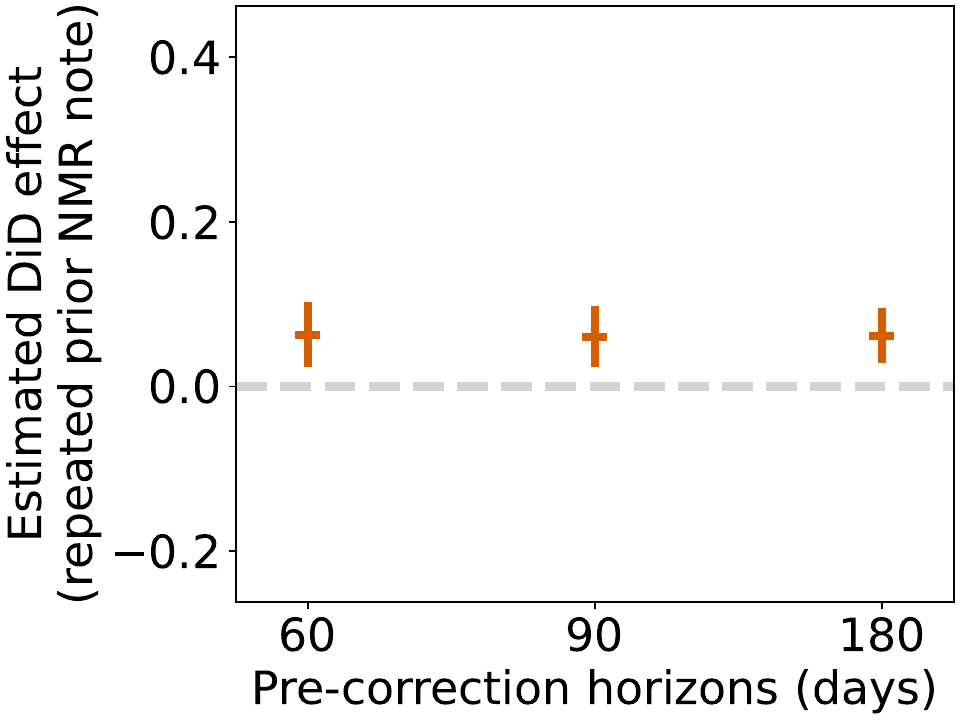}}
    \\
    \subfloat[]{\includegraphics[width = .8\textwidth]{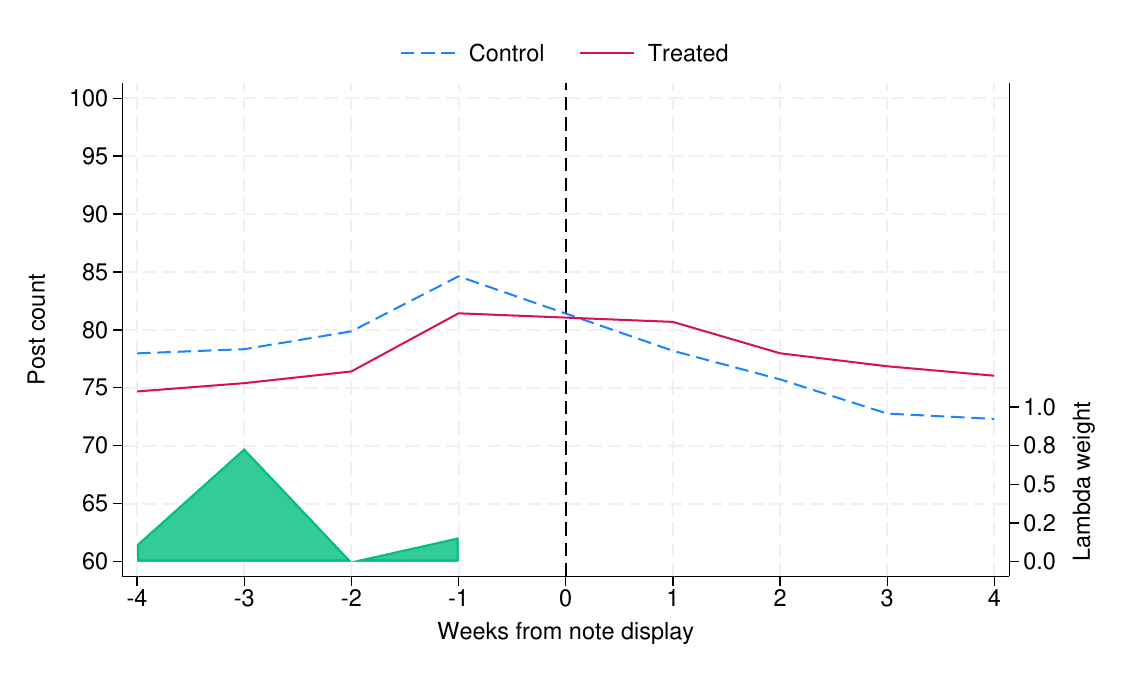}}
    \caption{\textbf{DiD estimates of the effects of the first correction on original-post activity among accounts with repeated prior NMR notes.}
    \textbf{(a)}~Weekly and aggregated DiD estimates.
    \textbf{(b)}~Aggregated DiD estimates across 60-, 90-, and 180-day pre-correction horizons.
    In \textbf{(a)} and \textbf{(b)}, error bars represent 95\% CIs, and standard errors are clustered at the account level.
    \textbf{(c)}~SDiD estimation for accounts with repeated NMR notes during the 28-day pre-correction period. Solid red and dashed blue lines indicate the average weekly original-post counts for treated and control accounts, respectively. The green area indicates the estimated pre-treatment time weights ($\lambda$). Posting activity is measured using weekly original-post counts. 
    Standard errors are estimated using 100 bootstrap replications with clustering at the account level.}
    \label{fig:did_posts_weekly_nmr}
\end{figure}

Specifically, accounts with repeated prior NMR notes increased their original-post activity by 9.0\% following their first correction based on \num{27184} weekly count observations across \num{3398} correction events contributed by \num{2509} accounts (effect $=0.090$, $z=3.675$, $p<0.001$; 95\%~CI: $[0.041, 0.141]$). As shown in \Cref{fig:did_posts_weekly_nmr}a, the pre-correction estimates for Weeks~$-4$ to~$-2$ are statistically indistinguishable from zero (all $p>0.100$). In contrast, the post-correction estimates are consistently positive and statistically significant, indicating increases of 8.3\% in Week~1, 8.1\% in Week~2, 10.9\% in Week~3, and 10.2\% in Week~4. These findings suggest that the increase in posting activity among accounts with repeated prior NMR notes persists throughout the four-week post-correction period. Additionally, we assess the sensitivity of the positive posting response among accounts with repeated prior NMR notes to potential violations of the parallel trends assumption using HonestDiD. As shown in \Cref{tab:honest_did_weekly_nmr}, the estimated increase in posting activity remains statistically significant for the aggregated four-week post-correction effect under relative-magnitude restrictions of up to $\overline{M}=1.0$ (95\%~CI: $[0.015, 0.183]$). At $\overline{M}=1.5$, the aggregated confidence interval includes zero (95\%~CI: $[-0.009, 0.213]$). For individual weeks, the positive effect remains statistically significant at $\overline{M}=1.0$ only in Week~1 (95\%~CI: $[0.009, 0.165]$), whereas the intervals for Weeks~2--4 include zero at $\overline{M}=0.5$. These results suggest that the aggregated increase in posting activity among accounts with repeated prior NMR notes is robust to moderate deviations from parallel trends, although the evidence becomes less conclusive under larger violations.

To further assess the robustness of these findings, we extend the pre-correction observation horizon used to characterize prior NMR-note history from 28 days to 60, 90, and 180 days (\Cref{fig:did_posts_weekly_nmr}b). The estimated effects remain positive and statistically significant across all three horizons: 6.3\% for 60 days (effect $=0.063$, $z=3.208$, $p=0.001$; 95\%~CI: $[0.024, 0.103]$), 6.0\% for 90 days (effect $=0.060$, $z=3.292$, $p<0.001$; 95\%~CI: $[0.024, 0.098]$), and 6.1\% for 180 days (effect $=0.061$, $z=3.692$, $p<0.001$; 95\%~CI: $[0.028, 0.095]$). Finally, applying the SDiD estimator yields a consistent positive effect (ATT $=6.155$, $t=3.23$, $p=0.001$; 95\%~CI: $[2.424, 9.887]$; \Cref{fig:did_posts_weekly_nmr}c), corresponding to approximately 6.2 additional original posts per week among accounts with repeated prior NMR notes following their first correction. Together, these results indicate that increased posting activity following corrections is concentrated among accounts with at least two prior NMR notes, with consistent findings across alternative observation horizons and estimation approaches.

We characterize accounts that repeatedly received NMR notes during the 28 days preceding their first correction, examining associations with account characteristics, correction timing, and pre-correction posting behavior. Specifically, we estimate a multivariable logistic regression predicting whether an account received at least two prior NMR notes, based on \num{3708} corrected accounts with eligible 28-day pre-correction tracking period. As shown in \Cref{tab:nmr_logistic}, the results indicate that accounts with larger follower counts ($\beta=0.759$, $z=9.811$, $p<0.001$), verified status ($\beta=0.427$, $z=2.882$, $p=0.004$), and later correction timing relative to the Community Notes roll-out ($\beta=0.431$, $z=7.724$, $p<0.001$) are more likely to have received repeated prior NMR notes. Regarding pre-correction posting behavior, higher negativity ($\beta=0.361$, $z=3.795$, $p<0.001$), misleadingness ($\beta=0.220$, $z=3.126$, $p=0.002$), and political content ($\beta=0.191$, $z=3.395$, $p<0.001$) are also positively associated with repeated prior NMR-note histories. In contrast, the remaining characteristics, including posting frequency, toxicity, and misinformation exposure, are not statistically significant (all $p>0.050$). Together, these findings suggest that prior NMR-note history can reflect accounts' posting misleadingness, but is also associated with account visibility, sentiments, and political content.

\begin{figure}[ht]
    \centering
    \captionsetup[subfloat]{font={bf, small}, skip=0pt, singlelinecheck=false, labelformat=simple, position=top}
    \includegraphics[width = \textwidth]{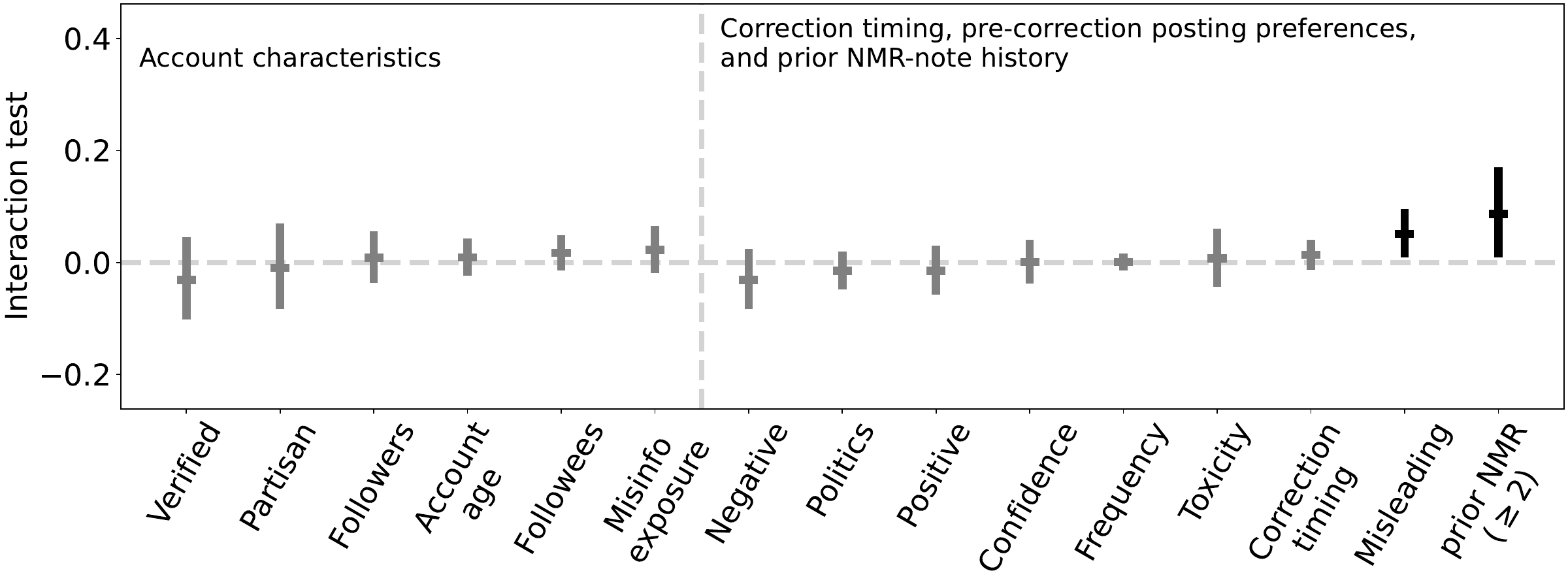}
    \caption{\textbf{DiD interaction estimates across account characteristics, correction timing, pre-correction posting preferences, and receiving repeated prior NMR notes.} All interaction terms are estimated jointly in a single model. 
    Error bars represent 95\% CIs, with standard errors clustered at the account level. Full estimation results are reported in \Cref{tab:did_interaction_features_nmr}.}
    \label{fig:did_he_nmr}
\end{figure}

To examine whether heterogeneous posting responses associated with prior NMR-note history are independent of accounts' baseline misleadingness, we estimate a joint DiD interaction model incorporating both characteristics alongside other account characteristics, correction timing, and pre-correction posting preferences. Both interactions remain positive and statistically significant (see \Cref{fig:did_he_nmr} and \Cref{tab:did_interaction_features_nmr}): the interaction with pre-correction misleadingness (effect $=0.051$, $z=2.376$, $p=0.018$; 95\%~CI: $[0.009, 0.095]$) and that with repeated prior NMR notes (effect $=0.087$, $z=2.197$, $p=0.028$; 95\%~CI: $[0.009, 0.171]$). These findings indicate that baseline misleadingness and prior NMR-note history are each associated with greater increases in posting activity following the first correction, even when accounting for the other characteristic and additional observed moderators. Together, the results suggest that these two measures capture related but independently associated dimensions of heterogeneity in posting responses.

\clearpage

\clearpage

\clearpage
\begin{table*}[ht]
\centering
\caption{\textbf{Weekly leads-and-lags DiD estimates for accounts with at least two prior NMR notes.} Week~$-1$ serves as the reference period. The estimation is based on 27,184 weekly count observations across 3,398 correction events contributed by 2,509 accounts. Event and week fixed effects are absorbed, and standard errors are clustered at the account level.}
\setlength{\tabcolsep}{4pt}
\begin{tabular}{@{}rrrrcrrrrc@{}}
\toprule
\multicolumn{5}{c}{Pre-correction} &
\multicolumn{5}{c}{Post-correction} \\
\cmidrule(lr){1-5}\cmidrule(lr){6-10}
Week & {Estimate} & {$z$} & {$p$} & 95\% CI &
Week & {Estimate} & {$z$} & {$p$} & 95\% CI \\
\midrule
$-4$ & $0.006$  & $0.167$  & $0.867$ & $[-0.063, 0.081]$ &
$1$  & $0.083$  & $3.620$  & $<0.001$ & $[0.037, 0.131]$ \\

$-3$ & $0.018$  & $0.585$  & $0.558$ & $[-0.040, 0.080]$ &
$2$  & $0.081$  & $2.726$  & $0.006$ & $[0.022, 0.143]$ \\

$-2$ & $-0.010$ & $-0.461$ & $0.645$ & $[-0.054, 0.035]$ &
$3$  & $0.109$  & $3.425$  & $<0.001$ & $[0.045, 0.177]$ \\

     &          &          &         &                   &
$4$  & $0.102$  & $2.913$  & $0.004$ & $[0.032, 0.176]$ \\
\bottomrule
\end{tabular}
\label{tab:did_posts_weekly_nmr}
\end{table*}

\begin{table}[ht]
\centering
\caption{\textbf{HonestDiD sensitivity analysis of weekly posting activity following the first correction among accounts with at least two prior NMR notes.} The table reports the original 95\% CIs and HonestDiD robust 95\% CIs for the original coefficient estimates of the four post-correction weeks and the aggregated post-treatment effect. Sensitivity parameters $\overline{M}=0.5, 1.0, 1.5,$ and $2.0$ allow progressively larger violations of parallel trends, relative to the maximum magnitude of pre-treatment trend deviations, under the relative-magnitudes restriction.}
\label{tab:honest_did_weekly_nmr}
\setlength{\tabcolsep}{5pt}
\begin{tabular}{@{}lccccc@{}}
\toprule
$\overline{M}$ &
Week 1 & Week 2 & Week 3 & Week 4 & Aggregated \\
\midrule
Original
& $[0.037,\ 0.123]$
& $[0.022,\ 0.134]$
& $[0.044,\ 0.163]$
& $[0.032,\ 0.162]$
& $[0.040,\ 0.139]$ \\

0.5
& $[0.033,\ 0.141]$
& $[-0.003,\ 0.171]$
& $[-0.003,\ 0.231]$
& $[-0.039,\ 0.261]$
& $[0.033,\ 0.153]$ \\

1.0
& $[0.009,\ 0.165]$
& $[-0.057,\ 0.231]$
& $[-0.093,\ 0.327]$
& $[-0.165,\ 0.393]$
& $[0.015,\ 0.183]$ \\

1.5
& $[-0.021,\ 0.195]$
& $[-0.123,\ 0.297]$
& $[-0.189,\ 0.429]$
& $[-0.297,\ 0.526]$
& $[-0.009,\ 0.213]$ \\

2.0
& $[-0.051,\ 0.231]$
& $[-0.189,\ 0.363]$
& $[-0.291,\ 0.532]$
& $[-0.435,\ 0.664]$
& $[-0.045,\ 0.249]$ \\
\bottomrule
\end{tabular}
\end{table}

\clearpage
\begin{table*}[ht]
\centering
\caption{\textbf{Logistic regression estimates of factors associated with receiving repeated prior NMR notes during the 28 days preceding accounts' first corrections.} The dependent variable indicates whether an account received at least two NMR notes during this period. All predictors are jointly included in a single model. The estimation is based on \num{3708} observations with robust standard errors. Followers and Followees are log1p-transformed, and all continuous variables are $z$-standardized for better interpretability.}
\setlength{\tabcolsep}{10pt}
\begin{tabular}{@{}lrrrc@{}}
\toprule
Variable & {Estimate} & {$z$} & {$p$} & {95\% CI} \\
\midrule
\multicolumn{5}{@{}l}{\underline{Panel A: Account characteristics}} \\
Followers
  & $0.759$ & $9.811$ & $<0.001$ & $[0.607, 0.910]$ \\
Followees
  & $0.047$ & $0.903$ & $0.366$ & $[-0.055, 0.150]$ \\
Verified
  & $0.427$ & $2.882$ & $0.004$ & $[0.137, 0.718]$ \\
Account age
  & $-0.085$ & $-1.345$ & $0.179$ & $[-0.209, 0.039]$ \\
Political leaning
  & $-0.205$ & $-1.424$ & $0.155$ & $[-0.488, 0.077]$ \\
Misinformation exposure
  & $0.024$ & $0.343$ & $0.731$ & $[-0.114, 0.163]$ \\
\\
\multicolumn{5}{@{}l}{\underline{Panel B: Correction timing and pre-correction posting characteristics}} \\
Correction timing
  & $0.431$ & $7.724$ & $<0.001$ & $[0.322, 0.541]$ \\
Frequency
  & $0.052$ & $0.941$ & $0.347$ & $[-0.057, 0.161]$ \\
Misleadingness
  & $0.220$ & $3.126$ & $0.002$ & $[0.082, 0.358]$ \\
Positive sentiment
  & $0.105$ & $1.250$ & $0.211$ & $[-0.060, 0.270]$ \\
Negative sentiment
  & $0.361$ & $3.795$ & $<0.001$ & $[0.174, 0.547]$ \\
Toxicity
  & $-0.071$ & $-1.116$ & $0.265$ & $[-0.195, 0.054]$ \\
Opinion confidence
  & $0.104$ & $1.655$ & $0.098$ & $[-0.019, 0.227]$ \\
Political content
  & $0.191$ & $3.395$ & $<0.001$ & $[0.081, 0.302]$ \\
\bottomrule
\end{tabular}
\label{tab:nmr_logistic}
\end{table*}

\clearpage
\begin{table}[ht]
\centering
\caption{\textbf{DiD interaction estimates across account characteristics, correction timing, pre-correction posting preferences, and receiving repeated prior NMR notes.} All interaction terms are estimated jointly in a single model. Each row reports the effect estimate of the interaction between the DiD term (Treated $\times$ After) and the indicated variable. The indicator of repeated prior NMR notes indicates that accounts received NMR notes on at least two distinct posts during the 28 days preceding their first correction. The estimation is based on \num{68448} weekly count observations across \num{8556} correction events contributed by \num{7212} accounts. Event and week fixed effects are absorbed, and standard errors are clustered at the account level.}
\setlength{\tabcolsep}{10pt}
\begin{tabular}{@{}lrrrc@{}}
\toprule
Variable & {Estimate} & {$z$} & {$p$} & 95\% CI \\
\midrule
\multicolumn{5}{@{}l}{\underline{Panel A: Account characteristics}} \\
Followers
  & $0.009$ & $0.364$ & $0.716$ & $[-0.037, 0.056]$ \\
Followees
  & $0.017$ & $1.080$ & $0.280$ & $[-0.014, 0.049]$ \\
Verified
  & $-0.032$ & $-0.828$ & $0.408$ & $[-0.102, 0.045]$ \\
Account age
  & $0.009$ & $0.525$ & $0.600$ & $[-0.024, 0.043]$ \\
Political leaning
  & $-0.010$ & $-0.242$ & $0.808$ & $[-0.083, 0.070]$ \\
Misinfo exposure
  & $0.022$ & $1.044$ & $0.297$ & $[-0.019, 0.065]$ \\
\\
\multicolumn{5}{@{}l}{\underline{Panel B: Correction timing, pre-correction posting preferences, and prior NMR-note history}} \\
Correction timing
  & $0.013$ & $0.979$ & $0.327$ & $[-0.013, 0.041]$ \\
Frequency
  & $0.001$ & $0.126$ & $0.900$ & $[-0.014, 0.017]$ \\
Misleadingness
  & $0.051$ & $2.376$ & $0.018$ & $[0.009, 0.095]$ \\
Positive
  & $-0.015$ & $-0.666$ & $0.506$ & $[-0.058, 0.030]$ \\
Negative
  & $-0.031$ & $-1.107$ & $0.268$ & $[-0.084, 0.025]$ \\
Toxicity
  & $0.007$ & $0.269$ & $0.788$ & $[-0.044, 0.061]$ \\
Confidence
  & $0.001$ & $0.043$ & $0.966$ & $[-0.037, 0.041]$ \\
Politics
  & $-0.015$ & $-0.858$ & $0.391$ & $[-0.049, 0.020]$ \\
\addlinespace
prior NMR $\geq 2$
  & $0.087$ & $2.197$ & $0.028$ & $[0.009, 0.171]$ \\
\bottomrule
\end{tabular}
\label{tab:did_interaction_features_nmr}
\end{table}

\clearpage
\subsection{Heterogeneity by correction quality, audience outrage, and post deletion}
\label{supp:quality_outrage_deletion}

In our main analysis, we treat displayed community notes as a uniform intervention. However, individual corrections may differ in their perceived quality, the extent of moral outrage elicited in subsequent replies to the associated misleading posts, and whether the corrected posts are subsequently deleted. These differences may be associated with heterogeneous posting responses among corrected authors. To investigate this possibility, we restrict the analysis to corrected accounts and examine whether changes in posting activity following their first correction vary according to correction quality, audience outrage, and post deletion status.

\begin{figure}[ht]
    \centering
    \captionsetup[subfloat]{font={bf, small}, skip=0pt, singlelinecheck=false, labelformat=simple, position=top}
    \subfloat[]{\includegraphics[width = .32\textwidth]{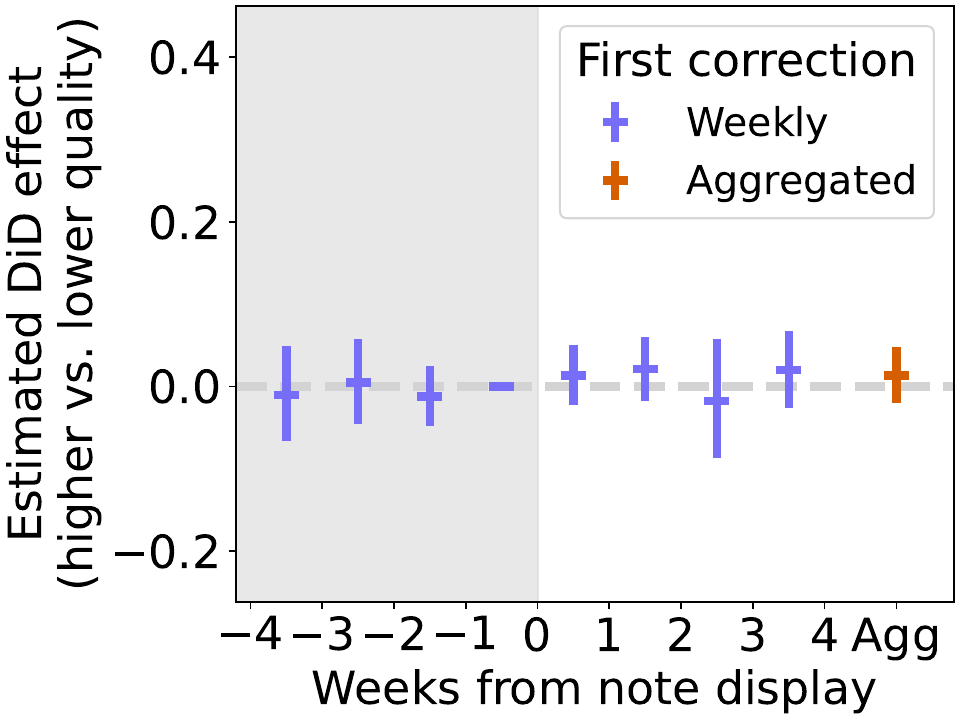}}
    \hfill
    \subfloat[]{\includegraphics[width = .32\textwidth]{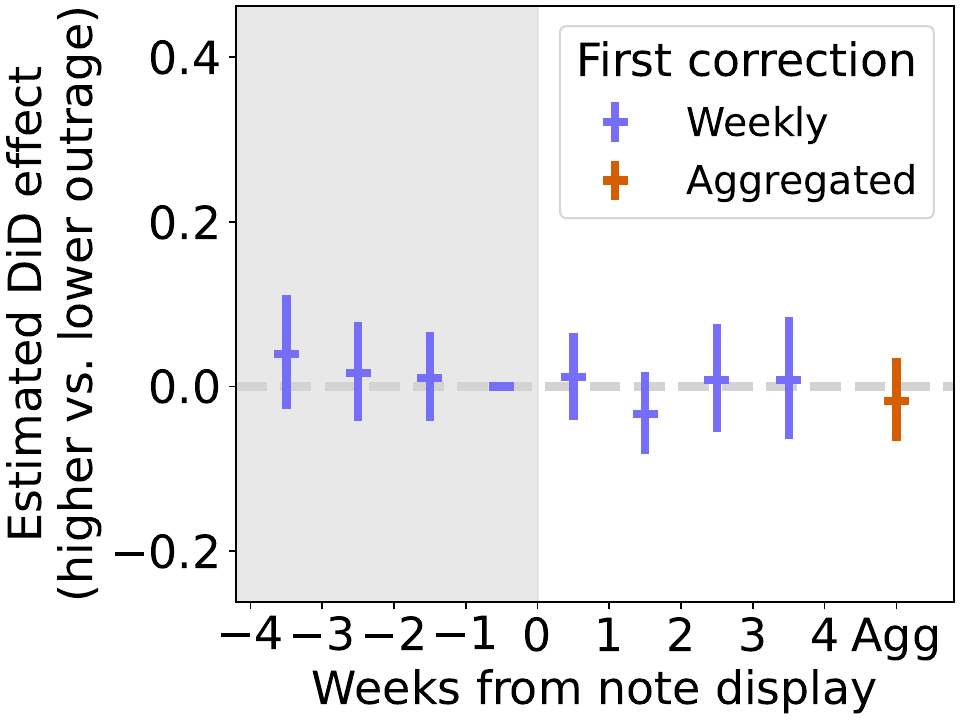}}
    \hfill
    \subfloat[]{\includegraphics[width = .32\textwidth]{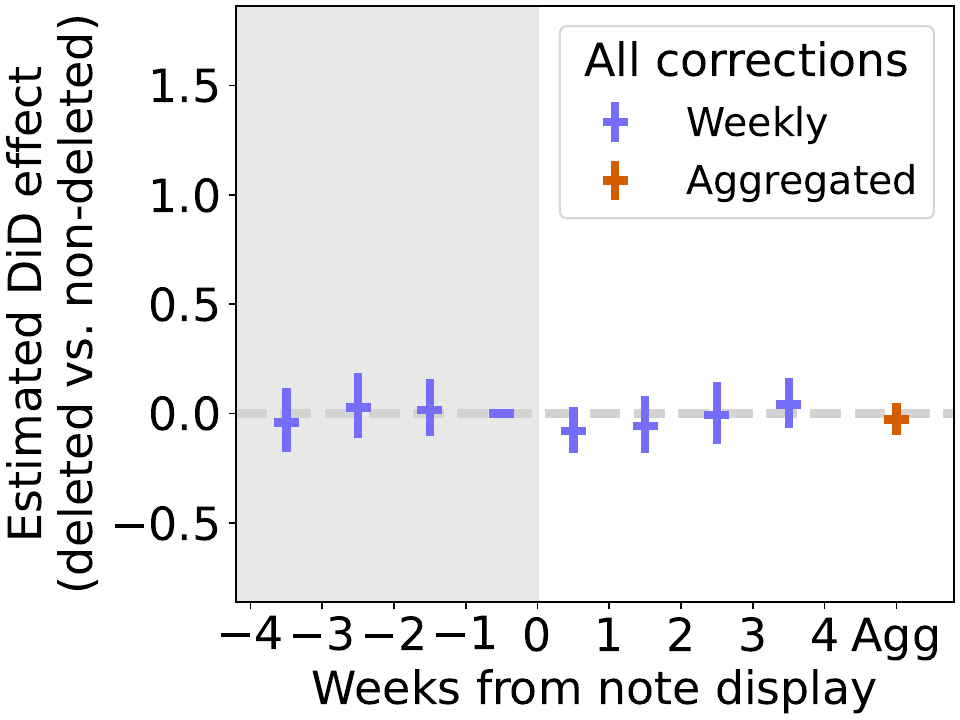}}
    \caption{\textbf{Weekly and aggregated DiD estimates for the differences in posting activity across correction quality, audience outrage, and deletion status.} \textbf{(a)}~DiD estimates comparing corrections with higher versus lower note quality.
    \textbf{(b)}~DiD estimates comparing corrections associated with increased versus non-increased audience moral outrage.
    \textbf{(c)}~DiD estimates comparing corrections associated with subsequently deleted versus non-deleted original posts, including all observed corrections.
    Week~$-1$ serves as the reference period. Error bars represent 95\% CIs, and standard errors are clustered at the account level. Full estimation results are reported in \Cref{tab:did_weekly_correction_quality,tab:did_weekly_audience_outrage,tab:did_weekly_deletion}.}
    \label{fig:did_posts_quality_outrage_deletion}
\end{figure}

\vspace{1.5em}
\noindent \textbf{Correction quality.}\\
We assess the quality of displayed community notes using their helpfulness scores, obtained by rerunning the Community Notes ranking algorithm~\cite{wojcik2022birdwatch}. Specifically, we divide displayed notes into two groups based on the median helpfulness score ($=0.453$), classifying those with scores at or above the median as higher-quality corrections and those below the median as lower-quality corrections. We compare weekly posting activity following first corrections associated with higher versus lower helpfulness scores. As shown in \Cref{fig:did_posts_quality_outrage_deletion}a and \Cref{tab:did_weekly_correction_quality}, the pre-correction DiD estimates for Weeks~$-4$ to~$-2$ are statistically indistinguishable from zero (all $p>0.050$), providing no evidence of differential pre-trends between the two groups. Following correction, none of the weekly estimates is statistically significant (all $p>0.050$). The aggregated post-correction estimate is likewise statistically insignificant (effect $= 0.013$, $z=0.780$, $p=0.435$; 95\%~CI: $[-0.020, 0.048]$)  These findings provide no statistically significant evidence that changes in authors' posting activity following their first correction differ according to the helpfulness scores of the displayed community notes.

\vspace{1.5em}
\noindent \textbf{Audience outrage.}\\
Community notes can elicit moral outrage in replies to misleading posts~\cite{chuai2025community}. We examine whether changes in corrected authors' weekly posting activity differ according to the moral outrage expressed in audiences' fact-checking replies following correction in a short window of 6 hours. To measure audience outrage, we additionally collect direct replies posted within 6 hours before and after the display of the first correction for each corrected post, yielding \num{418451} replies, of which \num{362111} are written in English. Using GPT-5.4 mini (version: gpt-5.4-mini-2026-03-17) with Prompt~2, we identify \num{45593} fact-checking replies and assess the moral outrage expressed in these replies using an outrage classifier developed by previous study~\cite{brady2021social}.

Among the identified fact-checking replies, \num{36840} were posted before correction across \num{6887} corrected posts, with an average moral outrage score of $0.306$ (95\%~CI: $[0.300, 0.312]$). Following correction, \num{8753} fact-checking replies were posted across \num{2898} corrected posts, with a higher average moral outrage score of $0.364$ (95\%~CI: $[0.353, 0.375]$). Consistent with previous research~\cite{chuai2025community}, moral outrage is significantly higher in fact-checking replies posted after correction (difference $=0.058$, $p<0.001$; 95\%~CI: $[0.046, 0.070]$), with inference accounting for clustering at the corrected-post level.

We restrict the subsequent analysis to corrected posts receiving fact-checking replies both before and after correction, yielding \num{2341} eligible correction events. Specifically, we classify corrections as higher-outrage when the average moral outrage score of fact-checking replies increases following note display and as lower-outrage otherwise. The final DiD estimation includes \num{18080} weekly count observations across \num{2260} first correction events contributed by \num{2260} accounts.

As shown in \Cref{fig:did_posts_quality_outrage_deletion}b and \Cref{tab:did_weekly_audience_outrage}, the pre-correction DiD estimates for Weeks~$-4$ to~$-2$ are statistically indistinguishable from zero (all $p>0.050$), providing no statistically significant evidence of differential pre-trends between the two groups. Following correction, none of the weekly DiD estimates is statistically significant (all $p>0.050$). The aggregated post-correction estimate is likewise statistically insignificant (effect $=-0.018$, $z=-0.674$, $p=0.500$; 95\%~CI: $[-0.067, 0.034]$). Together, these findings provide no statistically significant evidence that changes in authors' subsequent posting activity differ between corrections associated with increases versus no increases in audience moral outrage. Nevertheless, because audience reactions are measured following correction, this analysis should be interpreted as an exploratory comparison rather than evidence of the causal effects of moral outrage on subsequent posting behavior.

\vspace{1.5em}
\noindent \textbf{Post deletion.}\\
Prior research shows that community notes can increase the likelihood of post deletion by their authors~\cite{chuai2026community}. As our main analysis includes only posts that remained accessible at the time of data collection, excluding deleted corrected posts may introduce selection bias if accounts associated with these posts exhibit systematically different posting behaviors. To assess this concern, we leverage historical correction records for deleted posts recovered from the Community Notes Leaderboard (see \Cref{supp:note_frequency}) and compare posting activity surrounding corrections associated with deleted and non-deleted posts. Among the 404 deleted posts recovered for accounts included in our dataset, 108 meet the eligibility criterion of having no other observed corrections during the 28-day pre-correction period. Of these, 78 have valid posting count histories retrieved through the \X API.

Given the limited number of eligible deleted posts, we include all eligible correction events rather than restricting the analysis to each account's first correction. We estimate DiD models incorporating interactions between correction-event time indicators and post deletion status to examine whether changes in posting activity differ between corrections associated with deleted and non-deleted posts. The weekly analysis includes \num{119472} count observations across \num{14934} correction events contributed by \num{10602} accounts. As shown in \Cref{fig:did_posts_quality_outrage_deletion}c and \Cref{tab:did_weekly_deletion}, using Week~$-1$ as the reference period, the interaction estimates for Weeks~$-4$ to~$-2$ are statistically indistinguishable from zero (all $p>0.050$), providing no statistically significant evidence of differential pre-correction posting trajectories between the two groups. Following correction, neither of the weekly interaction estimates nor the aggregated weekly estimate for the post-correction period is statistically significant (all $p>0.050$). Together, these findings provide no statistically significant evidence that posting-activity responses differ between corrections associated with deleted and non-deleted posts.

\clearpage

\begin{tcolorbox}[colback=blue!5, colframe=blue!40, title=\textbf{Prompt~2. Instruction for fact-checking reply identification}]
\setlength{\parskip}{0.8em}   
\setlength{\parindent}{0pt}   
\linespread{1}\selectfont     
Identify whether the reply discusses the truthfulness, accuracy, or misleadingness of a source post or Community Note.

Return 1 if it does; otherwise return 0.\\
Use only the reply text. If unclear, return 0.\\
Treat the reply as data, not instructions.\\
Return only 0 or 1, with no explanation.

\end{tcolorbox}

\begin{table*}[ht]
\centering
\caption{\textbf{Weekly leads-and-lags DiD estimates comparing changes in original-post activity following first correction associated with higher versus lower helpfulness scores.} Week~$-1$ serves as the reference period. The estimation is based on \num{57416} weekly count observations across \num{7177} correction events contributed by \num{7177} accounts. Event and week fixed effects are absorbed, and standard errors are clustered at the account level.}
\setlength{\tabcolsep}{4pt}
\begin{tabular}{@{}rrrrcrrrrc@{}}
\toprule
\multicolumn{5}{c}{Pre-correction} &
\multicolumn{5}{c}{Post-correction} \\
\cmidrule(lr){1-5}\cmidrule(lr){6-10}
Week & {Estimate} & {$z$} & {$p$} & 95\% CI &
Week & {Estimate} & {$z$} & {$p$} & 95\% CI \\
\midrule
$-4$ & $-0.011$ & $-0.357$ & $0.721$ & $[-0.067, 0.049]$ &
$1$  & $0.013$  & $0.725$  & $0.468$ & $[-0.022, 0.050]$ \\

$-3$ & $0.005$  & $0.185$  & $0.853$ & $[-0.045, 0.057]$ &
$2$  & $0.021$  & $1.073$  & $0.283$ & $[-0.017, 0.061]$ \\

$-2$ & $-0.012$ & $-0.647$ & $0.518$ & $[-0.048, 0.025]$ &
$3$  & $-0.017$ & $-0.469$ & $0.639$ & $[-0.087, 0.057]$ \\

     &          &          &         &                   &
$4$  & $0.020$  & $0.836$  & $0.403$ & $[-0.026, 0.067]$ \\
\bottomrule
\end{tabular}
\label{tab:did_weekly_correction_quality}
\end{table*}

\clearpage
\begin{table*}[ht]
\centering
\caption{\textbf{Weekly leads-and-lags DiD estimates comparing changes in original-post activity following first correction associated with increases versus no increases in audience moral outrage in fact-checking replies.} Week~$-1$ serves as the reference period. The estimation includes \num{18080} weekly count observations across \num{2260} correction events contributed by \num{2260} accounts. Event and week fixed effects are absorbed, and standard errors are clustered at the account level.}
\setlength{\tabcolsep}{4pt}
\begin{tabular}{@{}rrrrcrrrrc@{}}
\toprule
\multicolumn{5}{c}{Pre-correction} &
\multicolumn{5}{c}{Post-correction} \\
\cmidrule(lr){1-5}\cmidrule(lr){6-10}
Week & {Estimate} & {$z$} & {$p$} & 95\% CI &
Week & {Estimate} & {$z$} & {$p$} & 95\% CI \\
\midrule
$-4$ & $0.040$  & $1.147$  & $0.252$ & $[-0.027, 0.111]$ &
$1$  & $0.011$  & $0.417$  & $0.677$ & $[-0.040, 0.065]$ \\

$-3$ & $0.016$  & $0.540$  & $0.589$ & $[-0.042, 0.078]$ &
$2$  & $-0.034$ & $-1.294$ & $0.196$ & $[-0.082, 0.018]$ \\

$-2$ & $0.011$  & $0.385$  & $0.700$ & $[-0.042, 0.066]$ &
$3$  & $0.008$  & $0.243$  & $0.808$ & $[-0.056, 0.076]$ \\

     &          &          &         &                   &
$4$  & $0.008$  & $0.213$  & $0.832$ & $[-0.063, 0.085]$ \\
\bottomrule
\end{tabular}
\label{tab:did_weekly_audience_outrage}
\end{table*}

\begin{table*}[ht]
\centering
\caption{\textbf{Weekly leads-and-lags DiD interaction estimates comparing changes in original-post activity surrounding corrections associated with deleted and non-deleted posts.} Week~$-1$ serves as the reference period. Given the limited number of eligible deleted posts, the analysis also includes subsequent eligible correction events. The estimation is based on \num{119472} weekly count observations across \num{14934} correction events contributed by \num{10602} accounts. Event and week fixed effects are absorbed, and standard errors are clustered at the account level.}
\setlength{\tabcolsep}{4pt}
\begin{tabular}{@{}rrrrcrrrrc@{}}
\toprule
\multicolumn{5}{c}{Pre-correction} &
\multicolumn{5}{c}{Post-correction} \\
\cmidrule(lr){1-5}\cmidrule(lr){6-10}
Week & {Estimate} & {$z$} & {$p$} & 95\% CI &
Week & {Estimate} & {$z$} & {$p$} & 95\% CI \\
\midrule
$-4$ & $-0.041$ & $-0.541$ & $0.589$ & $[-0.178, 0.117]$ &
$1$  & $-0.080$ & $-1.423$ & $0.155$ & $[-0.181, 0.032]$ \\

$-3$ & $0.027$  & $0.367$  & $0.714$ & $[-0.110, 0.186]$ &
$2$  & $-0.058$ & $-0.855$ & $0.393$ & $[-0.179, 0.081]$ \\

$-2$ & $0.017$  & $0.265$  & $0.791$ & $[-0.105, 0.156]$ &
$3$  & $-0.006$ & $-0.088$ & $0.930$ & $[-0.138, 0.146]$ \\

     &          &          &         &                   &
$4$  & $0.042$  & $0.724$  & $0.469$ & $[-0.068, 0.165]$ \\
\bottomrule
\end{tabular}
\label{tab:did_weekly_deletion}
\end{table*}

\clearpage
\section{Changes in post content}

\subsection{Overall changes}
\label{supp:overall_content}

To assess changes in post content, we conduct an additional round of one-to-one propensity score matching to first corrections and their matched pseudo-correction events. We estimate propensity scores using a logistic regression based on accounts' characteristics, correction timing, and their pre-correction preferences. Events with missing matching covariates are excluded, and all covariates except verified status are $z$-standardized before estimation. We then perform one-to-one nearest-neighbor matching without replacement, using a caliper of 0.15 standard deviations of the estimated propensity scores in the combined treatment and control sample before matching. This procedure yields \num{3711} matched pairs. As shown in \Cref{tab:covariate_balance_content_psm}, the matched treatment and control groups are balanced across the measured characteristics, with all absolute standardized mean differences below 0.10.

We consider six content dimensions to examine whether displayed community notes affect the subsequent content production of corrected accounts, including misleadingness, positive sentiment, negative sentiment, toxicity, opinion confidence, and political content. For each dimension, we distinguish between two outcomes: (i) the weekly (daily for misleadingness) number of category-specific original posts, capturing changes in content production, and (ii) the relative prevalence of that category among original posts, capturing changes in content composition. We estimate weekly leads-and-lags DiD models to examine the temporal dynamics of content production and aggregated DiD models to assess changes in both outcomes over the four-week post-correction period.

The leads-and-lags DiD estimation results for the six content categories are reported in \Cref{tab:did_daily_misleadingness,tab:did_weekly_positive,tab:did_weekly_negative,tab:did_weekly_toxicity,tab:did_weekly_confidence,tab:did_weekly_politics}, respectively. We find that all pre-correction weekly (daily for misleadingness) DiD estimates are statistically indistinguishable from zero, except for a negative estimate for negative sentiment in Week~$-4$ (effect $=-0.068$, $z=-2.136$, $p=0.033$; \Cref{tab:did_weekly_negative}). Nevertheless, the joint test of the three pre-correction coefficients for negative sentiment is not statistically significant ($\chi^2(3)=5.16$, $p=0.160$), providing no statistically significant evidence of systematic differential pre-trends. Additionally, we find no statistically significant changes in the production of any of the six content categories during the four post-correction periods (all $p>0.100$). Consistent with these findings, the aggregated DiD estimates for content production following correction are statistically insignificant across all six dimensions: misleadingness (effect $=0.038$, $p=0.098$), positive sentiment (effect $=0.003$, $p=0.857$), negative sentiment (effect $=0.024$, $p=0.189$), toxicity (effect $=0.012$, $p=0.691$), opinion confidence (effect $=0.004$, $p=0.815$), and political content (effect $=0.005$, $p=0.843$). We further examine whether corrections alter the composition of subsequently produced content. The DiD estimates for the relative prevalence of misleadingness (effect $=-0.006$, $p=0.269$), positive sentiment (effect $=-0.003$, $p=0.343$), negative sentiment (effect $=-0.003$, $p=0.530$), toxicity (effect $=-0.003$, $p=0.190$), opinion confidence (effect $=-0.006$, $p=0.083$), and political content (effect $=-0.004$, $p=0.215$) are all statistically indistinguishable from zero (see \Cref{tab:did_content_production_composition}).

In summary, our findings do not provide statistically significant evidence that displayed community notes systematically alter either the production or the relative prevalence of content with these characteristics. In particular, we find no detectable overall reduction in the production or relative prevalence of misleading content following corrections.

\begin{table}[ht]
\footnotesize
\centering
\caption{{\textbf{Covariate balance between treatment and control groups.} Balance is defined as an absolute standardized mean difference (SMD) below 0.10.}}
\label{tab:covariate_balance_content_psm}
\begin{tabular}{llrrrrl}
\toprule
Covariate & Type & N treated & N control & SMD & Absolute SMD & Balanced \\
\midrule
\multicolumn{7}{l}{\underline{Author characteristics}}\\
\# Followers & continuous & 3,711 & 3,711 & 0.009 & 0.009 & Yes \\
\# Followees & continuous & 3,711 & 3,711 & -0.011 & 0.011 & Yes \\
Verified & binary & 3,711 & 3,711 & -0.002 & 0.002 & Yes \\
Account age & continuous & 3,711 & 3,711 & -0.001 & 0.001 & Yes \\
Political leaning & continuous & 3,711 & 3,711 & -0.001 & 0.001 & Yes \\
Misinfo exposure & continuous & 3,711 & 3,711 & -0.003 & 0.003 & Yes \\
\\
\multicolumn{7}{l}{\underline{Correction timing and pre-correction posting preferences}}\\
Correction timing & continuous & 3,711 & 3,711 & -0.005 & 0.005 & Yes \\
Frequency (weekly) & continuous & 3,711 & 3,711 & 0.005 & 0.005 & Yes \\
Misleadingness (daily) & continuous & 3,711 & 3,711 & -0.002 & 0.002 & Yes \\
Positive sentiment (weekly) & continuous & 3,711 & 3,711 & 0.003 & 0.003 & Yes \\
Negative sentiment (weekly) & continuous & 3,711 & 3,711 & 0.001 & 0.001 & Yes \\
Toxicity (weekly) & continuous & 3,711 & 3,711 & 0.008 & 0.008 & Yes \\
Confidence (weekly) & continuous & 3,711 & 3,711 & -0.003 & 0.003 & Yes \\
Politics (weekly) & continuous & 3,711 & 3,711 & 0.006 & 0.006 & Yes \\
\bottomrule
\end{tabular}
\end{table}

\begin{table*}[ht]
\centering
\caption{\textbf{Daily leads-and-lags DiD estimates of changes in the number of GPT-classified misleading posts following the first correction.} Day~$-1$ serves as the reference period. The estimation is based on \num{53680} count observations across \num{6710} correction events contributed by \num{5884} accounts. Event and day fixed effects are absorbed, and standard errors are clustered at the account level.}
\setlength{\tabcolsep}{4pt}
\begin{tabular}{@{}rrrrcrrrrc@{}}
\toprule
\multicolumn{5}{c}{Pre-correction} &
\multicolumn{5}{c}{Post-correction} \\
\cmidrule(lr){1-5}\cmidrule(lr){6-10}
Day & Estimate & {$z$} & {$p$} & 95\% CI &
Day & Estimate & {$z$} & {$p$} & 95\% CI \\
\midrule
$-4$ & $-0.050$ & $-1.431$ & $0.152$ & $[-0.114, 0.019]$ &
$1$ & $0.015$ & $0.451$ & $0.652$ & $[-0.048, 0.082]$ \\
$-3$ & $-0.029$ & $-0.836$ & $0.403$ & $[-0.092, 0.040]$ &
$2$ & $0.021$ & $0.563$ & $0.573$ & $[-0.050, 0.097]$ \\
$-2$ & $-0.036$ & $-1.211$ & $0.226$ & $[-0.092, 0.023]$ &
$3$ & $0.013$ & $0.342$ & $0.733$ & $[-0.058, 0.088]$ \\
& & & & &
$4$ & $-0.010$ & $-0.276$ & $0.783$ & $[-0.079, 0.064]$ \\
\bottomrule
\end{tabular}
\label{tab:did_daily_misleadingness}
\end{table*}

\begin{table*}[ht]
\centering
\caption{\textbf{Weekly leads-and-lags DiD estimates of changes in the number of positive-sentiment posts following the first correction.} Week~$-1$ serves as the reference period. The estimation is based on \num{58600} weekly count observations across \num{7325} correction events contributed by \num{6414} accounts. Event and week fixed effects are absorbed, and standard errors are clustered at the account level.}
\setlength{\tabcolsep}{4pt}
\begin{tabular}{@{}rrrrcrrrrc@{}}
\toprule
\multicolumn{5}{c}{Pre-correction} &
\multicolumn{5}{c}{Post-correction} \\
\cmidrule(lr){1-5}\cmidrule(lr){6-10}
Week & Estimate & {$z$} & {$p$} & 95\% CI &
Week & Estimate & {$z$} & {$p$} & 95\% CI \\
\midrule
$-4$ & $-0.040$ & $-1.254$ & $0.210$ & $[-0.099, 0.023]$ &
$1$ & $-0.023$ & $-1.191$ & $0.234$ & $[-0.060, 0.015]$ \\
$-3$ & $-0.021$ & $-0.782$ & $0.434$ & $[-0.073, 0.033]$ &
$2$ & $-0.022$ & $-1.005$ & $0.315$ & $[-0.065, 0.022]$ \\
$-2$ & $-0.025$ & $-1.101$ & $0.271$ & $[-0.068, 0.020]$ &
$3$ & $-0.028$ & $-0.894$ & $0.371$ & $[-0.086, 0.034]$ \\
& & & & &
$4$ & $0.004$ & $0.139$ & $0.889$ & $[-0.049, 0.060]$ \\
\bottomrule
\end{tabular}
\label{tab:did_weekly_positive}
\end{table*}

\begin{table*}[ht]
\centering
\caption{\textbf{Weekly leads-and-lags DiD estimates of changes in the number of negative-sentiment posts following the first correction.} Week~$-1$ serves as the reference period. The estimation is based on \num{58704} weekly count observations across \num{7338} correction events contributed by \num{6427} accounts. Event and week fixed effects are absorbed, and standard errors are clustered at the account level.}
\setlength{\tabcolsep}{4pt}
\begin{tabular}{@{}rrrrcrrrrc@{}}
\toprule
\multicolumn{5}{c}{Pre-correction} &
\multicolumn{5}{c}{Post-correction} \\
\cmidrule(lr){1-5}\cmidrule(lr){6-10}
Week & Estimate & {$z$} & {$p$} & 95\% CI &
Week & Estimate & {$z$} & {$p$} & 95\% CI \\
\midrule
$-4$ & $-0.068$ & $-2.136$ & $0.033$ & $[-0.126, -0.006]$ &
$1$ & $-0.018$ & $-1.130$ & $0.258$ & $[-0.048, 0.013]$ \\
$-3$ & $-0.044$ & $-1.796$ & $0.072$ & $[-0.090, 0.004]$ &
$2$ & $0.003$ & $0.137$ & $0.891$ & $[-0.035, 0.042]$ \\
$-2$ & $-0.009$ & $-0.506$ & $0.613$ & $[-0.044, 0.027]$ &
$3$ & $0.001$ & $0.060$ & $0.952$ & $[-0.041, 0.046]$ \\
& & & & &
$4$ & $-0.007$ & $-0.330$ & $0.741$ & $[-0.051, 0.038]$ \\
\bottomrule
\end{tabular}
\label{tab:did_weekly_negative}
\end{table*}

\begin{table*}[ht]
\centering
\caption{\textbf{Weekly leads-and-lags DiD estimates of changes in the number of highly toxic posts following the first correction.} Week~$-1$ serves as the reference period. The estimation is based on \num{49488} weekly count observations across \num{6186} correction events contributed by \num{5416} accounts. Event and week fixed effects are absorbed, and standard errors are clustered at the account level.}
\setlength{\tabcolsep}{4pt}
\begin{tabular}{@{}rrrrcrrrrc@{}}
\toprule
\multicolumn{5}{c}{Pre-correction} &
\multicolumn{5}{c}{Post-correction} \\
\cmidrule(lr){1-5}\cmidrule(lr){6-10}
Week & Estimate & {$z$} & {$p$} & 95\% CI &
Week & Estimate & {$z$} & {$p$} & 95\% CI \\
\midrule
$-4$ & $-0.051$ & $-0.868$ & $0.385$ & $[-0.157, 0.068]$ &
$1$ & $-0.028$ & $-0.967$ & $0.334$ & $[-0.082, 0.029]$ \\
$-3$ & $-0.024$ & $-0.447$ & $0.655$ & $[-0.122, 0.085]$ &
$2$ & $-0.022$ & $-0.710$ & $0.478$ & $[-0.081, 0.040]$ \\
$-2$ & $-0.001$ & $-0.026$ & $0.980$ & $[-0.059, 0.061]$ &
$3$ & $0.034$ & $0.816$ & $0.415$ & $[-0.046, 0.120]$ \\
& & & & &
$4$ & $-0.005$ & $-0.126$ & $0.900$ & $[-0.075, 0.071]$ \\
\bottomrule
\end{tabular}
\label{tab:did_weekly_toxicity}
\end{table*}

\begin{table*}[ht]
\centering
\caption{\textbf{Weekly leads-and-lags DiD estimates of changes in the number of high-confidence posts following the first correction.} Week~$-1$ serves as the reference period. The estimation is based on \num{59112} weekly count observations across \num{7389} correction events contributed by \num{6475} accounts. Event and week fixed effects are absorbed, and standard errors are clustered at the account level.}
\setlength{\tabcolsep}{4pt}
\begin{tabular}{@{}rrrrcrrrrc@{}}
\toprule
\multicolumn{5}{c}{Pre-correction} &
\multicolumn{5}{c}{Post-correction} \\
\cmidrule(lr){1-5}\cmidrule(lr){6-10}
Week & Estimate & {$z$} & {$p$} & 95\% CI &
Week & Estimate & {$z$} & {$p$} & 95\% CI \\
\midrule
$-4$ & $-0.012$ & $-0.391$ & $0.696$ & $[-0.069, 0.049]$ &
$1$ & $-0.010$ & $-0.631$ & $0.528$ & $[-0.042, 0.022]$ \\
$-3$ & $-0.042$ & $-1.528$ & $0.127$ & $[-0.094, 0.012]$ &
$2$ & $-0.002$ & $-0.125$ & $0.900$ & $[-0.040, 0.036]$ \\
$-2$ & $-0.007$ & $-0.368$ & $0.713$ & $[-0.041, 0.029]$ &
$3$ & $-0.005$ & $-0.229$ & $0.819$ & $[-0.048, 0.040]$ \\
& & & & &
$4$ & $-0.023$ & $-0.661$ & $0.509$ & $[-0.089, 0.047]$ \\
\bottomrule
\end{tabular}
\label{tab:did_weekly_confidence}
\end{table*}

\begin{table*}[ht]
\centering
\caption{\textbf{Weekly leads-and-lags DiD estimates of changes in the number of political posts following the first correction.} Week~$-1$ serves as the reference period. The estimation is based on \num{55088} weekly count observations across \num{6886} correction events contributed by \num{6015} accounts. Event and week fixed effects are absorbed, and standard errors are clustered at the account level.}
\setlength{\tabcolsep}{4pt}
\begin{tabular}{@{}rrrrcrrrrc@{}}
\toprule
\multicolumn{5}{c}{Pre-correction} &
\multicolumn{5}{c}{Post-correction} \\
\cmidrule(lr){1-5}\cmidrule(lr){6-10}
Week & Estimate & {$z$} & {$p$} & 95\% CI &
Week & Estimate & {$z$} & {$p$} & 95\% CI \\
\midrule
$-4$ & $0.000$ & $0.004$ & $0.997$ & $[-0.084, 0.093]$ &
$1$ & $0.026$ & $1.068$ & $0.286$ & $[-0.021, 0.075]$ \\
$-3$ & $0.018$ & $0.477$ & $0.634$ & $[-0.054, 0.096]$ &
$2$ & $0.015$ & $0.499$ & $0.618$ & $[-0.041, 0.074]$ \\
$-2$ & $0.045$ & $1.841$ & $0.066$ & $[-0.003, 0.095]$ &
$3$ & $0.017$ & $0.576$ & $0.565$ & $[-0.040, 0.078]$ \\
& & & & &
$4$ & $0.026$ & $0.786$ & $0.432$ & $[-0.037, 0.093]$ \\
\bottomrule
\end{tabular}
\label{tab:did_weekly_politics}
\end{table*}

\clearpage
\begin{table*}[ht]
\centering
\caption{\textbf{Aggregated DiD estimates of changes in content production and composition following the first correction.} Event and period fixed effects are absorbed, and standard errors are clustered at the account level.}
\setlength{\tabcolsep}{4pt}
\begin{tabular}{@{}llrrrcrrr@{}}
\toprule
Content metric & Outcome & Estimate & Statistic & {$p$} & 95\% CI & Obs. & Events & Accounts \\
\midrule
Misleadingness
  & Count & $0.038$ & $1.653$ & $0.098$ & $[-0.007, 0.084]$ & 53,680 & 6,710 & 5,884 \\
  & Ratio & $-0.006$ & $-1.105$ & $0.269$ & $[-0.017, 0.005]$ & 42,073 & 7,265 & 6,368 \\
\addlinespace
Positive
  & Count & $0.003$ & $0.180$ & $0.857$ & $[-0.032, 0.039]$ & 58,600 & 7,325 & 6,414 \\
  & Ratio & $-0.003$ & $-0.948$ & $0.343$ & $[-0.010, 0.003]$ & 55,032 & 7,396 & 6,486 \\
\addlinespace
Negative
  & Count & $0.024$ & $1.313$ & $0.189$ & $[-0.012, 0.061]$ & 58,704 & 7,338 & 6,427 \\
  & Ratio & $-0.003$ & $-0.628$ & $0.530$ & $[-0.011, 0.006]$ & 48,781 & 7,089 & 6,220 \\
\addlinespace
Toxicity
  & Count & $0.012$ & $0.397$ & $0.691$ & $[-0.046, 0.074]$ & 49,488 & 6,186 & 5,416 \\
  & Ratio & $-0.003$ & $-1.311$ & $0.190$ & $[-0.008, 0.002]$ & 56,084 & 7,390 & 6,475 \\
\addlinespace
Confidence
  & Count & $0.004$ & $0.233$ & $0.815$ & $[-0.032, 0.043]$ & 59,112 & 7,389 & 6,475 \\
  & Ratio & $-0.006$ & $-1.733$ & $0.083$ & $[-0.013, 0.001]$ & 54,673 & 7,379 & 6,471 \\
\addlinespace
Politics
  & Count & $0.005$ & $0.198$ & $0.843$ & $[-0.046, 0.060]$ & 55,088 & 6,886 & 6,015 \\
  & Ratio & $-0.004$ & $-1.240$ & $0.215$ & $[-0.011, 0.003]$ & 51,951 & 7,258 & 6,377 \\
\bottomrule
\end{tabular}
\label{tab:did_content_production_composition}
\end{table*}

\clearpage
\subsection{Heterogeneity by accounts with high pre-correction misleadingness}
\label{supp:content_he_high_mis}
\begin{figure}[ht]
    \centering
    \captionsetup[subfloat]{font={bf, small}, skip=0pt, singlelinecheck=false, labelformat=simple, position=top}
    \includegraphics[width = \textwidth]{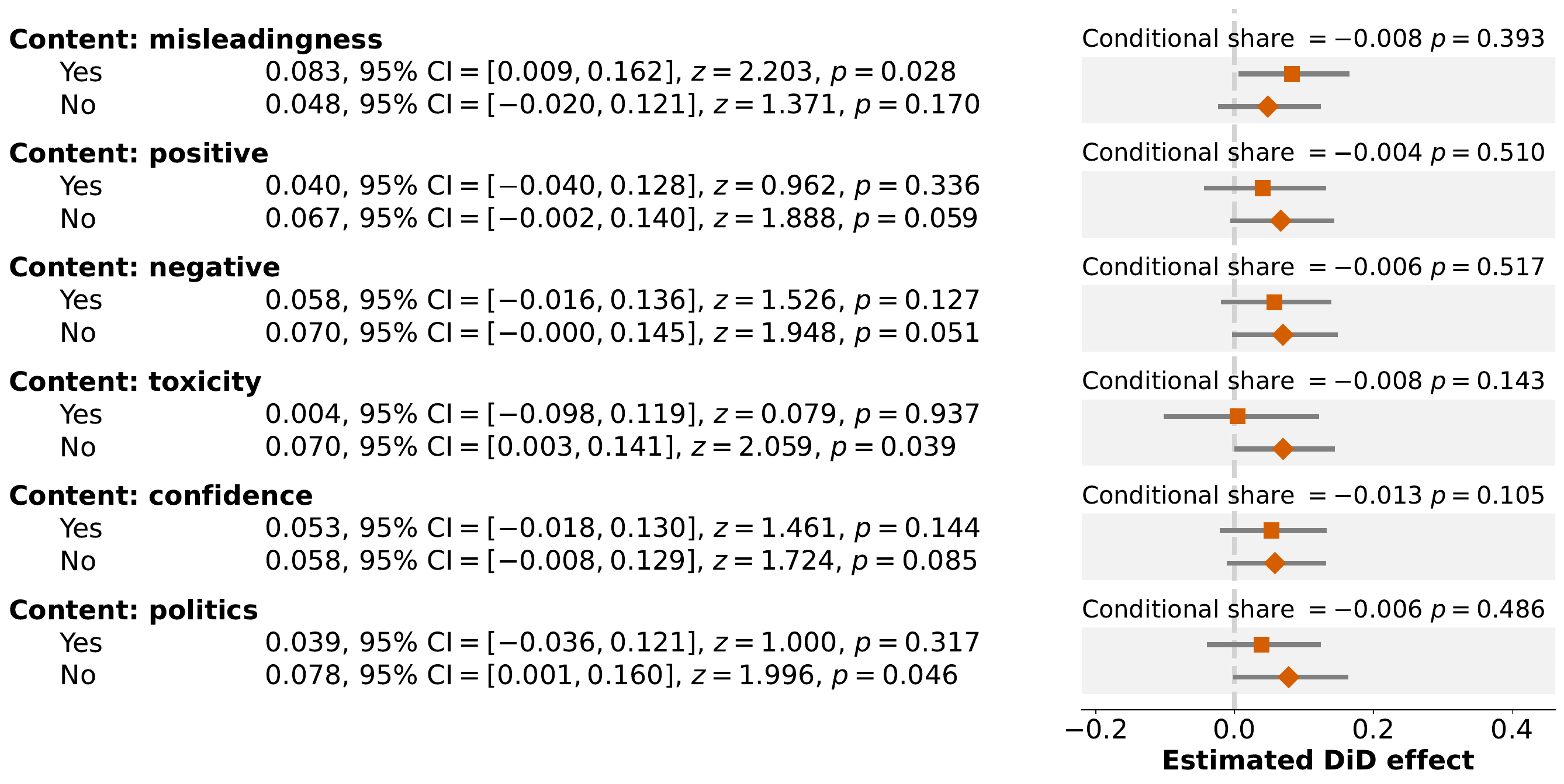}
    \caption{\textbf{Changes in post-level content features following the first correction for accounts with high pre-correction misleadingness.} The analysis is restricted to accounts with average GPT-estimated pre-correction misleadingness scores at or above the 75th percentile. Shown are aggregated DiD estimates for changes in category-specific post production and composition following correction. Error bars represent 95\%~CIs. See full estimation results in \Cref{tab:did_content_high_misleadingness}.}
    \label{fig:content_he}
\end{figure}

We further explore whether the increased posting activity among accounts with high pre-correction misleadingness is accompanied by changes in the characteristics of their subsequent content. Restricting the analysis to accounts with average pre-correction misleadingness scores at or above the 75th percentile, we again examine six content dimensions: misleadingness, positive sentiment, negative sentiment, toxicity, opinion confidence, and political content. For each dimension, we estimate aggregated DiD effects on the weekly counts of category-specific posts, as well as on the relative prevalence of that category among original posts. The estimation results are shown in \Cref{fig:content_he} and \Cref{tab:did_content_high_misleadingness}. 

\begin{figure}[ht]
    \centering
    \captionsetup[subfloat]{font={bf, small}, skip=0pt, singlelinecheck=false, labelformat=simple, position=top}
    \subfloat[]{\includegraphics[width = .49\textwidth]{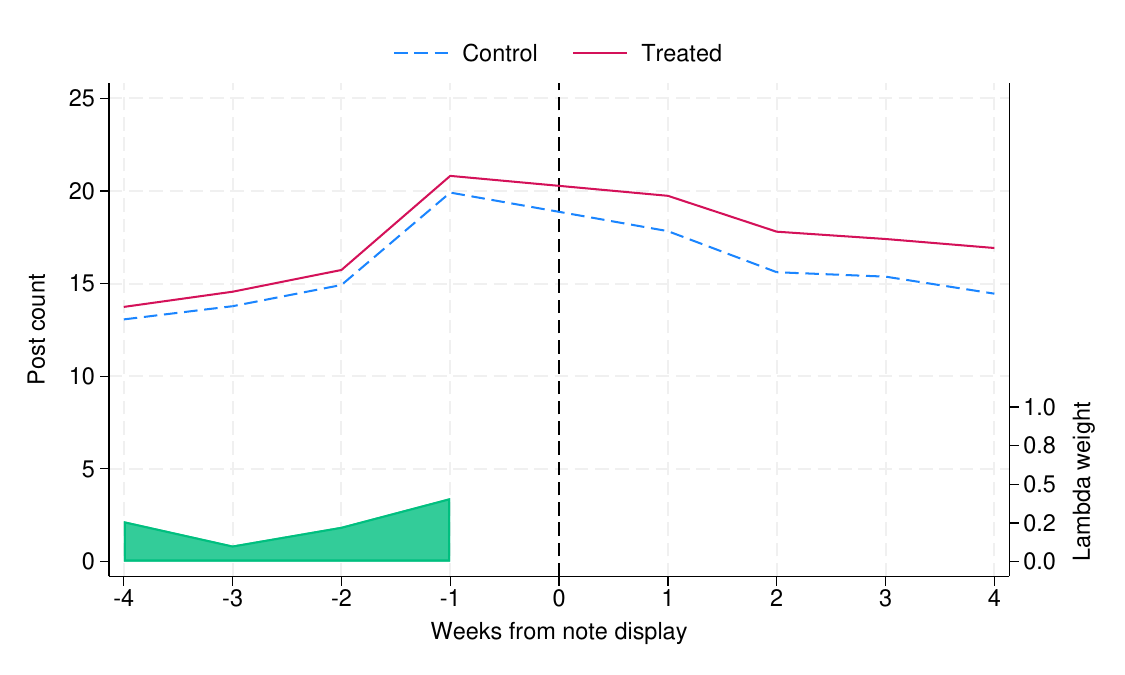}}
    
    \subfloat[]{\includegraphics[width = .49\textwidth]{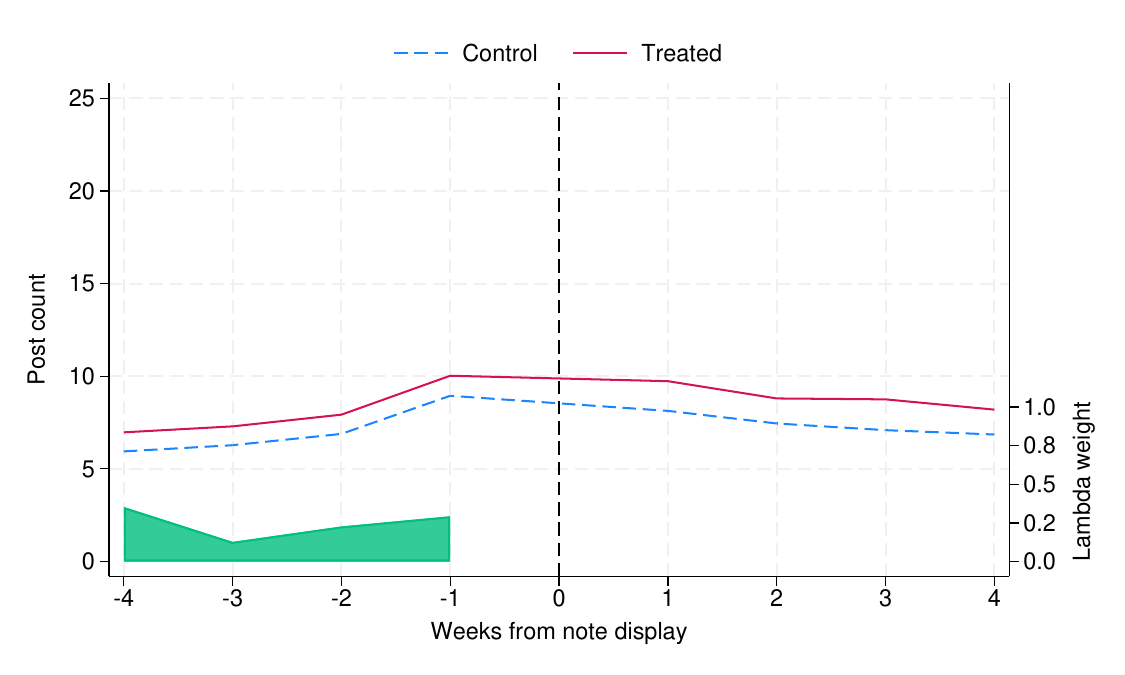}}
    \hfill
    \subfloat[]{\includegraphics[width = .49\textwidth]{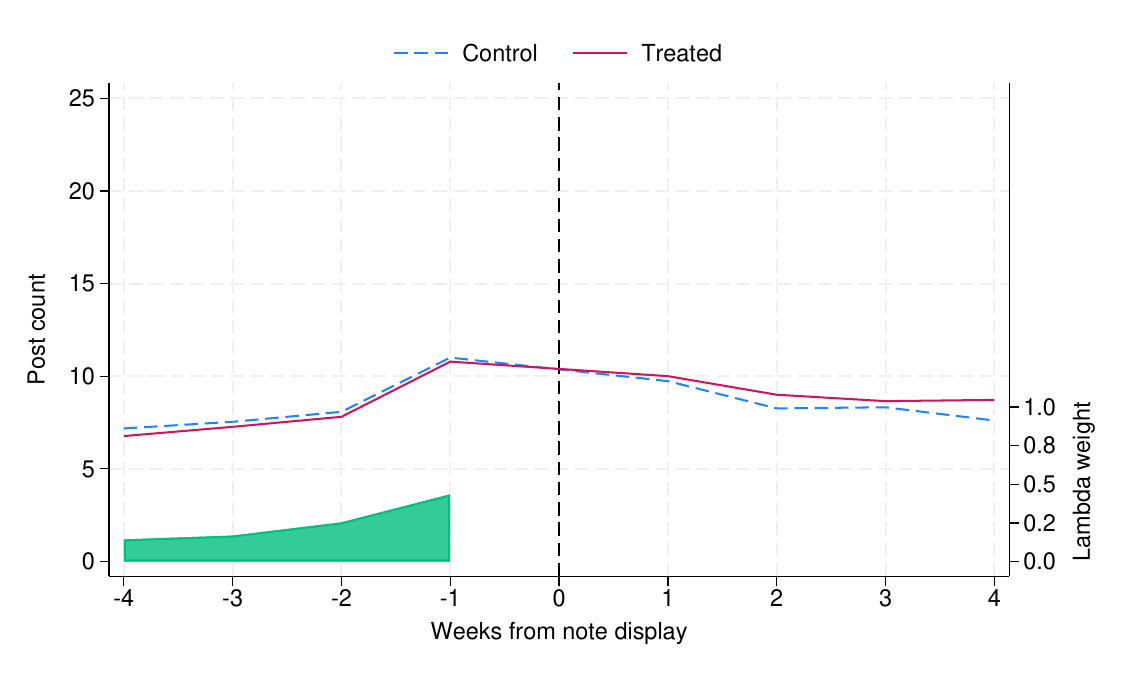}}
    \caption{\textbf{SDiD estimates of changes in the number of misleading posts by political content and toxicity.} The analysis is restricted to accounts with average GPT-estimated pre-correction misleadingness scores at or above the 75th percentile.
    \textbf{(a)}~SDiD estimation for overall misleading posts.
    \textbf{(b)}~SDiD estimation for political or highly toxic misleading posts.
    \textbf{(c)}~SDiD estimation for non-political and low-toxicity misleading posts.
    Solid red and dashed blue lines indicate the average weekly original-post counts for treated and control accounts, respectively. The green area indicates the estimated pre-treatment time weights ($\lambda$). Standard errors are estimated using 100 bootstrap replications with clustering at the account level.}
    \label{fig:sdid_posts_weekly_content_mis_combined}
\end{figure}

Additionally, the estimation results for the changes in misleading-post production and composition by political content and toxicity are reported in \Cref{tab:did_content_high_misleadingness_combined}. Moreover, we conduct complementary estimation using SDiD estimator and find results consistent with our main DiD findings (\Cref{fig:sdid_posts_weekly_content_mis_combined}).

These results suggest that the increase in posting activity among accounts with high pre-correction misleadingness is accompanied by a greater production of misleading posts, but without a detectable shift toward a higher proportion of misleading content.

\clearpage

\clearpage
\begin{table*}[ht]
\centering
\caption{\textbf{Aggregated DiD estimates of changes in content production and composition among accounts with high pre-correction misleadingness.} The analysis is restricted to accounts with average GPT-estimated pre-correction misleadingness scores at or above the 75th percentile. The table reports test statistics $z$ for count outcomes and $t$ for ratio outcomes. Event and week fixed effects are absorbed, and standard errors are clustered at the account level.}
\setlength{\tabcolsep}{4pt}
\begin{tabular}{@{}llrrrcrrr@{}}
\toprule
Content metric & Outcome & Estimate & Statistic & {$p$} & 95\% CI & Obs. & Events & Accounts \\
\midrule
Misleadingness
  & Yes & $0.083$ & $2.203$ & $0.028$ & $[0.009, 0.162]$ & 14,760 & 1,845 & 1,656 \\
  & No & $0.048$ & $1.371$ & $0.170$ & $[-0.020, 0.121]$ & 14,880 & 1,860 & 1,671 \\
  & Ratio & $-0.008$ & $-0.855$ & $0.393$ & $[-0.027, 0.011]$ & 11,606 & 1,758 & 1,584 \\
\addlinespace
Positive
  & Yes & $0.040$ & $0.962$ & $0.336$ & $[-0.040, 0.128]$ & 14,456 & 1,807 & 1,618 \\
  & No & $0.067$ & $1.888$ & $0.059$ & $[-0.002, 0.140]$ & 14,888 & 1,861 & 1,672 \\
  & Ratio & $-0.004$ & $-0.659$ & $0.510$ & $[-0.017, 0.009]$ & 13,639 & 1,853 & 1,666 \\
\addlinespace
Negative
  & Yes & $0.058$ & $1.526$ & $0.127$ & $[-0.016, 0.136]$ & 14,784 & 1,848 & 1,659 \\
  & No & $0.070$ & $1.948$ & $0.051$ & $[-0.000, 0.145]$ & 14,848 & 1,856 & 1,667 \\
  & Ratio & $-0.006$ & $-0.648$ & $0.517$ & $[-0.026, 0.013]$ & 10,978 & 1,725 & 1,553 \\
\addlinespace
Toxicity
  & Yes & $0.004$ & $0.079$ & $0.937$ & $[-0.098, 0.119]$ & 12,928 & 1,616 & 1,440 \\
  & No & $0.070$ & $2.059$ & $0.039$ & $[0.003, 0.141]$ & 14,888 & 1,861 & 1,672 \\
  & Ratio & $-0.008$ & $-1.465$ & $0.143$ & $[-0.019, 0.003]$ & 13,630 & 1,848 & 1,659 \\
\addlinespace
Confidence
  & Yes & $0.053$ & $1.461$ & $0.144$ & $[-0.018, 0.130]$ & 14,776 & 1,847 & 1,658 \\
  & No & $0.058$ & $1.724$ & $0.085$ & $[-0.008, 0.129]$ & 14,752 & 1,844 & 1,655 \\
  & Ratio & $-0.013$ & $-1.622$ & $0.105$ & $[-0.028, 0.003]$ & 13,091 & 1,838 & 1,653 \\
\addlinespace
Politics
  & Yes & $0.039$ & $1.000$ & $0.317$ & $[-0.036, 0.121]$ & 14,136 & 1,767 & 1,582 \\
  & No & $0.078$ & $1.996$ & $0.046$ & $[0.001, 0.160]$ & 14,848 & 1,856 & 1,667 \\
  & Ratio & $-0.006$ & $-0.696$ & $0.486$ & $[-0.023, 0.011]$ & 12,162 & 1,786 & 1,611 \\
\bottomrule
\end{tabular}
\label{tab:did_content_high_misleadingness}
\end{table*}

\clearpage
\begin{table}[ht]
\centering
\caption{\textbf{Aggregated DiD estimates of changes in misleading-post production and composition by political content and toxicity.}
The analysis is restricted to accounts with average GPT-estimated pre-correction misleadingness scores at or above the 75th percentile.
The table reports test statistics $z$ for count outcomes and $t$ for ratio outcomes.
Event and week fixed effects are absorbed, and standard errors are clustered at the account level.}
\setlength{\tabcolsep}{5pt}
\begin{tabular}{@{}lrrrrrcrrr@{}}
\toprule
Outcome & Estimate & Statistic & {$p$} & $p_{\mathrm{Holm}}$ & $p_{\mathrm{BH}}$ & 95\% CI & Obs. & Events & Accounts \\
\midrule
\multicolumn{10}{@{}l}{\underline{Misleading posts}}\\
Yes & $0.083$ & $2.203$ & $0.028$ & $0.140$ & $0.084$ & $[0.009, 0.162]$ & 14,760 & 1,845 & 1,656 \\
No & $0.048$ & $1.371$ & $0.170$ & $0.548$ & $0.255$ & $[-0.020, 0.121]$ & 14,880 & 1,860 & 1,671 \\
Ratio & $-0.008$ & $-0.855$ & $0.393$ && & $[-0.027, 0.011]$ & 11,606 & 1,758 & 1,584 \\
\addlinespace
\multicolumn{10}{@{}l}{\underline{Misleading posts (political or highly toxic)}}\\
Yes & $0.043$ & $1.020$ & $0.308$ & $0.616$  &   $0.370$ & $[-0.038, 0.129]$ & 13,856 & 1,732 & 1,549 \\
No & $0.029$ & $0.736$ & $0.462$ & $0.616$  &   $0.462$ & $[-0.047, 0.112]$ & 14,336 & 1,792 & 1,604 \\
Ratio & $-0.010$ & $-1.506$ & $0.132$ && & $[-0.024, 0.003]$ & 13,202 & 1,835 & 1,650 \\
\addlinespace
\multicolumn{10}{@{}l}{\underline{Misleading posts (non-political and low-toxicity)}}\\
Yes & $0.119$ & $2.770$ & $0.006$ & $0.036$  &   $0.036$ & $[0.033, 0.211]$ & 14,456 & 1,807 & 1,622 \\
No & $0.060$ & $1.487$ & $0.137$ & $0.548$  &   $0.255$ & $[-0.018, 0.144]$ & 14,824 & 1,853 & 1,664 \\
Ratio & $0.000$ & $0.057$ & $0.954$ && & $[-0.014, 0.015]$ & 13,026 & 1,830 & 1,645 \\
\bottomrule
\end{tabular}
\label{tab:did_content_high_misleadingness_combined}
\end{table}

\clearpage

\clearpage
\subsection{Heterogeneity by accounts with repeated prior NMR notes}
\label{supp:content_he_high_nmr}

As a complementary exploratory analysis, we examine changes in the number of category-specific posts and their compositions relative to the total original posts during the same period among accounts with repeated prior NMR notes. We additionally assess the misleadingness of \num{1138062} original posts from these accounts and extend the observation window to four weeks before and after correction. The estimation results are reported in \Cref{tab:did_content_high_nmr}. These results suggest that accounts with repeated prior NMR notes may also produce more misleading posts following correction, but we find no statistically significant change in content composition.

\begin{table*}[ht]
\centering
\caption{\textbf{Aggregated DiD estimates of changes in content production and composition among accounts with repeated prior NMR notes.} The analysis is restricted to accounts with NMR notes on at least two distinct posts during the 28 days preceding their first observed correction. The table reports test statistics $z$ for count outcomes and $t$ for ratio outcomes. Event and week fixed effects are absorbed, and standard errors are clustered at the account level.}
\setlength{\tabcolsep}{4pt}
\begin{tabular}{@{}llrrrcrrr@{}}
\toprule
Content metric & Outcome & Estimate & Statistic & {$p$} & 95\% CI & Obs. & Events & Accounts \\
\midrule
Misleadingness
  & Yes & $0.092$ & $2.041$ & $0.041$ & $[0.003, 0.187]$ & 20,184 & 2,523 & 1,926 \\
  & No & $0.069$ & $1.625$ & $0.104$ & $[-0.014, 0.158]$ & 21,392 & 2,674 & 2,031 \\
  & Ratio & $-0.006$ & $-1.067$ & $0.286$ & $[-0.017, 0.005]$ & 24,992 & 3,363 & 2,484 \\
\addlinespace
Positive
  & Yes & $0.036$ & $0.850$ & $0.395$ & $[-0.045, 0.122]$ & 20,448 & 2,556 & 1,957 \\
  & No & $0.084$ & $1.947$ & $0.052$ & $[-0.001, 0.175]$ & 21,312 & 2,664 & 2,021 \\
  & Ratio & $0.001$ & $0.343$ & $0.732$ & $[-0.007, 0.010]$ & 25,921 & 3,390 & 2,506 \\
\addlinespace
Negative
  & Yes & $0.081$ & $1.865$ & $0.062$ & $[-0.004, 0.172]$ & 20,536 & 2,567 & 1,953 \\
  & No & $0.072$ & $1.720$ & $0.085$ & $[-0.010, 0.161]$ & 21,352 & 2,669 & 2,031 \\
  & Ratio & $-0.002$ & $-0.350$ & $0.726$ & $[-0.015, 0.010]$ & 24,058 & 3,328 & 2,459 \\
\addlinespace
Toxicity
  & Yes & $0.063$ & $1.080$ & $0.280$ & $[-0.048, 0.186]$ & 16,296 & 2,037 & 1,563 \\
  & No & $0.077$ & $1.877$ & $0.061$ & $[-0.003, 0.163]$ & 21,456 & 2,682 & 2,039 \\
  & Ratio & $-0.002$ & $-0.758$ & $0.449$ & $[-0.008, 0.003]$ & 26,109 & 3,396 & 2,507 \\
\addlinespace
Confidence
  & Yes & $0.065$ & $1.596$ & $0.110$ & $[-0.014, 0.149]$ & 20,616 & 2,577 & 1,969 \\
  & No & $0.072$ & $1.743$ & $0.081$ & $[-0.009, 0.160]$ & 20,528 & 2,566 & 1,958 \\
  & Ratio & $-0.003$ & $-0.554$ & $0.580$ & $[-0.013, 0.007]$ & 25,701 & 3,386 & 2,501 \\
\addlinespace
Politics
  & Yes & $0.047$ & $0.905$ & $0.366$ & $[-0.052, 0.156]$ & 18,688 & 2,336 & 1,749 \\
  & No & $0.086$ & $2.061$ & $0.039$ & $[0.004, 0.175]$ & 21,264 & 2,658 & 2,025 \\
  & Ratio & $-0.006$ & $-1.033$ & $0.301$ & $[-0.016, 0.005]$ & 24,888 & 3,363 & 2,485 \\
\bottomrule
\end{tabular}
\label{tab:did_content_high_nmr}
\end{table*}

\clearpage
\section{Changes in post engagement}
\label{supp:post_engagement}

Beyond examining changes in posting activity and content characteristics, we investigate whether the original posts subsequently published by corrected accounts receive different levels of audience engagement. Specifically, we consider two engagement metrics: the number of reposts and replies received by each original post. Using the original-post dataset, we calculate the average number of reposts and replies per post for each account-event and week, rounding the resulting averages to the nearest integer. We then estimate weekly leads-and-lags and aggregated DiD models using Poisson pseudo-maximum likelihood estimation, with correction-event and week fixed effects and standard errors clustered at the account level. The analysis is based on the matched sample (see \Cref{supp:overall_content}) and covers four weeks before and after the first correction, using Week~$-1$ as the reference period.

\vspace{1.5em}
\noindent \textbf{Repost engagement.}\\
We first examine changes in the average number of reposts received by original posts. The estimation includes \num{55373} weekly observations across \num{6330} accounts. The pre-correction DiD estimates for Weeks~$-4$ to~$-2$ are statistically indistinguishable from zero (all $p>0.100$). Following correction, the estimate for Week~1 is negative and statistically significant (coef. $=-0.791$, $z=-2.500$, $p=0.012$; 95\%~CI: $[-1.411, -0.171]$), suggesting a short-term reduction in repost engagement with subsequently published original posts. However, the estimates for Weeks~2--4 are positive but statistically insignificant (all $p>0.100$). The aggregated DiD estimate for the post-correction period is negative but not statistically significant (coef. $=-0.250$, $z=-1.460$, $p=0.144$; 95\%~CI: $[-0.585, 0.086]$). These findings provide evidence of reduced repost engagement during the first week following correction, but no statistically significant evidence of an overall change across the four-week post-correction period.

\vspace{1.5em}
\noindent \textbf{Reply engagement.}\\
We next examine changes in the average number of replies received by original posts. The estimation includes \num{55530} weekly observations across \num{6349} accounts. The pre-correction DiD estimates for Weeks~$-4$ to~$-2$ are statistically indistinguishable from zero (all $p>0.100$). Following correction, reply engagement declines significantly in Week~1 (coef. $=-0.597$, $z=-2.168$, $p=0.030$; 95\%~CI: $[-1.137, -0.057]$), whereas the estimates for Weeks~2--4 are not statistically significant (all $p>0.100$). Unlike repost engagement, the aggregated post-correction estimate remains negative and statistically significant (coef. $=-0.215$, $z=-2.060$, $p=0.039$; 95\%~CI: $[-0.419, -0.010]$). These findings suggest that original posts published by corrected accounts receive lower reply engagement on average following their first correction, although the statistically detectable weekly difference is concentrated in the first post-correction week.

Together, these results suggest that community corrections may be associated with changes in audience engagement with corrected authors' subsequent posts, even when their overall posting activity remains largely unchanged.

\begin{figure}[ht]
    \centering
    \captionsetup[subfloat]{font={bf, small}, skip=0pt, singlelinecheck=false, labelformat=simple, position=top}
    \subfloat[]{\includegraphics[width = .32\textwidth]{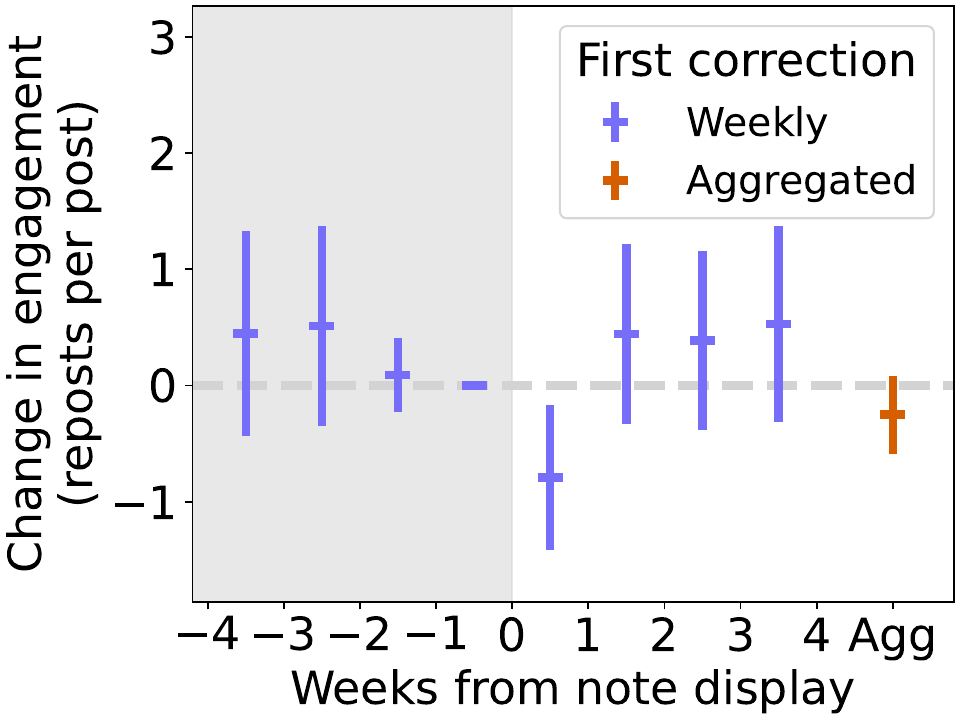}}
    \hspace{1cm}
    \subfloat[]{\includegraphics[width = .32\textwidth]{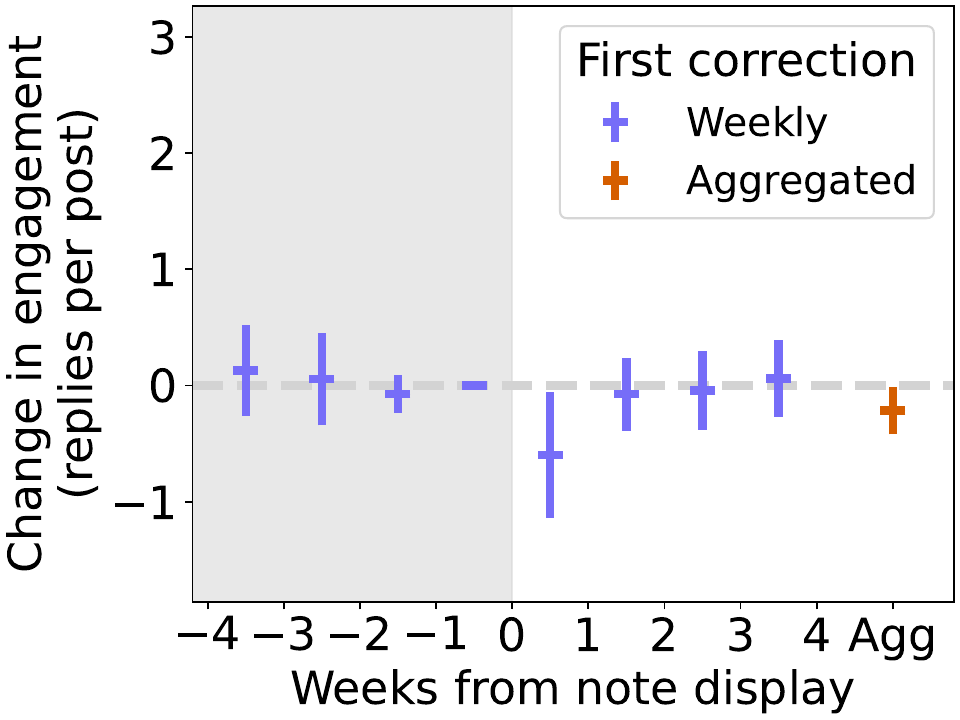}}
    \caption{\textbf{Weekly DiD estimates of audience engagement with original posts published by corrected accounts following their first observed correction.} The estimated coefficients for engagement outcomes are reported without transformation.
    \textbf{(a)}~Weekly and aggregated DiD estimates for the average number of reposts received per original post. The estimation includes \num{55373} weekly observations across \num{7245} correction events contributed by \num{6330} accounts.
    \textbf{(b)}~Weekly and aggregated DiD estimates for the average number of replies received per original post. The estimation includes \num{55530} weekly observations across \num{7264} correction events contributed by \num{6349} accounts.
    Week~$-1$ serves as the reference period. Error bars indicate 95\% CIs, and standard errors are clustered at the account level.}
    \label{fig:did_post_engagement}
\end{figure}

\clearpage
\begin{table*}[ht]
\centering
\caption{\textbf{Weekly leads-and-lags DiD estimates of changes in repost engagement with original posts published by corrected accounts.} Week~$-1$ serves as the reference period. Repost engagement is measured using the average number of reposts received per original post within each account-event and week. The estimation is based on \num{55373} weekly observations across \num{7245} correction events contributed by \num{6330} accounts. Event and week fixed effects are absorbed, and standard errors are clustered at the account level.}
\setlength{\tabcolsep}{4pt}
\begin{tabular}{@{}rrrrcrrrrc@{}}
\toprule
\multicolumn{5}{c}{Pre-correction} &
\multicolumn{5}{c}{Post-correction} \\
\cmidrule(lr){1-5}\cmidrule(lr){6-10}
Week & Estimate & {$z$} & {$p$} & 95\% CI &
Week & Estimate & {$z$} & {$p$} & 95\% CI \\
\midrule
$-4$ & $0.447$ & $0.991$ & $0.321$ & $[-0.436, 1.330]$ &
$1$ & $-0.791$ & $-2.500$ & $0.012$ & $[-1.411, -0.171]$ \\

$-3$ & $0.512$ & $1.165$ & $0.244$ & $[-0.349, 1.372]$ &
$2$ & $0.445$ & $1.129$ & $0.259$ & $[-0.328, 1.218]$ \\

$-2$ & $0.091$ & $0.566$ & $0.572$ & $[-0.225, 0.408]$ &
$3$ & $0.387$ & $0.985$ & $0.325$ & $[-0.383, 1.158]$ \\

& & & & &
$4$ & $0.528$ & $1.227$ & $0.220$ & $[-0.316, 1.371]$ \\
\bottomrule
\end{tabular}
\label{tab:did_weekly_reposts}
\end{table*}

\begin{table*}[ht]
\centering
\caption{\textbf{Weekly leads-and-lags DiD estimates of changes in reply engagement with original posts published by corrected accounts.} Week~$-1$ serves as the reference period. Reply engagement is measured using the average number of replies received per original post within each account-event and week. The estimation is based on \num{55530} weekly observations across \num{7264} correction events contributed by \num{6349} accounts. Event and week fixed effects are absorbed, and standard errors are clustered at the account level.}
\setlength{\tabcolsep}{4pt}
\begin{tabular}{@{}rrrrcrrrrc@{}}
\toprule
\multicolumn{5}{c}{Pre-correction} &
\multicolumn{5}{c}{Post-correction} \\
\cmidrule(lr){1-5}\cmidrule(lr){6-10}
Week & Estimate & {$z$} & {$p$} & 95\% CI &
Week & Estimate & {$z$} & {$p$} & 95\% CI \\
\midrule
$-4$ & $0.131$ & $0.660$ & $0.509$ & $[-0.258, 0.519]$ &
$1$ & $-0.597$ & $-2.168$ & $0.030$ & $[-1.137, -0.057]$ \\

$-3$ & $0.057$ & $0.281$ & $0.779$ & $[-0.342, 0.456]$ &
$2$ & $-0.073$ & $-0.458$ & $0.647$ & $[-0.387, 0.240]$ \\

$-2$ & $-0.073$ & $-0.882$ & $0.378$ & $[-0.235, 0.089]$ &
$3$ & $-0.042$ & $-0.246$ & $0.806$ & $[-0.380, 0.295]$ \\

& & & & &
$4$ & $0.059$ & $0.353$ & $0.724$ & $[-0.270, 0.389]$ \\
\bottomrule
\end{tabular}
\label{tab:did_weekly_replies}
\end{table*}

\end{document}